\documentclass[aps,prl,reprint,showpacs,longbibliography,floatfix,footnoteinbib]{revtex4-1}
\usepackage[latin9]{inputenc}
\usepackage{xcolor}
\usepackage{float}
\usepackage{bm}
\usepackage{amsmath}
\usepackage{amssymb}
\usepackage{graphicx}
\usepackage[bookmarks=false,
 breaklinks=false,pdfborder={0 0 1},backref=false,colorlinks=true]
 {hyperref}
\hypersetup{pdftitle={Title},
 plainpages=false,pdfpagelabels,linkcolor=blue,urlcolor=blue,citecolor=blue,pdfdisplaydoctitle=true,pdfduplex=DuplexFlipLongEdge}

\makeatletter

\newcommand{\lyxdot}{.}

\usepackage{units}\usepackage{wasysym}

\definecolor{orange}{rgb}{0.50, 0.20, 0.0}

\usepackage{bm}
\usepackage{braket}

\makeatother

\begin{document}
\noindent\begin{minipage}[t]{1\columnwidth}%
\global\long\def\ket#1{\left| #1\right\rangle }%

\global\long\def\bra#1{\left\langle #1 \right|}%

\global\long\def\kket#1{\left\Vert #1\right\rangle }%

\global\long\def\bbra#1{\left\langle #1\right\Vert }%

\global\long\def\braket#1#2{\left\langle #1\right. \left| #2 \right\rangle }%

\global\long\def\bbrakket#1#2{\left\langle #1\right. \left\Vert #2\right\rangle }%

\global\long\def\av#1{\left\langle #1 \right\rangle }%

\global\long\def\tr{\text{tr}}%

\global\long\def\Tr{\text{Tr}}%

\global\long\def\pd{\partial}%

\global\long\def\im{\text{Im}}%

\global\long\def\re{\text{Re}}%

\global\long\def\sgn{\text{sgn}}%

\global\long\def\Det{\text{Det}}%

\global\long\def\abs#1{\left|#1\right|}%

\global\long\def\up{\uparrow}%

\global\long\def\down{\downarrow}%

\global\long\def\vc#1{\mathbf{#1}}%

\global\long\def\bs#1{\boldsymbol{#1}}%

\global\long\def\t#1{\text{#1}}%
\end{minipage}
\title{Quasiperiodic Quadrupole Insulators}
\author{Raul Liquito$^{1}$, Miguel Gonçalves$^{1,2}$, Eduardo V. Castro$^{1,3}$}
\affiliation{$^{1}$Centro de F\'{\i}sica da Universidade do Porto, Departamento
de F\'{\i}sica e Astronomia, Faculdade de Ciências, Universidade do
Porto, 4169-007 Porto, Portugal}
\affiliation{$^{2}$CeFEMA, Instituto Superior Técnico, Universidade de Lisboa,
Av. Rovisco Pais, 1049-001 Lisboa, Portugal}
\affiliation{$^{3}$Beijing Computational Science and Research Center, Beijing
100084, China}
\begin{abstract}
Higher-order topological insulators are an intriguing new family of
topological states that host lower-dimensional boundary states. Concurrently,
quasiperiodic systems have garnered significant interest due to their
complex localization and topological properties. In this work we study
the impact of chiral symmetry preserving quasiperiodic modulations
on the paradigmatic Benalcazar-Bernevig-Hughes model, which hosts
topological insulating phases with zero-energy sublattice-polarized
modes. We find that the topological properties are not only robust
to the quasiperiodic modulation, but can even be enriched. In particular,
we unveil the first instance of a quasiperiodic induced second-order
topological insulating phase. Furthermore, in contrast with disorder,
we find that quasiperiodic modulations can induce multiple reentrant
topological transitions, showing an intricate sequence of localization
properties. Our results open a promising avenue for exploring the
rich interplay between higher-order topology and quasiperiodicity.
\end{abstract}
\maketitle

\section{Introduction}

\label{sec:intro}

Higher order topological insulators (HOTIs) have been established
as an intriguing novel topological state of matter. Unlike the more
conventional first order topological insulators, for which the bulk-boundary
correspondence guarantees the existence of $D-1$ dimensional boundary
states, HOTIs are characterized by boundary states with lower dimensionality.
In particular, a $m^{\text{th}}$ order topological insulator has
$D-m$ dimensional gapless boundary modes for a $D$-dimensional system
\citep{Benalcazar2017,benalcazar2017_prb}. These gapless boundary
modes are a manifestation of the boundary non-trivial topology and
can only be removed through a closure of the bulk or edge energy gap
\citep{Benalcazar2017,PhysRevResearch.3.013239,tao2023quadrupole,benalcazar2017_prb}.
The stability of these gapless boundary modes requires the existence
of local symmetries such as time reversal, particle-hole and chiral
symmetry, or crystalline symmetries . Since their proposal several
higher order topological phases have been characterized and observed
in insulators \citep{Benalcazar2017,Ghosh2020,Schindler2018,Neupert2018,Wang2021,Xu2017,Ezawa2018},
semimetals \citep{Lin2018}, superconductors \citep{PhysRevB.97.205136,PhysRevB.102.094503}
and fractal lattices \citep{PhysRevB.105.L201301}. Considerable efforts
have been made to unveil the symmetries behind the protection of these
topological phases and to understand their robustness against weak
disorder. HOTIs have been experimentally realized in phononic metamaterials
\citep{10.1038/nature25156}, electric circuits \citep{PhysRevB.100.201406,Zhang2021}
and even real solid-state materials \citep{Schindler2018Bismuth,Shumiya2022,Lee2023}.
Disordered HOTIs have been studied \citep{hugo2023,PhysRevResearch.2.033521,PhysRevLett.126.206404,Peng2021}
and a classical analog has been experimentally realized in electric
circuits \citep{Zhang2021}.

A different class of systems that break translational invariance,
where more exotic localization properties can occur, are quasiperiodic
systems. Contrary to disordered systems, extended, localized and critical
multifractal phases can arise even in one dimension (1D) \citep{PhysRevB.43.13468,PhysRevLett.104.070601,PhysRevLett.113.236403,Liu2015,PhysRevB.91.235134,PhysRevLett.114.146601,10.21468/SciPostPhys.13.3.046,anomScipost,PhysRevB.108.L100201,PhysRevLett.131.186303}.
In higher dimensions, these systems have received considerable attention,
including on the interplay between moiré physics and localization
\citep{Huang2016a,PhysRevLett.120.207604,Park2018,PhysRevX.7.041047,PhysRevB.100.144202,Fu2020,PhysRevB.101.235121,Goncalves_2022_2DMat}.
quasiperiodic systems are also known to display intrinsic topological
properties and associated edge physics characterized by topological
invariants defined in higher dimensions \citep{Kraus2012,PhysRevB.91.245104,Zilberberg21,PhysRevLett.111.226401,PhysRevB.101.041112,PhysRevResearch.4.013028}.
Systems with quasiperiodic modulations can be realized in widely different
platforms, including optical lattices \citep{PhysRevA.75.063404,Roati2008,Modugno_2009,Schreiber842,Luschen2018,PhysRevLett.123.070405,PhysRevLett.125.060401,PhysRevLett.126.110401,PhysRevLett.126.040603,PhysRevLett.122.170403},
photonic \citep{Lahini2009,Kraus2012,Verbin2013,PhysRevB.91.064201,Wang2020,https://doi.org/10.1002/adom.202001170,Wang2022}
and phononic \citep{PhysRevLett.122.095501,Ni2019,PhysRevLett.125.224301,PhysRevApplied.13.014023,PhysRevX.11.011016,doi:10.1063/5.0013528}
metamaterials, and more recently using moiré materials \citep{Balents2020,Andrei2020,uri2023superconductivity}.
The impact of quasiperiodic modulations on parent first-order topological
systems has also been previously studied \citep{Fu_2021,cheng2022_december,PhysRevB.106.224505,goncalves2022topological,liquito2023fate},
and was found to give rise to interesting topological phases with
richer localization properties than in the disordered cases. The effects
of quasiperiodicity in higher order topolgical systems remains poorly
unexplored. Higher order topological phases have been shown to appear
in quasicrystals with five-fold, eight-fold, and twelve-fold rotational
symmetries without crystalline counterparts \citep{PhysRevLett.124.036803,acs.nanolett.1c02661,PhysRevB.102.241102,PhysRevLett.123.196401},
and low energy descriptions are available for higher order topological
quasicrystals \citep{PhysRevLett.129.056403,PhysRevResearch.2.033071}.
Nonetheless, the focus of these works as been on the role of rotation
symmetries without crystalline counterparts in higher order topological
phases.

\begin{figure}[h]
\centering{}\includegraphics[width=0.5\columnwidth]{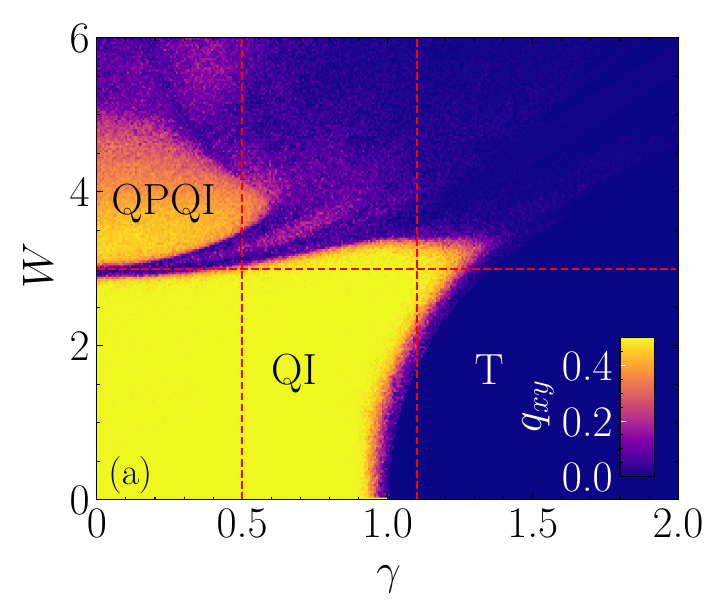}\includegraphics[width=0.5\columnwidth]{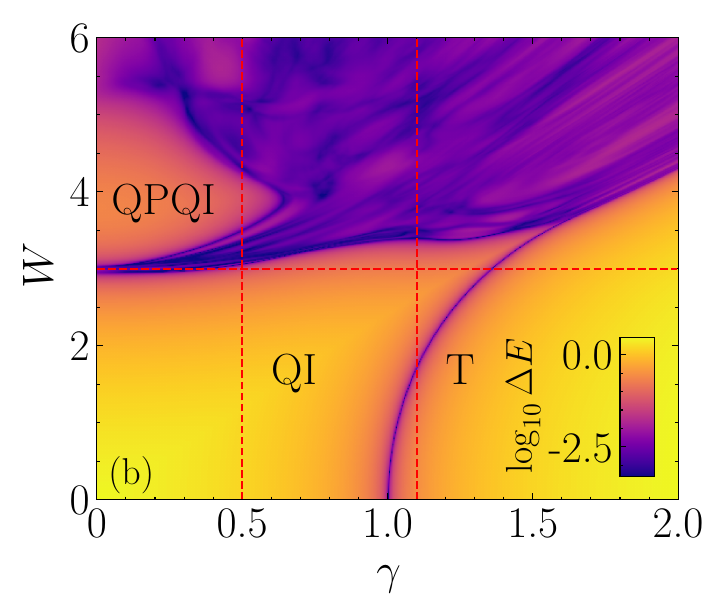}\caption{Phase diagrams in the plane $(\gamma,W)$. (a) The quadrupole moment
was computed for a system size $L=13$ with $\theta_{i}=\pi$ and
$100$ averages over phase shifts were realized. (b) Spectral gap
($\Delta E$) in the $(\gamma,W)$ plane for a system size $L=34$
with $\theta_{i}=0$ and $50$ realizations of phase shifts. A minimum
value of the spectral gap was chosen ($\Delta E<10^{-3}$) to highlight
all the gapped and gapless phases appearing in the phase diagram.
The dotted white lines are constant $\gamma$ and $W$ cuts for $\gamma=0.5,1.1$
and $W=3$ respectively.\label{fig:phase-diagram}}
\end{figure}

In this work we study a different class of non crystalline system,
a quasiperiodic (QP) chiral quadrupole insulator presenting a full
characterization of topological, spectral and localization properties.
The main results are shown in Fig.~\ref{fig:phase-diagram}, where
we plot the quadrupole moment $q_{xy}$ on {[}\ref{fig:phase-diagram}(a){]}
and the bulk energy gap {[}\ref{fig:phase-diagram}(b){]} in the plane
of intracell hoppings strength, $\gamma$, and QP modulation, $W$.
As seen in Fig.~\ref{fig:phase-diagram}(a), the clean limit quadrupole
insulator (QI) and the trivial insulator (T) are robust to QP modulations.
Starting from the clean limit trivial regime ($\gamma>1$), these
modulations may induce a topological phase transition (TPT) into a
QI phase adiabatically connected to the clean limit QI for $\gamma\gtrsim1$.
For the topological case ($\gamma<1$), the QP modulations eventually
induce a TPT into a gapless critical metal. Further increasing the
strength of the QP modulation leads to a novel reentrant topological
regime which we entitle quasiperiodic QI (QPQI). The QPQI phase display
similar boundary signatures as their clean limit counterparts, such
as four fold zero energy corner modes that give rise to fractional
corner charges. However, in this regime an intricate interplay between
QP induced edge states and corner modes emerges, which can lead to
edge-corner hybridization. This is a unique feature of the QPQI phase.
Overall, the observed TPT between the various regimes occur in gap
closing/opening as seen in Fig.~\ref{fig:phase-diagram}(b).

This paper is organized as follows: in Sec.~\ref{sec:model_methods}
we introduce the quasiperiodic Benalcazar-Bernevig-Hughes model and
present the methods used to numerically study the diverse phases that
emerge under QP modulations; in Sec.~\ref{sec:results} we display
and discuss our results in a organized fashion; in Sec.~\ref{sec:Conclusion}
we discuss and reinforce our main conclusions; in Appendix~\ref{Sec:DOS-edge}
we show calculations of the density of states (DOS) obtained via kernel
polynomial methods (KPM) for each relevant observed phases and in
Appendix.~\ref{Sec:constat_W_cut} we perform the same analysis done
throughout this work but for a constant QP modulation strength ($W$)
cut (varying the inter unit cell hopping $\gamma$); in Appendix~\ref{sec:Phase-Twist-and-spectral-gap}
some remarks are made regarding phase twists; and in Appendix.~\ref{sec:phase shift dependences}
we expand on the results of phase shifts dependence in the QPQI phase.

\section{Model and Methods}

\label{sec:model_methods}

\subsubsection{Model}

We consider the BBH model \citep{Benalcazar2017} with a QP modulation
defined by the following Hamiltonian:

\begin{equation}
\mathcal{H}=\sum_{\mathbf{R}}\left(\mathbf{\Psi}_{\mathbf{R}}^{\dagger}\Delta_{\mathbf{R}}\mathbf{\Psi}_{\mathbf{R}}+\sum_{i\in\{x,y\}}\left(\mathbf{\Psi}_{\mathbf{R}+\mathbf{\hat{e}}_{i}}^{\dagger}\Lambda_{i}\mathbf{\Psi}_{\mathbf{R}}+h.c.\right)\right),
\end{equation}
where $\mathbf{\Psi}_{\mathbf{R}}^{\dagger}=\left(\begin{array}{cc}
c_{\mathbf{R},1}^{\dagger} & c_{\mathbf{R},2}^{\dagger}\end{array},c_{\mathbf{R},3}^{\dagger},c_{\mathbf{R},4}^{\dagger}\right)$ and $c_{\mathbf{R},\alpha}^{\dagger}$ creates an electron on the
orbital $\alpha$ at site $\mathbf{R}=(x,y)$. The model is defined
by the hopping matrices:

\begin{align}
\Delta_{\mathbf{R}} & =\delta\sigma_{z}\otimes\sigma_{0}+\gamma_{\mathbf{R}}\left(\sigma_{x}\otimes\sigma_{0}-\sigma_{y}\otimes\sigma_{y}\right)\nonumber \\
\Lambda_{x} & =\frac{\lambda}{2}\left(\sigma_{x}\otimes\sigma_{0}-\sigma_{y}\otimes\sigma_{y}\right)\\
\Lambda_{y} & =i\frac{\lambda}{2}\sigma_{y}\otimes\left(\sigma_{x}+\sigma_{z}\right),\nonumber 
\end{align}
where $\left\{ \sigma_{i}\right\} $ are the set of Pauli matrices,
and $\delta$ is a staggered mass term in each sublattice which opens
a gap proportional to $\delta$\footnote{In calculations with OBC we set the $\delta\approx10^{-5}$ to break
the degeneracy of the corner states.}. Throughout this paper we set $\lambda=1$ such that $\gamma$ is
displayed in units of $\lambda$. The QP modulation is introduced
on the inter UC hopping's, 
\begin{equation}
\gamma_{\mathbf{R}}=\gamma+\text{\ensuremath{\frac{W}{2}\cos\left(2\pi\beta x+\phi_{x}\right)}\ensuremath{\cos\left(2\pi\beta y+\phi_{y}\right)}},\label{eq:modHopp}
\end{equation}
where $\beta$ is an irrational number. This choice of QP modulation
ensures that chiral symmetry remains preserved for any QP modulation
strength ($W$). Furthermore, this model also preserves time reversal
and charge conjugation symmetries. In periodic finite systems, a generic
phase shift $\bm{\phi}=(\phi_{x}\mod{2\pi},\phi_{y}\mod{2\pi})$,
 breaks the mirror and $C_{4}$ crystalline symmetries. However, for
a set of phase shifts $\phi_{i}\in\left\{ \pi\beta(2m_{i}+1)\mod{2\pi}|m_{i}\in\mathbb{Z}\right\} $
(for $L$ even) or $\phi_{i}\in\left\{ 2\pi\beta m_{i}\mod{2\pi}|m_{i}\in\mathbb{Z}\right\} $
(for $L$ odd) the potential is shifted by $\mathbf{R}=m_{x}\hat{\mathbf{e}}_{x}+m_{y}\hat{\mathbf{e}}_{y}$,
and the original crystalline symmetries are preserved. .  Nonetheless,
in the thermodynamic limit the mirror, inversion, and $C_{4}$ symmetries
are recovered for a generic $\bm{\phi}=(\phi_{x},\phi_{y})$. In contrast,
finite open systems only retain said symmetries around the center
of the lattice. In a finite open quasiperiodic system, the crystalline
symmetries are only compatible with the potential for a single choice
of $\boldsymbol{\phi}=\left(\phi_{x},\phi_{y}\right)$, which aligns
the geometric centers of the lattice and the potential \footnote{For a lattice with geometric center at $(m_{x},m_{y})$, symmetry
is retained for $\left(\phi_{x},\phi_{y}\right)=(\pi\beta(2m_{x}+1),\pi\beta(2m_{y}+1))\mod{2\pi}$
(for even $L$), $\left(\phi_{x},\phi_{y}\right)=(2\pi\beta m_{x},2\pi\beta m_{y})\mod{2\pi}$
(for odd $L$).}.  

The BBH model can be interpreted as having fractional corner charges
protected by mirror symmetries and no chiral symmetry or sublattice-polarized
zero modes protected by chiral symmetry. In this work, we choose the
latter interpretation since chiral symmetry is the only preserved
symmetry for a general choice of potential parameters and open or
periodic boundaries.

We carried out numerical simulations for finite systems with $L_{x}=L_{y}=L$
(with $L$ the number of unit cells in each direction) and periodic/twisted
boundary conditions. In order to avoid boundary defects the system
sizes were chosen to be $L=F_{n}$, where $F_{n}$ is the $n$-th
order Fibonacci number. In Eq.~\eqref{eq:modHopp}, $\beta$ was
taken as a rational approximant of the golden ratio $\beta\rightarrow\beta_{n}=F_{n+1}/F_{n}$.
This choice ensures that the system's unit cell is of size $L$, guaranteeing
that the system remains incommensurate as $L$ increases.

In finite systems the introduction of phase shifts $\boldsymbol{\phi}=\left(\phi_{x},\phi_{y}\right)$
and phase twists $\boldsymbol{\theta}=(\theta_{x},\theta_{y})$ can
alter the energy spectra of the system. The dependence of the energy
spectra on $\boldsymbol{\phi}$ and $\boldsymbol{\theta}$ has a direct
correlation with the localization properties of the corresponding
eigenfunctions, which remain unchanged by different choices of $(\boldsymbol{\phi},\boldsymbol{\theta})$,
as discussed in Ref.~\citep{PhysRevB.108.L100201}. In fact, taking
simultaneous averages over $(\boldsymbol{\phi},\boldsymbol{\theta})$
in IPR and IPRk calculations, culminates in smoother results with
better scalings with system size $N=L_{x}L_{y}$. Regarding topological
and spectral properties the same argument does not hold, and in finite
systems a spectral gap can appear for particular choices of a boundary
twist. In this manner, for spectral gap dependent results\footnote{These results include direct calculations of the spectral gap and
topological properties} we fixed a twist of $0$ or $\pi$ depending if the system size is
even en or odd respectively (see Appendix~\ref{sec:Phase-Twist-and-spectral-gap}).
In the thermodynamic limit, the $\boldsymbol{\phi}$-dependence vanishes
for the bulk properties. For a quasiperiodic potential in a square
lattice, the phase shifts appear as twists in the reciprocal space
Hamiltonian, thus phase shifts are transformed to boundary conditions
in reciprocal space and have no impact on the thermodynamic limit.
More generally, in an infinite quasiperiodic system, all local configurations
of the quasiperiodic potential are realized, and introducing a phase
shift leads to no change in the bulk states. In finite open systems,
localized and critical regimes depend on the choice of phase shifts,
which set the local configuration of the quasiperiodic potential.
Moreover, edge properties, such as the edge spectrum, retain a dependence
on phase shifts even in the thermodynamic limit since different shifts
change the local configurations at the boundaries, no matter the size
of the system. A simple way of probing all the possible configurations
in the infinite system is to average over different phase shifts.
However, this strategy in finite open systems should never preserve
the symmetry on average. The averaging mechanism in quasiperiodic
systems is comparable to that performed in disordered systems, where
average symmetries of disordered ensembles can be similar to the exact
ones \citep{chaou2024disordered}. It is plausible that the same holds
in quasiperiodic systems, even though the unique interplay between
quasiperiodic potentials and edge physics may raise some concerns.

To apply phase twists, the boundaries are periodically closed (as
for periodic boundary conditions), but with an additional twist, such
that:

\begin{equation}
\psi_{\alpha}(\boldsymbol{R}+L\mathbf{a}_{i})=e^{i\text{\ensuremath{\theta_{i}}}}\psi_{\alpha}(\boldsymbol{R}),\,i=x,y
\end{equation}
where $\psi_{\alpha}(\boldsymbol{R})=c_{\mathbf{R},\alpha}^{\dagger}\ket 0$.
 In the clean limit ($W=0$) the system preserves translational invariance
and we re-obtain the BBH model \citep{Benalcazar2017} , described
by the following Bloch Hamiltonian:

\begin{multline}
\mathcal{H}_{\mathbf{k}}=\left[\gamma+\lambda\cos(k_{x})\right]\Gamma_{4}+\lambda\sin(k_{x})\Gamma_{3}+\\
+\left[\gamma+\lambda\cos(k_{y})\right]\Gamma_{2}+\lambda\sin(k_{y})\Gamma_{1}+\delta\Gamma_{0}
\end{multline}
where $\Gamma$ are $4\times4$ matrices ($\Gamma_{0}=\tau_{3}\otimes\sigma_{0}$,
$\Gamma_{k}=-\tau_{2}\otimes\sigma_{k}$, $\Gamma_{4}=\tau_{1}\otimes\sigma_{0}$)
that define the internal degrees of freedom within a unit cell.

For $\left|\gamma\right|<1$($\left|\gamma\right|>1$) the system
is in a topological (trivial) phase defined by $q_{xy}=0.5$ ($q_{xy}=0$).
As for the topological phase, the system displays fourfold zero energy
corner modes that give rise to fractional corner charges $Q_{i}=\pm0.5$.
Since this quadrupole insulating phase is a second order topological
insulator (SOTI), the boundary is a first order topological insulator.
In fact, from the boundary perspective the corner modes appear as
localized edge states. In this manner, we can characterize the topological
phase by calculating the boundary polarizations with the nested Wilson
loop approach introduced and discussed in Refs.~\citep{Benalcazar2017,benalcazar2017_prb}.
It was shown that in the presence of quantizing crystalline symmetries
($C_{4}$, and reflection along $x$ and $y$, $M_{x}$ and $M_{y}$),
the quadrupole insulating phase obeys $q_{xy}=p_{x}=p_{y}=0.5$, and
that anti-commuting mirror symmetries are needed to have a Wannier
gap. For any given finite value of $W$ these crystalline symmetries
are broken for a finite system with generic $\bm{\phi}$, however,
it has recently been shown that the quadrupole moment is equally quantized
by chiral symmetry \citep{Yang2021,PhysRevLett.125.166801} and that
a new set of $\mathbb{Z}$ topological invariants arise, known as
the multipole chiral numbers (MCNs) \citep{Benalcazar2022}. Although
this new characterization is shown for systems falling in the AIII
symmetry class, it should be valid for any of the chiral symmetric
classes (AIII, BDI, CII), since BDI and CII classes display more symmetry
than the AIII class. Chiral symmetric quadrupole phases are higher
order topological phases and can exhibit   zero energy corner modes,
quantized edge polarizations and bulk quadrupole moments.

\subsubsection{Methods}

\emph{Non-trivial topology} -- To compute $q_{xy}$ for a disordered
system the multipole operators are considered \citep{Kang2019,Wheeler2019},
which are generalizations of the Resta's formula \citep{Resta1998},

\begin{equation}
q_{xy}=\left[\frac{1}{2\pi}\text{Im}\log\left\langle \Psi_{0}\right|e^{2\pi i\sum_{\mathbf{R}}\hat{q}_{xy}(\mathbf{R})}\left|\Psi_{0}\right\rangle -q_{xy}^{0}\right]\mod 1,
\end{equation}
where $\hat{q}_{xy}(\mathbf{R})=\frac{xy\hat{n}(\mathbf{R})}{L_{x}L_{y}}$
is the quadrupole moment density at site $\mathbf{R}$, $\hat{n}(\mathbf{R})=\sum_{\alpha}c_{\mathbf{R}\alpha}^{\dagger}c_{\mathbf{R}\alpha}$
is the charge density at site $\mathbf{R}$, $\left|\Psi_{0}\right\rangle $
is the ground state of the system and $q_{xy}^{0}$ the background
positive charge contribution for the quadrupole moment. To study boundary
topology, we also compute the boundary polarizations making use of
the effective boundary Hamiltonian defined through $\mathcal{H}_{\text{Bound}}=G_{\text{N}}^{-1}(E=0)$
\citep{Peng2017}, where $G_{\text{N}}(E)$ is the $Nth$ block of
the full Green function of the system. To obtain $G_{\text{N}}(E=0)$,
we take a transfer matrix approach and divide the system in 1D strips
described by the Hamiltonian $\mathbf{h}_{n}$, with each consecutive
strips connected by hopping matrices $\mathbf{V}_{n}$. Due to the
simple Jacobi form of the Hamiltonian, we can perform a blockwise
matrix inversion to reach the following Dyson equation

\begin{equation}
\mathbf{G}_{n}(E)=\left(E\mathcal{I}-\mathbf{h}_{n}-\mathbf{V}_{n-1}\mathbf{G}_{n-1}\mathbf{V}_{n-1}^{\dagger}\right)^{-1}
\end{equation}
To compute the boundary polarizations, we resort to Resta's formula:

\begin{equation}
p_{i}=\left[\frac{1}{2\pi}\text{Im}\log\left\langle \Psi_{c}\right|e^{2\pi i\sum_{\mathbf{R}}\hat{p}_{i}(\mathbf{R})}\left|\Psi_{c}\right\rangle -p_{i}^{0}\right]\mod 1
\end{equation}
where $\hat{p}_{i}(\mathbf{R})=\frac{x_{i}\hat{n}(\mathbf{R})}{L_{x_{i}}}$
is the polarization in site $x_{i}$, $\left|\Psi_{c}\right\rangle $
is the boundary ground state obtained with exact diagonalization of
the boundary Hamiltonian and $p_{i}^{0}$ the background positive
charge contribution to the polarization. With these polarizations
we can define a boundary invariant $P=4\left|p_{x}p_{y}\right|$ such
that $P=0$ for $q_{xy}=0$ and $P=1$ for $q_{xy}=0.5$ as in Ref.~\citep{Yang2021}\footnote{This holds for this particular model. It has been shown \citep{Benalcazar2022}
that in the case of chiral symmetric quadrupole insulators, topological
phases defined by $q_{xy}=P=0$ can occur, where zero energy corner
modes are present with much higher degeneracy. Although these are
topologically non trivial, they display zero quadrupole moment.}. We note that both $q_{xy}$ and $P$ are quantized for every configuration,
even if the system is gapless. However, the average over phase shifts
$\bm{\phi}$ and phase twists $\bm{\theta}$ can lead to non quantized
$q_{xy}$ and $P$.

\emph{Spectral Methods - }To study the spectral properties of our
system we compute the spectral gap ($\Delta E$) via sparse diagonalization
with shift invert and cross check the results with the density of
states (DOS) at Fermi level $\rho(E=0)$. The DOS is defined as $\frac{1}{L_{x}L_{y}}\sum_{i}\delta(E-E_{i})$
and it was computed with an implementation of KPM as presented in
Ref.~\citep{Weisse2006} making use of the Jackson kernel. We also
define the corner occupation probability,

\begin{equation}
P_{occ}=\sum_{\mathbf{R}\in\text{corner}}\sum_{\alpha=1}^{4}\left|\psi_{n}^{\alpha}(\mathbf{R})\right|^{2},\label{eq:P_occ_def}
\end{equation}
where $\psi_{n}^{\alpha}\left(\mathbf{R}\right)$ represents the wavefunction
amplitude in the $\alpha^{\text{th}}$ orbital at site $\mathbf{R}$
for the eigenstate with energy $E_{n}$. For the calculations that
follow, we average $P_{occ}$ over the four eigenstates closest to
the Fermi level obtained with Lanczos decomposition and shift invert.
In a QI phase, zero energy corner modes occur and thus $P_{occ}\approx1$.
For a trivial phase, no corner modes exist and $P_{occ}<1$ is expected.
Since the corner states display a localization length dependent on
$W$ and $\gamma$, we compute $P_{occ}$ over an $l\times l$ region
which we assume to be the corner, i.e. we restrict the sum in Eq.~\eqref{eq:P_occ_def}
to that region. Therefore, $l$ is effectively an estimation of the
localization length of the corner modes. Since the zero energies corner
modes give rise to fractional corner charges, we also compute the
corner charge $\bar{Q}=\sum_{i}\left|Q_{i}\right|$ as a function
of $W$, obtained from $Q_{i}=\sum_{\mathbf{R}\in\text{corner}}\rho\left(\mathbf{R}\right)$,
where $\rho(\mathbf{R})=2-\sum_{n\in\text{occ.}}\sum_{\alpha=1}^{4}\left|\psi_{n}^{\alpha}\left(\mathbf{R}\right)\right|^{2}$
is the charge density. The first and second terms arise from the atomic
positive charges and from the electronic density, respectively. Unlike
the computations for $P_{occ}$, the system is partitioned into four
quadrants, each of the size of a quarter of the system. Then, the
corner charge is computed by integrating $\rho\left(\mathbf{R}\right)$
over each of the quadrants. By definition, the corner charge requires
complete knowledge of the systems spectra, obtained by full diagonalization
of the systems Hamiltonian, limiting the maximum system sizes to $L=55$.
In quadrupole insulating phases $\bar{Q}=0.5$.

\emph{Localization Methods - }To tackle localization properties several
quantities are computed such as the inverse participation ratio ($\textrm{IPR}$),
$k$-space inverse participation ratio ($\textrm{IPR}_{k}$), fractal
dimension ($D_{2}$) and the localization length. The normalized localization
length $\xi/M$ where $\xi$ is the localization length and $M$ the
transversal system size, was obtained via transfer matrix method (TMM)
introduced in Ref.~\citep{MacKinnon1983,MK81} for a choice of longitudinal
size that yields an error $\epsilon\approx1\%$. Similarly to the
boundary Green's function method, the system is divided in 1D strips
with $M$ unit cells described by the Hamiltonian $\mathbf{h}_{n}$.
Each consecutive strip is connected by hopping matrices $\mathbf{V}_{n}$
with which the transfer matrix of the system $\mathbf{T}_{n}$ is
constructed. The localization length $\xi=1/\gamma_{i}$ is calculated
from Lyapunov exponent $\gamma_{i}$, where $e^{\gamma_{i}}$ is the
smallest positive eigenvalue of the full transfer matrix $\mathbf{M}_{N}=\prod_{m=1}^{N}\mathbf{T}_{n}$,
that connects the first and last strip of the system. For localized
states, $\xi/M\to0$, for extended states $\xi/M\to\infty$ and for
critical states $\xi/M$ remains constant.

The $\textrm{IPR}$ \citep{wegner1980} and $\textrm{IPR}_{k}$ are
defined as:

\begin{align}
\textrm{IPR} & =\frac{1}{N}\sum_{\mathbf{R},\alpha}\left|\psi_{\alpha}(\mathbf{R})\right|^{4}\\
\textrm{IPR}_{k} & =\frac{1}{N}\sum_{\mathbf{k},\alpha}\left|\psi_{\alpha}(\mathbf{k})\right|^{4}
\end{align}
where $\psi_{\alpha}(\mathbf{R})=\left\langle \mathbf{R},\alpha|\psi\right\rangle $,
and $\psi_{\alpha}(\mathbf{k})=\left\langle \mathbf{k},\alpha|\psi\right\rangle $
for a normalized eigenstate $\left\langle \psi|\psi\right\rangle =1$.
From the scaling $\textrm{IPR}\propto L^{-D_{2}}$, we extract the
fractal dimension $D_{2}$ by fitting the results for different system
sizes. Thus for $D_{2}=2$ the eigenstates are extended, for $D_{2}=0$
the eigenstates are localized, while for $0<D_{2}<1$ the eigenstates
display critical character.

\section{Results}

\label{sec:results}

The phase diagram of the system is presented Fig.$\,$\ref{fig:phase-diagram}
on the $(\gamma,W)$ plane, where topological properties are shown
in panel~\ref{fig:phase-diagram}(a) and spectral properties in panel~\ref{fig:phase-diagram}(b).
In what follows, we study in detail the cuts indicated in Fig.$\,$\ref{fig:phase-diagram},
focusing on topological, spectral, localization and edge properties.

\subsection{Topological Properties}

\begin{figure}
\begin{centering}
\includegraphics[width=1\columnwidth]{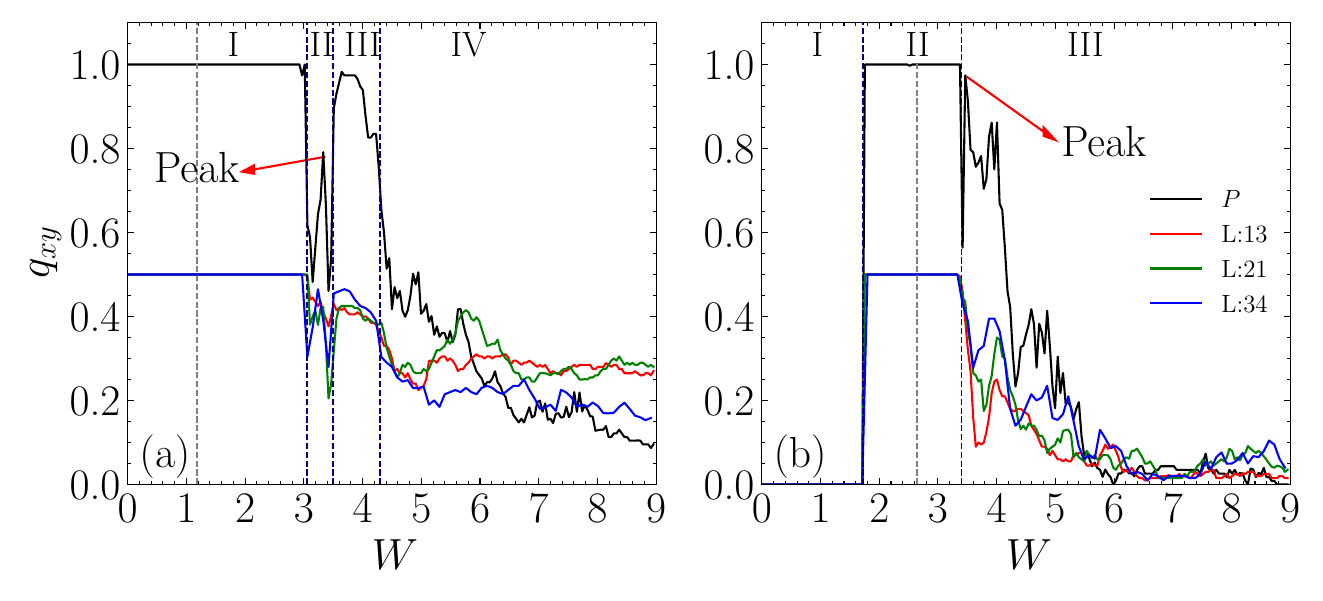}
\par\end{centering}
\begin{centering}
\includegraphics[width=1\columnwidth]{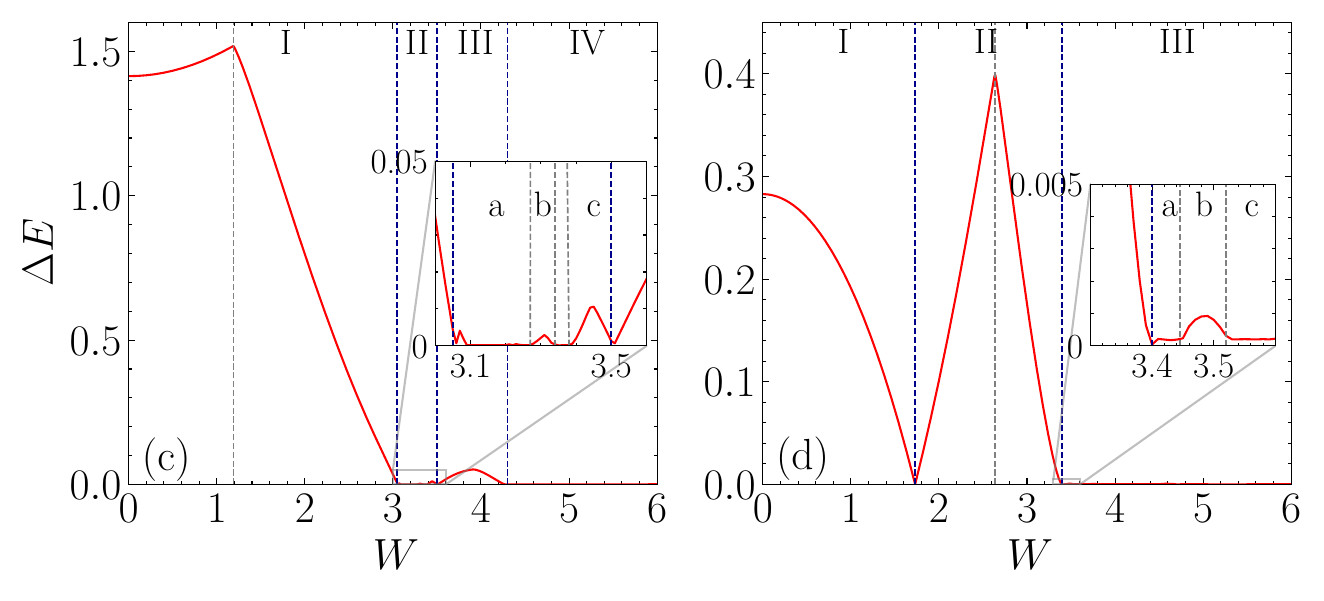}
\par\end{centering}
\caption{(a-b) $q_{xy}$ and $P$ as a function of $W$. The $P$ invariant
was computed for a system size of $L=610$ with $115$ averages over
phase shifts. $q_{xy}$ was obtained via real space methods with $100$
averages over phase shifts. For even $L$, $\theta_{x}=\theta_{y}=0$;
for odd $L$ $\theta_{x}=\theta_{y}=\pi$. (a) $\gamma=0.5$. (b)
$\gamma=1.1$. (c-d) Spectral gap ($\Delta E$) as a function of $W$
computed for an even system size $L=144\protect\implies\theta_{i}=0$
and averaged over $50$ phase shifts realizations. (c) $\gamma=0.5$.
(d) $\gamma=1.1$. \label{fig:qxy_P_vs_W_and spectral gap}}
\end{figure}

\emph{Starting Topological} - The results obtained by increasing $W$
starting at a topological phase are shown in Figs.$\,$\ref{fig:qxy_P_vs_W_and spectral gap}(a)
and~\ref{fig:qxy_P_vs_W_and spectral gap}(c). For $\gamma=0.5$
and $W=0$ the system is in a topological phase. The quadrupole moment
and the boundary invariant are quantized respectively to $0.5$ and
$1$ for each realization of phase shifts, remaining quantized with
increasing $W$, demonstrating that the quadrupole phase is robust
to quasiperiodic modulations. Eventually, at $W_{c}^{I\to II}\approx3.05$
the gap closes {[}see panel~\ref{fig:qxy_P_vs_W_and spectral gap}(c){]}
and a TPT occurs, with the system entering a critical metal phase
{[}$\text{II}a$, see inset of Fig.$\,$\ref{fig:qxy_P_vs_W_and spectral gap}(c){]},
which we motivate below after analyzing the localization properties.
Further increasing $W$, the gap reopens into a topological regime
{[}$\textrm{II}b$ in panel~\ref{fig:qxy_P_vs_W_and spectral gap}(c){]}.
Focusing on the $P$ invariant (or the quadrupole moment $q_{xy}^{L=34}$),
a sharp peak can be observed in this region. The fact that this peak
approaches $1$ ($0.5$ for $q_{xy}^{L=34}$) with increasing system
size, while the value of $P$ (or $q_{xy}^{L=34}$) in $\textrm{II}a$
and $\textrm{II}c$ go to zero, hints that $\textrm{II}b$ is a quadrupole
insulating phase. Furthermore, since this is a very narrow $W$ window
and the gap is small ($\Delta E\approx10^{-3}$) we attribute the
lack of quantization to finite size effects. Increasing $W$ even
more induces a trivial insulating phase ($\text{II}c$) before another
quadrupole insulating regime is reached ($\text{III}$) that suffers
from the same finite size effects as $\text{II}b$. In this manner,
QP modulations can induce a novel type of SOTI phases which we entitle
quasiperiodic Quadrupole Insulator (QPQI) phases. At high $W$, the
gap closes (at $W_{c}^{III\to IV}\approx4.3$) and the system reaches
a gapless regime ($\textrm{IV}$) with a finite, although not quantized,
quadrupole moment ($q_{xy}\neq0$).

\emph{Starting Trivial }- The results obtained by increasing $W$
starting at a trivial phase are shown in Figs.$\,$\ref{fig:qxy_P_vs_W_and spectral gap}(b)
and~\ref{fig:qxy_P_vs_W_and spectral gap}(d). For $\gamma=1.1$
and $W=0$ the system is in a gapped trivial phase, which is stable
to the application of a QP modulation ($\textrm{I}$). Increasing
$W$ induces a TPT into a gapped QI phase ($\textrm{II}$) with the
gap closing and reopening at $W_{c}^{I\to II}\approx1.73$. Furthermore,
the gap reaches its maximum, with a level crossing between energy
levels in the gap edge occurring at $W\approx2.64$ (dotted \textcolor{gray}{grey
line}). Away from the QI phase, the gap closes $(W_{c}^{II\to III}\approx3.4)$
and reopens in $\textrm{IIIb}$ with a sharp peak in $P$ that almost
quantizes to $1$. Just as in the previous case ($\gamma=0.5$), we
attribute this lack of quantization to finite size effects and classify
$\textrm{IIIb}$ as QPQI phase. Eventually, at the high $W$ regime,
the system undergoes a TPT into a gapless regime ($\textrm{IIIc}$).

We note that there are regions where there could be an even more intricate
structure of gap and gapless regions with reentrant topological transitions
(as unveiled by the complex behavior of $P$ and $q_{xy}$) which
would only be possible to capture with much larger system sizes.

\subsection{Density of States}

\begin{figure}[h]
\centering{}\includegraphics[width=1\columnwidth]{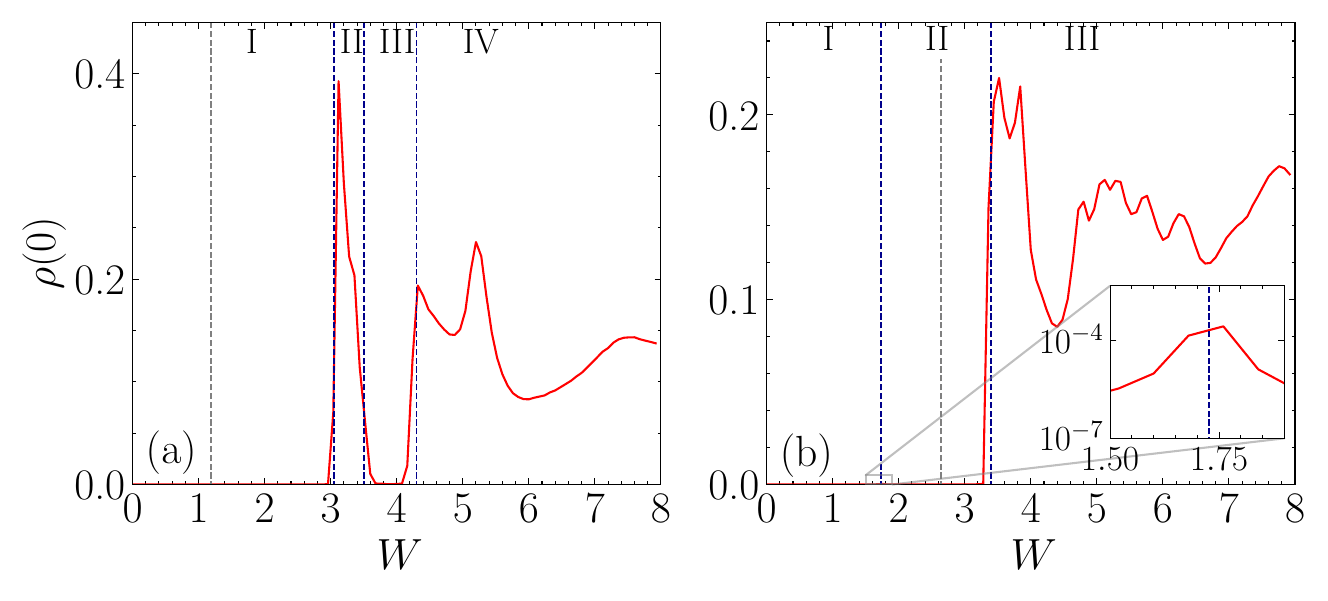}\caption{The DOS at zero energy ($\rho(0)$) as a function of $W$. The following
KPM parameters were used: $L=987$ $M=5000$ Chebyshev moments, and
$R=10$ stochastic traces. (a) $\gamma=0.5$. (b) $\gamma=1.1$. \label{fig:DOS_E0}}
\end{figure}

In Fig.~\ref{fig:DOS_E0} we plot the DOS at $E=0$ as a function
of $W$ with the intent of verifying the results of the spectral gap
presented in Fig.~\ref{fig:qxy_P_vs_W_and spectral gap}. In gapped
regions $\rho(0)\approx0$, while in gapless regions the system should
have a finite DOS. All gapless phases ($\textrm{II}$ and $\textrm{IV}$
for $\gamma=0.5$ and $\textrm{III}$ for $\gamma=1.1$) display a
finite DOS at Fermi level. Furthermore, at the transition $\textrm{I}\to\textrm{II}$
($\gamma=1.1$) the gap closes and reopens at a critical QP potential
strength ($W_{c}^{I\to II}\approx1.73$), which is signaled by a small
increase of $\rho(0)$ around it, as shown in the inset of Fig.~\ref{fig:DOS_E0}(b).
The Jackson kernel broadens the Dirac delta as a Gaussian with spread
$\sigma\propto1/M$, where $M$ is the number of Chebyshev moments.
For this reason, it was not possible to reach enough resolution to
capture the gapped $\textrm{IIb}$ and $\textrm{IIc}$ phases, that
are filled by the tails of the Gaussians. The same occurs for $\gamma=1.1$
in phase $\textrm{IIIb}$. These narrow phases are only expected to
be resolved for larger system sizes and number of moments.

\subsection{Localization Properties}

\begin{figure}
\centering{}\includegraphics[width=1\columnwidth]{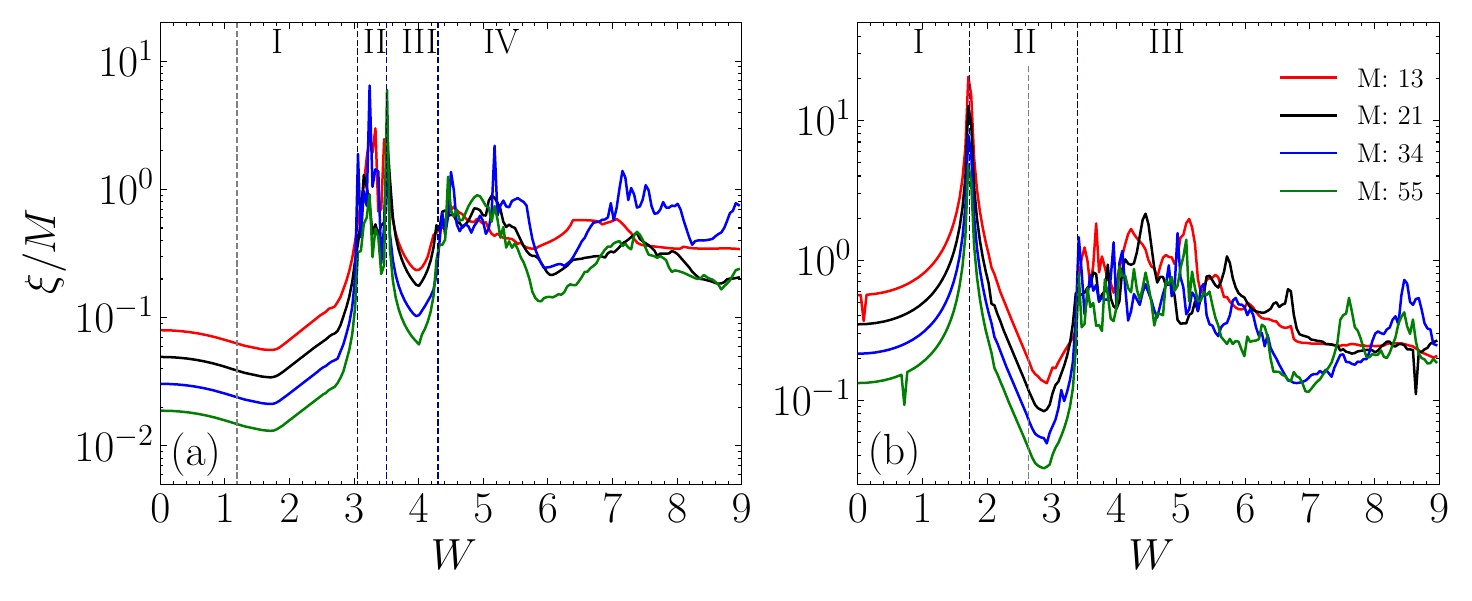}\caption{TMM at Fermi level ($E=0$) for several transversal system sizes ($M$).
For odd $M$ we add a twits of $\theta_{M}=\pi$. The size of the
longitudinal direction was chosen such that the relative error $\epsilon<1\%$.
(a) $\gamma=0.5$. (b) $\gamma=1.1$. \label{fig:TMM_E0}}
\end{figure}

To complete the characterization of the phase diagram, we turn to
the study of the localization properties. In Fig.~\ref{fig:TMM_E0}
we show the normalized localization length ($\xi/M$) at the Fermi
level ($E=0$), obtained using the TMM. For the regions with larger
gap, it is clear that $\xi/M$ scales down to zero, since TMM captures
localized evanescent wave solutions of the Schrödinger equation. At
gapless regions, or regions with small gaps, the results are noisier
and in some cases larger system sizes would be needed to understand
whether $\xi/M$ converges with $M$, showing critical behavior, or
decreases with $M$, signaling gapless localized states. This will
become clearer from the eigenstate analysis based on exact diagonalization
that we carry out below. Noteworthy, the topological phase transition
occurring through gap closing and reopening as seen in Fig.$\,$\ref{fig:qxy_P_vs_W_and spectral gap}(d)
is well captured by a peak in $\xi/M$ in Fig.$\,$\ref{fig:TMM_E0}(b),
which is expected to not decrease with $M$ precisely at the critical
point.

\begin{figure}[h]
\centering{}\includegraphics[width=1\columnwidth]{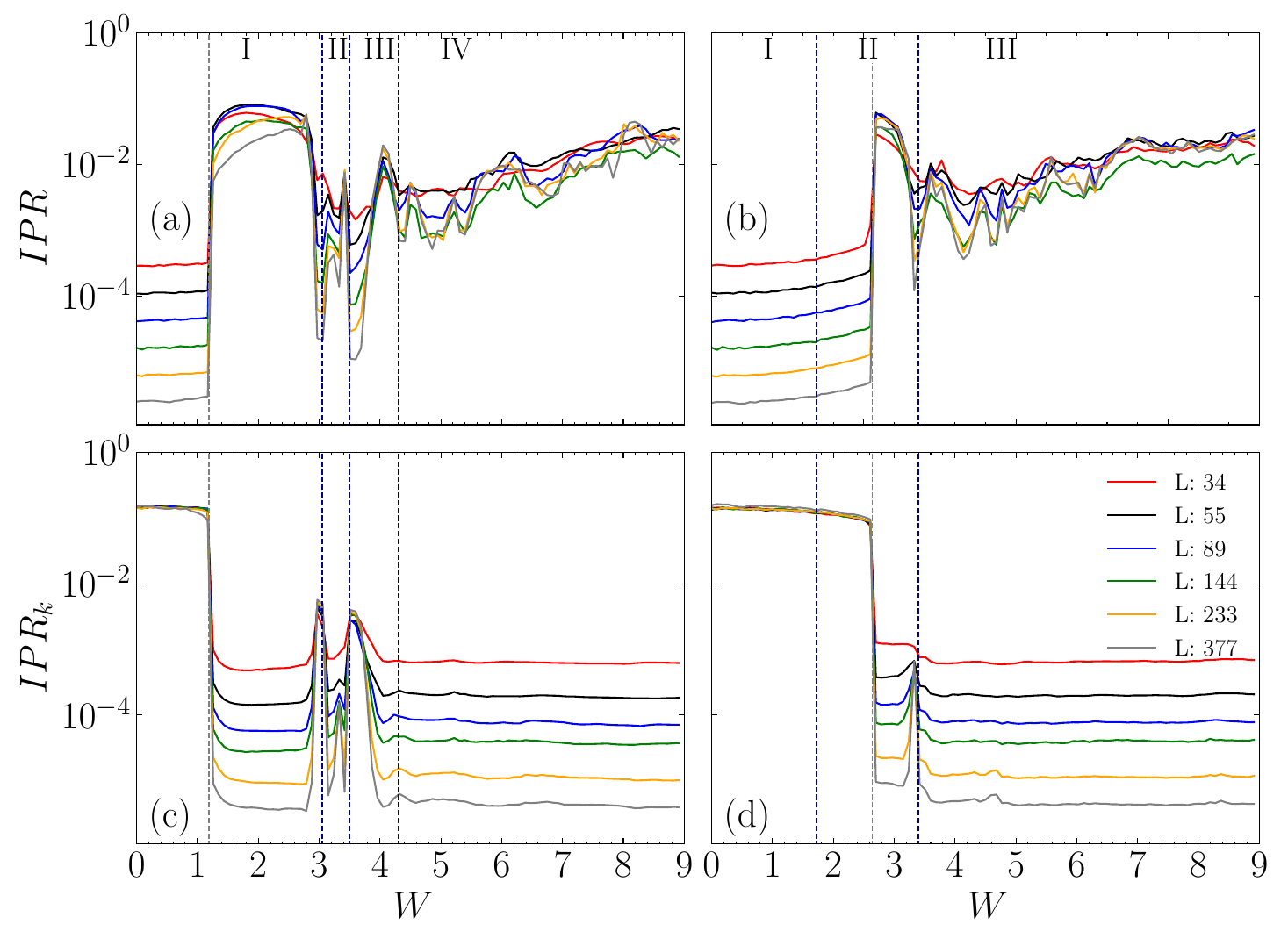}\caption{(a-b) $\textrm{IPR}$ and (c-d) $\textrm{IPR}_{k}$ for different
system sizes obtained with exact diagonalization methods. The $\textrm{IPR}$
and $\textrm{IPR}_{k}$ are averaged over $100$ phase twists, shifts
and over the first eight eigenstates with energies closest to $E=0$.
(a,c) $\gamma=0.5$. (b,d) $\gamma=1.1$.\label{fig:IPR_IPRk_gapedge}}
\end{figure}

\begin{figure}
\centering{}\includegraphics[width=1\columnwidth]{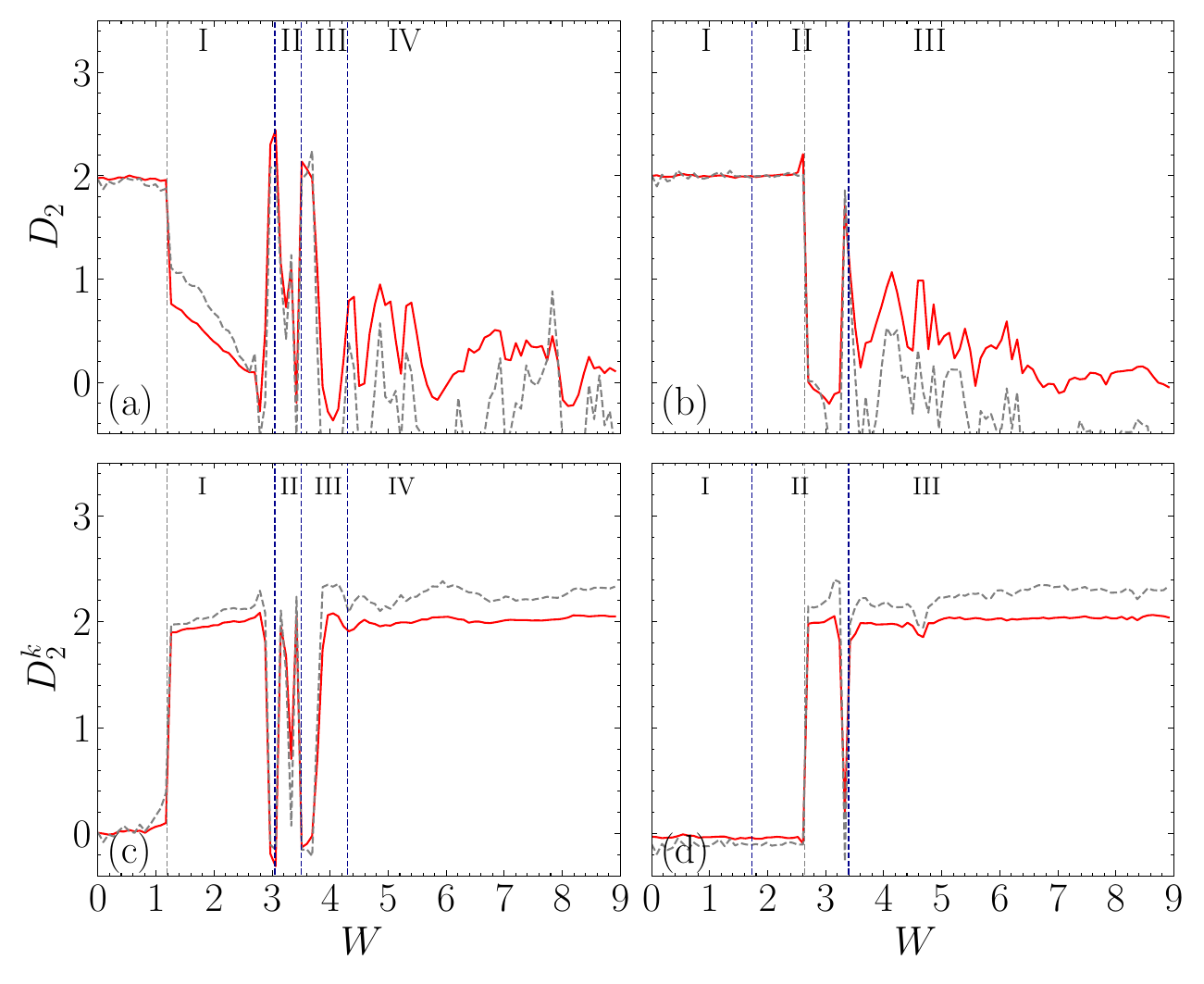}\caption{$D_{2}$ (a,b) and $D_{2}^{k}$(c,d) for different system sizes obtained
from the $\text{IPR}$ and $I\text{IPR}$ results respectively. (a,c)
$\gamma=0.5$. (b,d) $\gamma=1.1$. The red curves were estimated
making a fit in $\log-\log$ scale for all $L=\{34,55,89,144,233,377\}$,
while the dotted grey lines were obtained only considering the biggest
three systems sizes $L=\{144,233,377\}$.\label{fig:fractal_dimensions}}
\end{figure}

To further characterize the localization properties of the eigenstates
with energies closer to $E=0$, we compute the IPR and the $\text{IPR}{}_{k}$.
In the gapped phases, these states correspond to eigenstates at the
gap edge that are not captured by the TMM at Fermi level. In Fig.~\ref{fig:IPR_IPRk_gapedge},
we plot the average $\textrm{IPR}$ (a-b) and $\textrm{IPR}_{k}$
(c-d) of the eight eigenstates closer to $E=0$ as a function of $W$.
We also plot the fractal dimensions $D_{2}$ and $D_{2}^{k}$ in Fig.~\ref{fig:fractal_dimensions}
estimated, respectively, through the $\textrm{IPR}$ and $\text{IPR}_{k}$.

\emph{Starting Topological} - For $\gamma=0.5$, in the gapped phase~$\textrm{I}$,
the eigenstates at the gap edge start off ballistic as seen in Fig.~\ref{fig:fractal_dimensions},
with $D_{2}\to2$ and $D_{2}^{k}\to0$. They undergo a ballistic$\to$localized
($D_{2}\to0$ and $D_{2}^{k}\to2$) transition at $W\approx1.19$
(\textcolor{gray}{grey line}) further ascertaining that a level crossing
between gap edge states occurs. This transition is not captured by
the TMM since the system remains gapped. At the TPT ($\text{I\ensuremath{\to}II}$)
the gap-edge states are ballistic, however, these states quickly turn
critical as seen by $D_{2}\approx1$, indicating that the gapless
critical metal regime is reached ($\text{IIa}$). The normalized localization
length ($\xi/M$) obtained via TMM does not diverge at the transition,
instead, it goes to a constant value {[}see Fig.~\ref{fig:TMM_E0}(a){]},
in agreement with the critical metal regime. In the QPQI phase ($\text{IIb}$),
the gap-edge states remain critical, localizing in the $\text{IIc}$
phase. In these latter regimes TMM fails to capture such narrow dynamics,
due to the lack of discretization in TMM calculations. At the transition
from $\text{II\ensuremath{\to}III}$, $\text{IPR}$, $\text{IPR}_{k}$
(Fig.~\ref{fig:IPR_IPRk_gapedge}) and $D_{2}$ (Fig.~\ref{fig:fractal_dimensions})
indicate a ballistic regime with the the gap-edge states undergoing
a ballistic-localized transitions halfway through phase $\text{III}$.

Phase $\text{IV}$ is of difficult classification with any of the
methods at our disposal. At the transition $\text{III\ensuremath{\to}IV}$
the $\text{IPR}$ results suggest that the eigenstates are critical.
After this transition, TMM and $\text{IPR}$ results are noisy. However,
considering the fractal dimension computed with the last three system
sizes, $D_{2}$ approaches zero indicating that the system has reached
an Anderson insulating phase ($D_{2}\to0$).

\emph{Starting Trivial} - Starting at the trivial phase with $\gamma=1.1$,
the gap edge eigenstates are ballistic within phase~I and remain
ballistic even after the transition into phase $\text{II}$ {[}$D_{2}\to2$
and $D_{2}^{k}\to0$, see Fig.~\ref{fig:fractal_dimensions}{]}.
At the TPT ($\text{I\ensuremath{\to}II}$), a gap closing and reopening
topological transition with ballistic gap edge states occurs, as in
the absence of quasiperiodicity. This explains why this particular
gap closure is highly dependent on the choice of twists, for small
system sizes \citep{10.21468/SciPostPhys.13.3.046}. As in the previous
case, the ballistic regime is maintained until the level crossing
between ballistic and localized gap edge levels occurs (\textcolor{gray}{grey
line}), where we observe $D_{2}\to0$. For higher $W$, slightly before
the gap closing at $W_{c}^{II\to III}\approx3.4$, gap edge states
delocalize and then become critical at the transition ($D_{2}\approx1.2$).

The results for the final phase ($\text{III}$) are similar to phase
$\text{IV}$ of the previous cut, leading us to conclude that an Anderson
insulating regime is reached.

\subsection{Opening the Boundaries}

\begin{figure}[h]
\begin{centering}
\includegraphics[width=1\columnwidth]{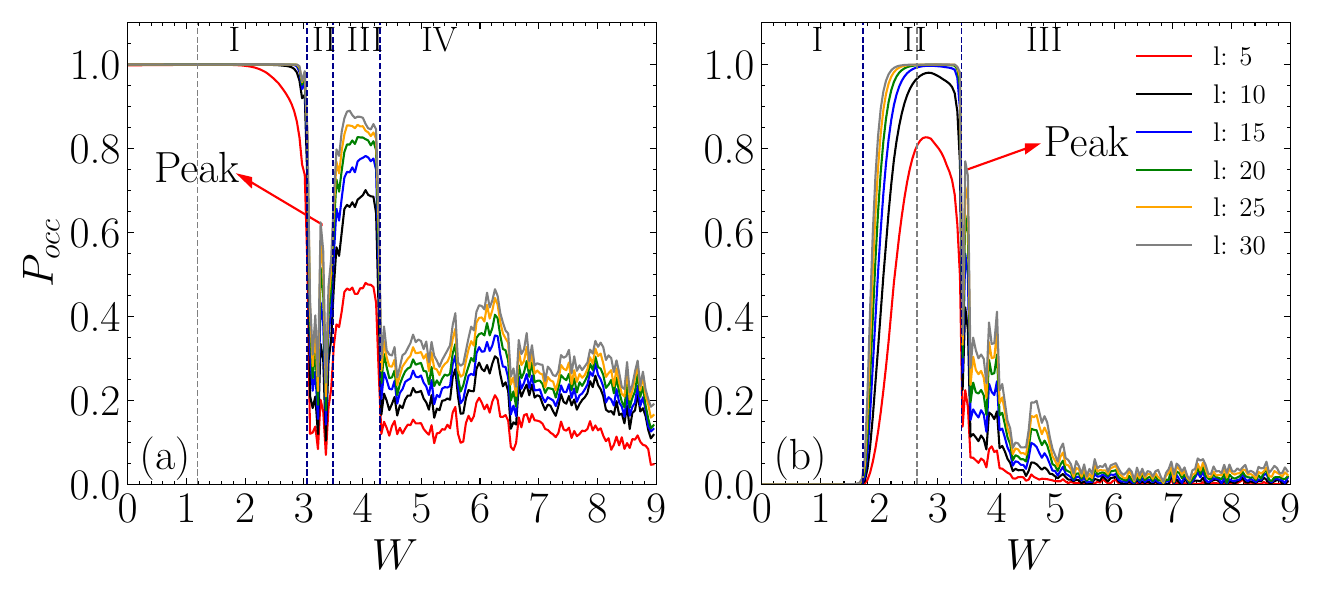}
\par\end{centering}
\begin{centering}
\includegraphics[width=1\columnwidth]{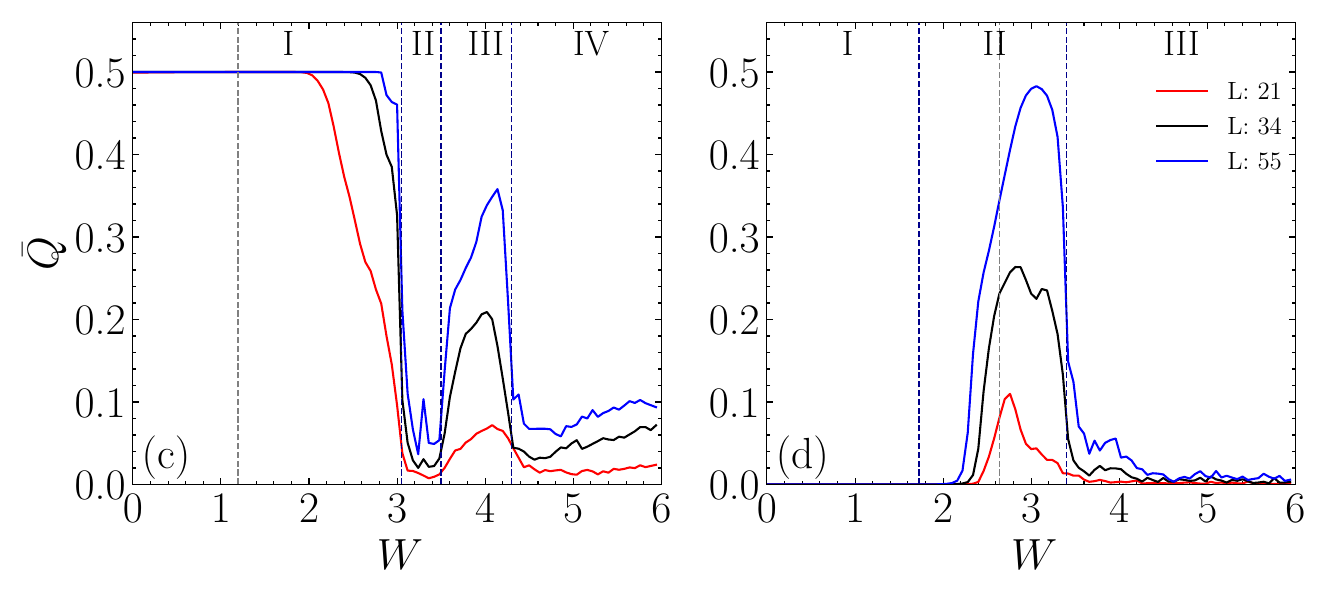}
\par\end{centering}
\centering{}\caption{(a-b) Corner occupation probability ($P_{occ}$) as a function of
$W$ for the first two zero energy states. The results were obtained
for a system with size $L=377$ via Lanczos decomposition and shift
invert and averaged over $200$ phase shifts. (c-d) Corner charge
($\bar{Q}$) as a function of $W$ for several system sizes calculated
via full diagonalization of the systems Hamiltonian and averaged over
$100$ phase shifts and over the four corners. A small $\delta=10^{-5}$
was used in both cases to split the degeneracy of the four fold corner
modes. (a,c) $\gamma=0.5$. (b,d) $\gamma=1.1$.\label{fig:Corner-occupation-probabilit}}
\end{figure}

An important hallmark of topological quadrupole insulators is that
quadrupole phases defined by $q_{xy}=0.5$ have zero energy corner
modes that give rise to fractional corner charges when the boundaries
of the system are opened. For this reason we now study the system
with open boundary conditions. In Fig.~\ref{fig:Corner-occupation-probabilit}(a-b),
we show the $P_{occ}$ as a function of $W$ and in Fig.~\ref{fig:Corner-occupation-probabilit}(c-d)
we show the averaged corner charge $\bar{Q}=\sum_{i}\left|Q_{i}\right|$
($Q_{i}$ is the corner charge of each individual corner) as a function
of $W$.

\emph{Starting topological -} For $\gamma=0.5$, it can be seen in
Fig.~~\ref{fig:Corner-occupation-probabilit}(a) that $P_{occ}$
has perfect plateaus at $P_{occ}=1$ for every studied $l$, indicating
the existence of highly localized corner states ($\xi_{l}<10$ where
$\xi_{l}$ is the localization length of the corner modes). Similarly,
the corner charge also quantizes at $\bar{Q}=0.5$ throughout the
topological phase, as shown in Fig.~\ref{fig:Corner-occupation-probabilit}(c).
In phase~$\textrm{II}$, the critical metal ($\textrm{IIa}$) and
trivial insulator ($\textrm{IIc}$) display a finite, but significantly
smaller $P_{occ}$, that arises from contributions of edge or bulk
states (below we will provide a detailed discussion on edge states).
In the narrow topological regime observed ($\textrm{IIb}$), $P_{occ}$
sharply peaks, scaling closer to $1$ as indicated by the arrow in
Fig.~\ref{fig:Corner-occupation-probabilit}(a). However, the peak
does not properly quantize to~$1$ for the studied $l$ values, which
we attribute to the large localization length of the corner modes,
$\xi_{l}>30$, due to the very small gap in this phase. A peak can
also be observed in the corner charge for $L=55$ {[}Fig.~~\ref{fig:Corner-occupation-probabilit}(c){]},
albeit smaller due to the considerably smaller system size. Regarding
phase $\textrm{III}$, $P_{occ}$ has a noisy plateau that scales
to~$1$ as $l$ increases, indicating again that the corner states
display a larger localization length than the maximum considered corner
($l=30$). The corner charge also scales to $0.5$ with increasing
size, further corroborating that~$\textrm{III}$ is indeed topological
with corner modes characterized by a larger localization length. Finally,
as expected, phase~$\textrm{IV}$ has a small $P_{occ}$. However,
we also see a growth with increasing corner/system size. This increase
is explained by the gapless nature of this phase. As $l$ increases,
more and more bulk spectral weight fall within the considered $l\times l$
corner, leading to a visible increase in $P_{occ}$. Regarding the
corner charge, although there are no quantized corner charges, for
some realizations of phase shifts the electron density displays localized
peaks around the average bulk electron density which leads to finite
contributions to the corner charge. This occurs due to the breaking
of chiral symmetry by the small $\delta$ introduced to lift the degeneracy
of the corner modes. Overall the $P_{occ}$ and $\bar{Q}$ of each
phase agree well with the topological phase diagram.

\emph{Starting trivial -} For $\gamma=1.1$ the trivial phases~$\textrm{I}$
and~$\textrm{IIIc}$ display $P_{occ}\approx0$ as seen in Fig.~\ref{fig:Corner-occupation-probabilit}(b).
For phase~$\text{II}$, the corner occupation probability, though
initially suffers from finite size effects, eventually reaches~$1$.
Its dependence on $l$ after the TPT ($\textrm{I}\to\textrm{II}$)
indicates a diverging localization length of the corner states as
the critical point is approached. Deeper into the QI phase, the corner
modes become more localized. Regarding~$\textrm{IIIb}$, $P_{occ}$
has a small peak in this region as indicated by the arrow in Fig.~\ref{fig:Corner-occupation-probabilit}(b).
Regarding the corner charge shown in Fig.~\ref{fig:Corner-occupation-probabilit}(d),
it displays scaling with increasing system size, however a plateau
cannot be seen even for the biggest considered size $L=55$. Moreover,
the corner charge calculations do not have enough resolution to catch
narrow topological regimes like the $\textrm{IIIb}$~phase. We have
checked for $P_{occ}$ in phase~$\text{II}$ that taking $l=30$
for the linear size of the corner is not enough in a system with size
$L=377$ if we want to measure all the weight of the corner state.
This is a clear indication that the corner states display a localization
length $\xi_{l}>30$. Therefore, simulating a system of size $L=55$
\footnote{Notice that for $\xi_{l}>30$, the localization length of the corner
state is of the order of the linear size of the entire system.}, leads to the hybridization of the corner modes, which in turn leads
to a non quantized corner charge.

\emph{Properties of QPQI phases - }Until now a proper characterization
of the QPQI phases has not been provided. In the paragraphs that follow,
we aim to fully define these quasiperiodic induced regime. We start
with their definition: a QPQI phase is a topological quadrupole insulating
phase that is not adiabatically connected to any of the clean limit
phases while displaying a quasiperiodic induced bulk-spectral gap
(or is gapless but displays localized states at the Fermi level, as
discussed in Appendix~\ref{Sec:constat_W_cut}).
\begin{figure}[h]
\centering{}\includegraphics[width=1\columnwidth]{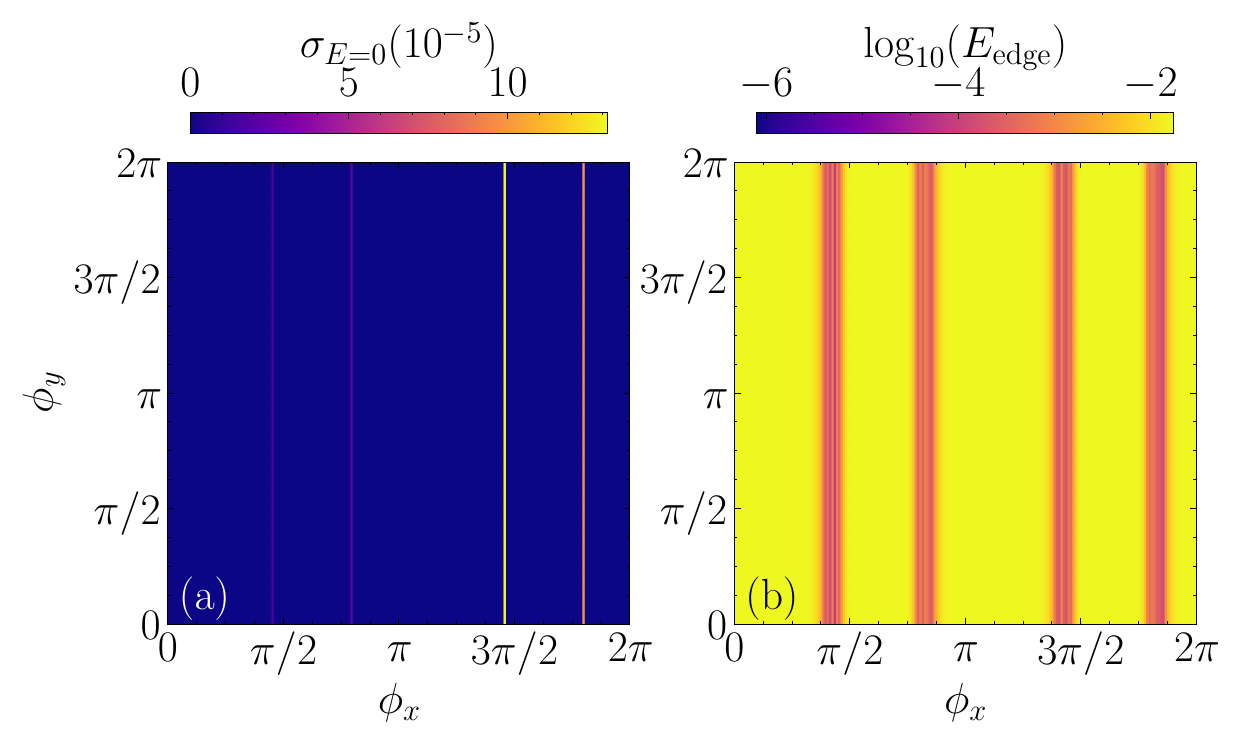}\caption{(a) Standard deviation ($\sigma_{E=0}$) of the energies of the corner
modes as a function of the phase shifts ($\phi_{x},\phi_{y}$). (b)
Logarithm of edge state energy ($\log_{10}(E_{edge})$) as a function
of the phase shifts ($\phi_{x},\phi_{y}$). Results obtained for $W=3$,
$\gamma=0.5$ for a system size $L=233$. \label{fig:(a)-Standard-deviation-1}}
\end{figure}

We have found that quasiperiodic induced gapped phases usually display
edge states that disperse with the phase shifts. This phenomena is
common in QP systems and can be observed in the Aubry-André model
\citep{Kraus_2012}. As $\phi$ changes, the energy of the edge states
can move closer or even cross the Fermi level. This phenomena is particularly
relevant in topological phases since the edge states can hybridize
with zero energy topological modes, thus pulling them away from zero
energy. When these crossings occurs the system can transition into
a trivial phase characterized by $P=0$ and $q_{xy}=0.$ Furthermore,
$P_{occ}$ is also impacted since the corner modes are hybridized
with edge states that exhibit spectral weight outside of the corners.

In order to check the existence of edge states and study edge-corner
hybridization we resort to exact diagonalization of the open boundary
system Hamiltonian, calculating the six eigenvalues closest to the
Fermi level. Assuming we choose a point of the phase diagram that
resides in a SOTI phase, out of the six energies, four correspond
to zero energy corner modes with the remaining two corresponding to
the edge states with energy closest to the Fermi level. We then proceed
to compute the standard deviation of the corner modes energy $\left(\sigma_{E=0}=\sqrt{\frac{1}{N}\sum_{i}E_{i}^{2}}\right)$
and the energy of the edge state closest to the Fermi level ($E_{edge}$).
In this manner, when the edge states energy approach the Fermi level
($E_{edge}\to0$) a peak should be seen in $\sigma_{E=0}$, indicating
that the corner modes hybridized with the edge states, since they
moved away from zero energy.

\begin{figure}[H]
\begin{centering}
\includegraphics[width=1\columnwidth]{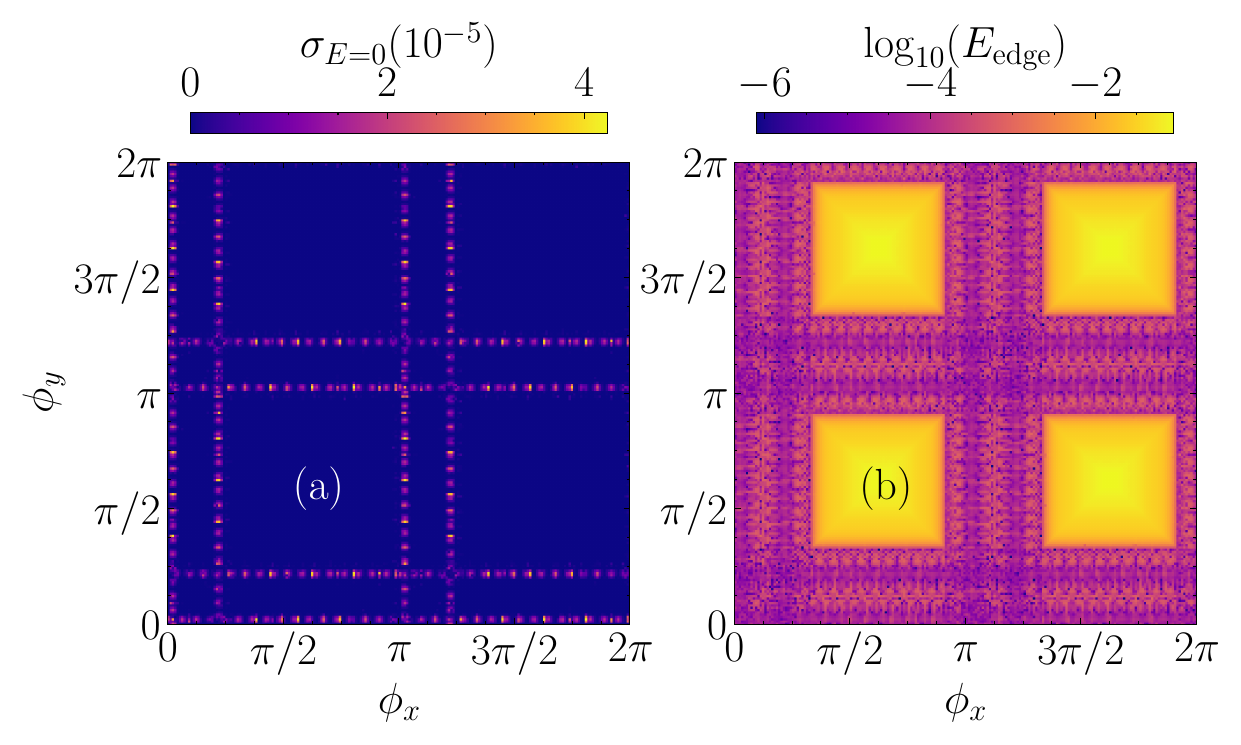}
\par\end{centering}
\caption{(a) Standard deviation ($\sigma_{E=0}$) of the energies of the corner
modes as a function of the phase shifts ($\phi_{x},\phi_{y}$). (b)
Logarithm of edge state energy ($\log_{10}(E_{edge})$) as a function
of the phase shifts ($\phi_{x},\phi_{y}$). Results obtained for $W=3.8$
and $\gamma=0.2$ in the QPQI phase for a system size $L=377$. \label{fig:QPQI_phi_phase_diagram}}
\end{figure}

Before tackling QPQI phases that display complex dynamics, we consider
a simpler case. In Fig.~\ref{fig:qxy_P_vs_W_and spectral gap}(a),
just before the TPT from $\text{I\ensuremath{\to}II}$, a small dip
can be seen in the $P$ invariant ($W\approx3$ and $\gamma\approx0.5$),
suggesting that some dependence on the phase shifts might exist. In
this manner, in Fig.~\ref{fig:(a)-Standard-deviation-1}(a) we show
the density plot of $\sigma_{E=0}$ and in Fig.~\ref{fig:(a)-Standard-deviation-1}(b)
the density plot for $E_{edge}$ in the $(\phi_{x},\phi_{y})$ plane.
Four lines of constant $\phi_{x}$ can be observed where $\sigma_{E=0}>0$
{[}Fig.~\ref{fig:(a)-Standard-deviation-1}(a){]} for $E_{edge}\to0$
{[}Fig.~\ref{fig:(a)-Standard-deviation-1}(b){]}, indicating that
the corner modes hybridized away from zero energy. Moreover, not all
four corner modes hybridize with the edge states. In fact, only two
of them move away from zero energy through hybridization, allowing
us to distinguish two types of corner states. We attribute the term
\emph{inner corner modes} to the two corner modes that remain at $E=0$
and \emph{outer corner modes} to the remaining corner states that
move away from zero energy.

\begin{figure*}[!t]
\begin{centering}
\includegraphics[scale=0.5]{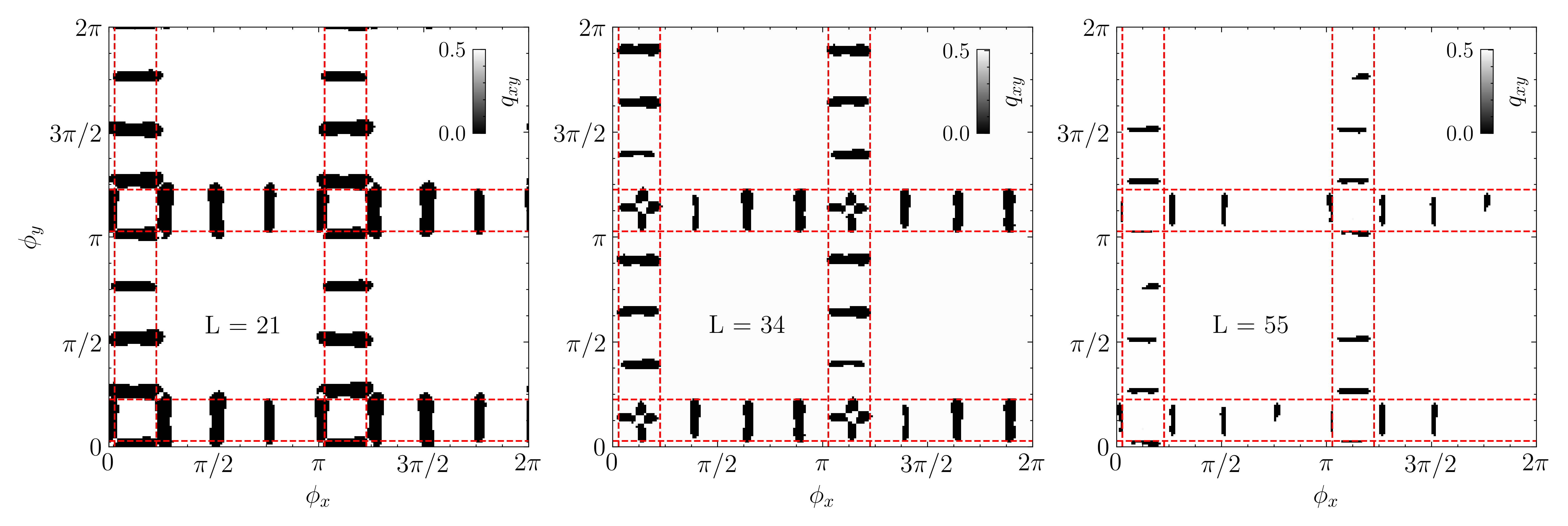}
\par\end{centering}
\caption{Quadrupole moment ($q_{xy}$) as a function of the phase shifts ($\phi_{x},\phi_{y}$)
for different system sizes. Results obtained for $W=3.8$ and $\gamma=0.2$
in the QPQI phase. For even $L$ we use $\theta_{x}=\theta_{y}=0$
and for odd $L$ we use $\theta_{x}=\theta_{y}=\pi$. The dotted red
lines are regions where $\sigma_{E=0}>0$. \label{fig:Quadrupole-moment-phi-dependence}}
\end{figure*}

We now focus on boundary effects on the QPQI phases. A distinct feature
of QPQI phases is the much stronger dependence on phase shifts. In
Fig.~\ref{fig:QPQI_phi_phase_diagram} we show $\sigma_{E=0}$ and
$E_{edge}$ for the choice of parameters $\gamma=0.2$ and $W=3.8$,
well inside the QPQI phase. Just as before, regions with $\sigma_{E=0}>0$
{[}Fig.~\ref{fig:QPQI_phi_phase_diagram}(a){]} are correlated to
regions where $E_{edge}\to0$ {[}Fig.~\ref{fig:QPQI_phi_phase_diagram}(b){]}.
In fact, for a finite system, the edge states hybridize with the corner
modes, whenever their energy $E_{edge}$ is smaller than the mean
level spacing of the system. Intuitively, increasing system size reduces
the mean level spacing and confines these regions to singular points
(for smaller system sizes see Appendix~\ref{sec:phase shift dependences}).
This behavior suggests that all the quantities of relevance ($P$,$P_{occ}$,
$\bar{Q}$ and $q_{xy}$) quantize to their respective values as system
size increases. This can be seen in Fig.~\ref{fig:Quadrupole-moment-phi-dependence}
where we plot $q_{xy}$ as a function of the phase shifts. As the
system size increases, the black regions ($q_{xy}=0$) become smaller.
Moreover, these regions occur for phase shifts values where corner-edge
hybridization is more strongly observed.

\begin{figure}
\begin{centering}
\includegraphics[width=1\columnwidth]{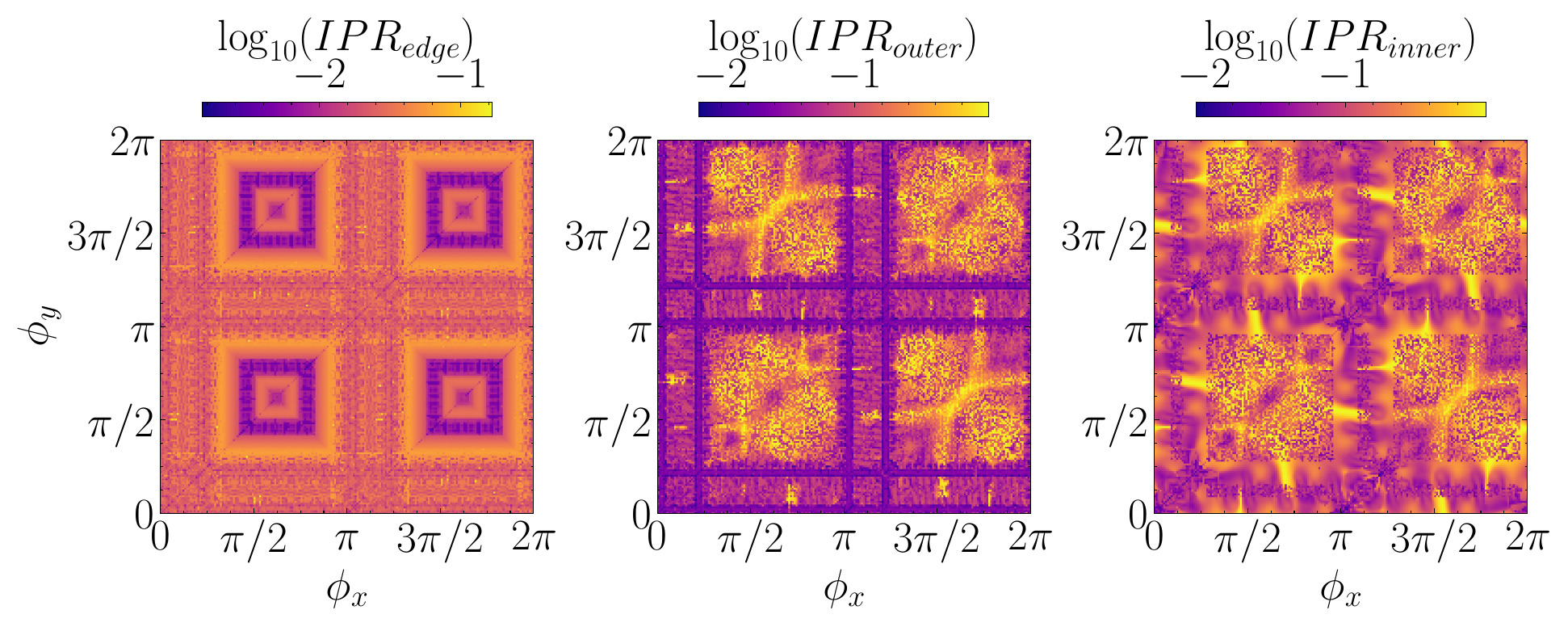}
\par\end{centering}
\caption{$\text{IPR}$ of the edge states, outer and inner corner modes, plotted
from left to right respectively, for $L=377$.\label{fig:IPR_edge_outer_inner}}
\end{figure}

As $E_{edge}\to0$, we expect the localization length of the corner
modes to increase as the states hybridize. In Fig.~\ref{fig:IPR_edge_outer_inner}
we plot the $\text{IPR}$ of the edge states, outer and inner corner
modes for $L=377$. As discussed previously, only the outer corner
modes hybridize, as can be seen by the clear pattern emerging in the
middle plot of Fig.~\ref{fig:IPR_edge_outer_inner}, corresponding
to the $\text{IPR}$ of the outer corner modes. As the edge states
cross the Fermi energy ($E_{edge}\to0$), the $\text{IPR}$ of the
outer corner modes decreases, indicating that the localization length
increases. Although $\text{IPR}$ alone is not enough to conclude
about the localization length of a state, for fully delocalized states
$\text{IPR}\approx\frac{1}{N}$, thus for the considered system size
($L=377$) we expect $\text{IPR}\approx10^{-5}$. Therefore, an $\text{IPR}$
of the order of $10^{-2}$ is a good indication that the corner modes
remain localized throughout the $(\phi_{x},\phi_{y})$ plane.

Overall, QPQI phases exhibit strong correlation between the localization
length of the corner modes and the edge gap (which is $\boldsymbol{\phi}$
dependent). As the edge states energy approach the Fermi level, they
localize (as opposed to the general behavior where edge states are
delocalized along the edge) moving towards the corners and hybridizing
with the corner modes. This hybridization ``pulls'' the outer corner
modes away from zero energy while increasing their localization length.
In the QI regime, the corner modes display oscillating localization
length with varying phase shifts, however to a far lesser degree than
in the QPQI phases. Furthermore, unlike the QPQI regime, there is
no apparent correlation between the localization length of the corner
modes and the energy of the edge states, suggesting that the oscillations
on the localization length have a different origin than corner-edge
hybridization.

\section{Conclusions}

\label{sec:Conclusion}

We studied the phase diagram of a quasiperiodic quadrupole chiral
insulator. The higher-order topological phases of the parent model
were found to be stable upon the addition of quasiperiodicity. More
interestingly, quasiperiodicity induces multiple reentrant topological
transitions into quadrupole insulating phases, expanding the known
class of chiral symmetric quadrupole insulators introduced in Ref.~\citep{PhysRevResearch.3.013239,Benalcazar2022}

Similarly to disorder-induced transitions into higher-order topological
Anderson insulators \citep{PhysRevLett.125.166801,Yang2021,hugo2023},
quasiperiodic-induced topological transitions occur under the usual
bulk gap closing mechanism were observed compatible an intrinsic HOTI.
However, in contrast to the disordered case, the eigenstates in the
gap edge were found to be ballistic across the transition, as in the
homogeneous limit, and multiple quasiperiodic-induced topological
transitions were found.

The main result of this work is the identification of a reentrant
topological phase in which corner states coexist with gapless edge
states, showcasing the rich interplay between quasiperiodicity and
intrinsic higher-order topology. Lifting the degeneracy of the corner
modes leads to the emergence of quantized fractional corner charges,
characteristic of a canonical quadrupole moment. The gapless edge
states can interact nontrivially with the zero-energy corner modes,
suggesting a complex edge--corner interplay that warrants further
investigation.

Our findings can be observed experimentally in different tunable platforms,
including electric circuits \citep{Imhof2018,PhysRevB.100.201406},
mechanical metamaterials \citep{10.1038/nature25156} and photonics
\citep{Mittal2019}.

\section*{Acknowledgements}

The authors acknowledge partial support from FCT-Portugal through
Grant No. UIDB/04650/2020. M. G. acknowledges partial support from
Fundação para a Ciência e Tecnologia (FCT-Portugal) through Grant
No. UID/CTM/04540/2019. M. G. acknowledges further support from FCT-Portugal
through the Grant No. SFRH/BD/145152/2019. We finally acknowledge
the Tianhe-2JK cluster at the Beijing Computational Science Research
Center (CSRC) and the OBLIVION supercomputer through Projects No.
2022.15834.CPCA.A1 and No. 2022.15910.CPCA.A1 (based at the High Performance
Computing Center---University of Évora) funded by the ENGAGE SKA
Research Infrastructure (Reference No. POCI-01-0145-FEDER-022217---COMPETE
2020 and the Foundation for Science and Technology, Portugal) and
by the BigData@UE project (Reference No. ALT20-03- 0246-FEDER-000033---FEDER)
and the Alentejo 2020 Regional Operational Program. Computer assistance
was provided by CSRC and the OBLIVION support team.

\bibliographystyle{apsrev4-1}
\bibliography{refs,library}

%merlin.mbs apsrev4-1.bst 2010-07-25 4.21a (PWD, AO, DPC) hacked
%Control: key (0)
%Control: author (72) initials jnrlst
%Control: editor formatted (1) identically to author
%Control: production of article title (-1) disabled
%Control: page (0) single
%Control: year (1) truncated
%Control: production of eprint (0) enabled
\begin{thebibliography}{104}%
\makeatletter
\providecommand \@ifxundefined [1]{%
 \@ifx{#1\undefined}
}%
\providecommand \@ifnum [1]{%
 \ifnum #1\expandafter \@firstoftwo
 \else \expandafter \@secondoftwo
 \fi
}%
\providecommand \@ifx [1]{%
 \ifx #1\expandafter \@firstoftwo
 \else \expandafter \@secondoftwo
 \fi
}%
\providecommand \natexlab [1]{#1}%
\providecommand \enquote  [1]{``#1''}%
\providecommand \bibnamefont  [1]{#1}%
\providecommand \bibfnamefont [1]{#1}%
\providecommand \citenamefont [1]{#1}%
\providecommand \href@noop [0]{\@secondoftwo}%
\providecommand \href [0]{\begingroup \@sanitize@url \@href}%
\providecommand \@href[1]{\@@startlink{#1}\@@href}%
\providecommand \@@href[1]{\endgroup#1\@@endlink}%
\providecommand \@sanitize@url [0]{\catcode `\\12\catcode `\$12\catcode
  `\&12\catcode `\#12\catcode `\^12\catcode `\_12\catcode `\%12\relax}%
\providecommand \@@startlink[1]{}%
\providecommand \@@endlink[0]{}%
\providecommand \url  [0]{\begingroup\@sanitize@url \@url }%
\providecommand \@url [1]{\endgroup\@href {#1}{\urlprefix }}%
\providecommand \urlprefix  [0]{URL }%
\providecommand \Eprint [0]{\href }%
\providecommand \doibase [0]{http://dx.doi.org/}%
\providecommand \selectlanguage [0]{\@gobble}%
\providecommand \bibinfo  [0]{\@secondoftwo}%
\providecommand \bibfield  [0]{\@secondoftwo}%
\providecommand \translation [1]{[#1]}%
\providecommand \BibitemOpen [0]{}%
\providecommand \bibitemStop [0]{}%
\providecommand \bibitemNoStop [0]{.\EOS\space}%
\providecommand \EOS [0]{\spacefactor3000\relax}%
\providecommand \BibitemShut  [1]{\csname bibitem#1\endcsname}%
\let\auto@bib@innerbib\@empty
%</preamble>
\bibitem [{\citenamefont {Benalcazar}\ \emph
  {et~al.}(2017{\natexlab{a}})\citenamefont {Benalcazar}, \citenamefont
  {Bernevig},\ and\ \citenamefont {Hughes}}]{Benalcazar2017}%
  \BibitemOpen
  \bibfield  {author} {\bibinfo {author} {\bibfnamefont {W.~A.}\ \bibnamefont
  {Benalcazar}}, \bibinfo {author} {\bibfnamefont {B.~A.}\ \bibnamefont
  {Bernevig}}, \ and\ \bibinfo {author} {\bibfnamefont {T.~L.}\ \bibnamefont
  {Hughes}},\ }\href {\doibase 10.1126/science.aah6442} {\bibfield  {journal}
  {\bibinfo  {journal} {Science}\ }\textbf {\bibinfo {volume} {357}},\ \bibinfo
  {pages} {61} (\bibinfo {year} {2017}{\natexlab{a}})}\BibitemShut {NoStop}%
\bibitem [{\citenamefont {Benalcazar}\ \emph
  {et~al.}(2017{\natexlab{b}})\citenamefont {Benalcazar}, \citenamefont
  {Bernevig},\ and\ \citenamefont {Hughes}}]{benalcazar2017_prb}%
  \BibitemOpen
  \bibfield  {author} {\bibinfo {author} {\bibfnamefont {W.~A.}\ \bibnamefont
  {Benalcazar}}, \bibinfo {author} {\bibfnamefont {B.~A.}\ \bibnamefont
  {Bernevig}}, \ and\ \bibinfo {author} {\bibfnamefont {T.~L.}\ \bibnamefont
  {Hughes}},\ }\href {\doibase 10.1103/PhysRevB.96.245115} {\bibfield
  {journal} {\bibinfo  {journal} {Physical Review B}\ }\textbf {\bibinfo
  {volume} {96}},\ \bibinfo {pages} {245115} (\bibinfo {year}
  {2017}{\natexlab{b}})}\BibitemShut {NoStop}%
\bibitem [{\citenamefont {Khalaf}\ \emph {et~al.}(2021)\citenamefont {Khalaf},
  \citenamefont {Benalcazar}, \citenamefont {Hughes},\ and\ \citenamefont
  {Queiroz}}]{PhysRevResearch.3.013239}%
  \BibitemOpen
  \bibfield  {author} {\bibinfo {author} {\bibfnamefont {E.}~\bibnamefont
  {Khalaf}}, \bibinfo {author} {\bibfnamefont {W.~A.}\ \bibnamefont
  {Benalcazar}}, \bibinfo {author} {\bibfnamefont {T.~L.}\ \bibnamefont
  {Hughes}}, \ and\ \bibinfo {author} {\bibfnamefont {R.}~\bibnamefont
  {Queiroz}},\ }\href {\doibase 10.1103/PhysRevResearch.3.013239} {\bibfield
  {journal} {\bibinfo  {journal} {Phys. Rev. Res.}\ }\textbf {\bibinfo {volume}
  {3}},\ \bibinfo {pages} {013239} (\bibinfo {year} {2021})}\BibitemShut
  {NoStop}%
\bibitem [{\citenamefont {Tao}\ \emph {et~al.}(2023)\citenamefont {Tao},
  \citenamefont {Wang},\ and\ \citenamefont {Xu}}]{tao2023quadrupole}%
  \BibitemOpen
  \bibfield  {author} {\bibinfo {author} {\bibfnamefont {Y.-L.}\ \bibnamefont
  {Tao}}, \bibinfo {author} {\bibfnamefont {J.-H.}\ \bibnamefont {Wang}}, \
  and\ \bibinfo {author} {\bibfnamefont {Y.}~\bibnamefont {Xu}},\ }\href@noop
  {} {\bibfield  {journal} {\bibinfo  {journal} {arXiv preprint
  arXiv:2307.00486}\ } (\bibinfo {year} {2023})}\BibitemShut {NoStop}%
\bibitem [{\citenamefont {Ghosh}\ \emph {et~al.}(2020)\citenamefont {Ghosh},
  \citenamefont {Paul},\ and\ \citenamefont {Saha}}]{Ghosh2020}%
  \BibitemOpen
  \bibfield  {author} {\bibinfo {author} {\bibfnamefont {A.~K.}\ \bibnamefont
  {Ghosh}}, \bibinfo {author} {\bibfnamefont {G.~C.}\ \bibnamefont {Paul}}, \
  and\ \bibinfo {author} {\bibfnamefont {A.}~\bibnamefont {Saha}},\ }\href
  {\doibase 10.1103/PhysRevB.101.235403} {\bibfield  {journal} {\bibinfo
  {journal} {Physical Review B}\ }\textbf {\bibinfo {volume} {101}},\ \bibinfo
  {pages} {235403} (\bibinfo {year} {2020})}\BibitemShut {NoStop}%
\bibitem [{\citenamefont {Schindler}\ \emph
  {et~al.}(2018{\natexlab{a}})\citenamefont {Schindler}, \citenamefont {Cook},
  \citenamefont {Vergniory}, \citenamefont {Wang}, \citenamefont {Parkin},
  \citenamefont {Bernevig},\ and\ \citenamefont {Neupert}}]{Schindler2018}%
  \BibitemOpen
  \bibfield  {author} {\bibinfo {author} {\bibfnamefont {F.}~\bibnamefont
  {Schindler}}, \bibinfo {author} {\bibfnamefont {A.~M.}\ \bibnamefont {Cook}},
  \bibinfo {author} {\bibfnamefont {M.~G.}\ \bibnamefont {Vergniory}}, \bibinfo
  {author} {\bibfnamefont {Z.}~\bibnamefont {Wang}}, \bibinfo {author}
  {\bibfnamefont {S.~S.~P.}\ \bibnamefont {Parkin}}, \bibinfo {author}
  {\bibfnamefont {B.~A.}\ \bibnamefont {Bernevig}}, \ and\ \bibinfo {author}
  {\bibfnamefont {T.}~\bibnamefont {Neupert}},\ }\href {\doibase
  10.1126/sciadv.aat0346} {\bibfield  {journal} {\bibinfo  {journal} {Science
  Advances}\ }\textbf {\bibinfo {volume} {4}} (\bibinfo {year}
  {2018}{\natexlab{a}}),\ 10.1126/sciadv.aat0346}\BibitemShut {NoStop}%
\bibitem [{\citenamefont {Neupert}\ and\ \citenamefont
  {Schindler}(2018)}]{Neupert2018}%
  \BibitemOpen
  \bibfield  {author} {\bibinfo {author} {\bibfnamefont {T.}~\bibnamefont
  {Neupert}}\ and\ \bibinfo {author} {\bibfnamefont {F.}~\bibnamefont
  {Schindler}},\ }in\ \href {\doibase 10.1007/978-3-319-76388-0_2} {\emph
  {\bibinfo {booktitle} {Topological Matter: Lectures from the Topological
  Matter School 2017}}},\ \bibinfo {editor} {edited by\ \bibinfo {editor}
  {\bibfnamefont {D.}~\bibnamefont {Bercioux}}, \bibinfo {editor}
  {\bibfnamefont {J.}~\bibnamefont {Cayssol}}, \bibinfo {editor} {\bibfnamefont
  {M.~G.}\ \bibnamefont {Vergniory}}, \ and\ \bibinfo {editor} {\bibfnamefont
  {M.}~\bibnamefont {{Reyes Calvo}}}}\ (\bibinfo  {publisher} {Springer
  International Publishing},\ \bibinfo {address} {Cham},\ \bibinfo {year}
  {2018})\ pp.\ \bibinfo {pages} {31--61}\BibitemShut {NoStop}%
\bibitem [{\citenamefont {Wang}\ \emph
  {et~al.}(2021{\natexlab{a}})\citenamefont {Wang}, \citenamefont {Zhou},
  \citenamefont {Lin},\ and\ \citenamefont {Bansil}}]{Wang2021}%
  \BibitemOpen
  \bibfield  {author} {\bibinfo {author} {\bibfnamefont {B.}~\bibnamefont
  {Wang}}, \bibinfo {author} {\bibfnamefont {X.}~\bibnamefont {Zhou}}, \bibinfo
  {author} {\bibfnamefont {H.}~\bibnamefont {Lin}}, \ and\ \bibinfo {author}
  {\bibfnamefont {A.}~\bibnamefont {Bansil}},\ }\href {\doibase
  10.1103/PhysRevB.104.L121108} {\bibfield  {journal} {\bibinfo  {journal}
  {Physical Review B}\ }\textbf {\bibinfo {volume} {104}},\ \bibinfo {pages}
  {L121108} (\bibinfo {year} {2021}{\natexlab{a}})}\BibitemShut {NoStop}%
\bibitem [{\citenamefont {Xu}\ \emph {et~al.}(2017)\citenamefont {Xu},
  \citenamefont {Xue},\ and\ \citenamefont {Wan}}]{Xu2017}%
  \BibitemOpen
  \bibfield  {author} {\bibinfo {author} {\bibfnamefont {Y.}~\bibnamefont
  {Xu}}, \bibinfo {author} {\bibfnamefont {R.}~\bibnamefont {Xue}}, \ and\
  \bibinfo {author} {\bibfnamefont {S.}~\bibnamefont {Wan}},\ }\href@noop {}
  {\bibfield  {journal} {\bibinfo  {journal} {arXiv}\ } (\bibinfo {year}
  {2017})}\BibitemShut {NoStop}%
\bibitem [{\citenamefont {Ezawa}(2018)}]{Ezawa2018}%
  \BibitemOpen
  \bibfield  {author} {\bibinfo {author} {\bibfnamefont {M.}~\bibnamefont
  {Ezawa}},\ }\href {\doibase 10.1103/PhysRevLett.120.026801} {\bibfield
  {journal} {\bibinfo  {journal} {Physical Review Letters}\ }\textbf {\bibinfo
  {volume} {120}},\ \bibinfo {pages} {026801} (\bibinfo {year}
  {2018})}\BibitemShut {NoStop}%
\bibitem [{\citenamefont {Lin}\ and\ \citenamefont {Hughes}(2018)}]{Lin2018}%
  \BibitemOpen
  \bibfield  {author} {\bibinfo {author} {\bibfnamefont {M.}~\bibnamefont
  {Lin}}\ and\ \bibinfo {author} {\bibfnamefont {T.~L.}\ \bibnamefont
  {Hughes}},\ }\href {\doibase 10.1103/PhysRevB.98.241103} {\bibfield
  {journal} {\bibinfo  {journal} {Physical Review B}\ }\textbf {\bibinfo
  {volume} {98}},\ \bibinfo {pages} {241103} (\bibinfo {year}
  {2018})}\BibitemShut {NoStop}%
\bibitem [{\citenamefont {Khalaf}(2018)}]{PhysRevB.97.205136}%
  \BibitemOpen
  \bibfield  {author} {\bibinfo {author} {\bibfnamefont {E.}~\bibnamefont
  {Khalaf}},\ }\href {\doibase 10.1103/PhysRevB.97.205136} {\bibfield
  {journal} {\bibinfo  {journal} {Phys. Rev. B}\ }\textbf {\bibinfo {volume}
  {97}},\ \bibinfo {pages} {205136} (\bibinfo {year} {2018})}\BibitemShut
  {NoStop}%
\bibitem [{\citenamefont {Zhang}\ \emph {et~al.}(2020)\citenamefont {Zhang},
  \citenamefont {Hsu},\ and\ \citenamefont {Das~Sarma}}]{PhysRevB.102.094503}%
  \BibitemOpen
  \bibfield  {author} {\bibinfo {author} {\bibfnamefont {R.-X.}\ \bibnamefont
  {Zhang}}, \bibinfo {author} {\bibfnamefont {Y.-T.}\ \bibnamefont {Hsu}}, \
  and\ \bibinfo {author} {\bibfnamefont {S.}~\bibnamefont {Das~Sarma}},\ }\href
  {\doibase 10.1103/PhysRevB.102.094503} {\bibfield  {journal} {\bibinfo
  {journal} {Phys. Rev. B}\ }\textbf {\bibinfo {volume} {102}},\ \bibinfo
  {pages} {094503} (\bibinfo {year} {2020})}\BibitemShut {NoStop}%
\bibitem [{\citenamefont {Manna}\ \emph {et~al.}(2022)\citenamefont {Manna},
  \citenamefont {Nandy},\ and\ \citenamefont {Roy}}]{PhysRevB.105.L201301}%
  \BibitemOpen
  \bibfield  {author} {\bibinfo {author} {\bibfnamefont {S.}~\bibnamefont
  {Manna}}, \bibinfo {author} {\bibfnamefont {S.}~\bibnamefont {Nandy}}, \ and\
  \bibinfo {author} {\bibfnamefont {B.}~\bibnamefont {Roy}},\ }\href {\doibase
  10.1103/PhysRevB.105.L201301} {\bibfield  {journal} {\bibinfo  {journal}
  {Phys. Rev. B}\ }\textbf {\bibinfo {volume} {105}},\ \bibinfo {pages}
  {L201301} (\bibinfo {year} {2022})}\BibitemShut {NoStop}%
\bibitem [{\citenamefont {Serra-Garcia}\ \emph {et~al.}(2018)\citenamefont
  {Serra-Garcia}, \citenamefont {Marc}, \citenamefont {Peri}, \citenamefont
  {Valerio}, \citenamefont {Süsstrunk}, \citenamefont {Roman}, \citenamefont
  {Bilal}, \citenamefont {R.}, \citenamefont {Larsen}, \citenamefont {Tom},
  \citenamefont {Villanueva}, \citenamefont {Guillermo}, \citenamefont
  {Huber},\ and\ \citenamefont {D.}}]{10.1038/nature25156}%
  \BibitemOpen
  \bibfield  {author} {\bibinfo {author} {\bibnamefont {Serra-Garcia}},
  \bibinfo {author} {\bibnamefont {Marc}}, \bibinfo {author} {\bibnamefont
  {Peri}}, \bibinfo {author} {\bibnamefont {Valerio}}, \bibinfo {author}
  {\bibnamefont {Süsstrunk}}, \bibinfo {author} {\bibnamefont {Roman}},
  \bibinfo {author} {\bibnamefont {Bilal}}, \bibinfo {author} {\bibfnamefont
  {O.}~\bibnamefont {R.}}, \bibinfo {author} {\bibnamefont {Larsen}}, \bibinfo
  {author} {\bibnamefont {Tom}}, \bibinfo {author} {\bibnamefont {Villanueva}},
  \bibinfo {author} {\bibfnamefont {L.}~\bibnamefont {Guillermo}}, \bibinfo
  {author} {\bibnamefont {Huber}}, \ and\ \bibinfo {author} {\bibfnamefont
  {S.}~\bibnamefont {D.}},\ }\href {\doibase 10.1038/nature25156} {\bibfield
  {journal} {\bibinfo  {journal} {Nature}\ }\textbf {\bibinfo {volume} {555}},\
  \bibinfo {pages} {342} (\bibinfo {year} {2018})}\BibitemShut {NoStop}%
\bibitem [{\citenamefont {Bao}\ \emph {et~al.}(2019)\citenamefont {Bao},
  \citenamefont {Zou}, \citenamefont {Zhang}, \citenamefont {He}, \citenamefont
  {Sun},\ and\ \citenamefont {Zhang}}]{PhysRevB.100.201406}%
  \BibitemOpen
  \bibfield  {author} {\bibinfo {author} {\bibfnamefont {J.}~\bibnamefont
  {Bao}}, \bibinfo {author} {\bibfnamefont {D.}~\bibnamefont {Zou}}, \bibinfo
  {author} {\bibfnamefont {W.}~\bibnamefont {Zhang}}, \bibinfo {author}
  {\bibfnamefont {W.}~\bibnamefont {He}}, \bibinfo {author} {\bibfnamefont
  {H.}~\bibnamefont {Sun}}, \ and\ \bibinfo {author} {\bibfnamefont
  {X.}~\bibnamefont {Zhang}},\ }\href {\doibase 10.1103/PhysRevB.100.201406}
  {\bibfield  {journal} {\bibinfo  {journal} {Phys. Rev. B}\ }\textbf {\bibinfo
  {volume} {100}},\ \bibinfo {pages} {201406} (\bibinfo {year}
  {2019})}\BibitemShut {NoStop}%
\bibitem [{\citenamefont {Zhang}\ \emph {et~al.}(2021)\citenamefont {Zhang},
  \citenamefont {Zou}, \citenamefont {Pei}, \citenamefont {He}, \citenamefont
  {Bao}, \citenamefont {Sun},\ and\ \citenamefont {Zhang}}]{Zhang2021}%
  \BibitemOpen
  \bibfield  {author} {\bibinfo {author} {\bibfnamefont {W.}~\bibnamefont
  {Zhang}}, \bibinfo {author} {\bibfnamefont {D.}~\bibnamefont {Zou}}, \bibinfo
  {author} {\bibfnamefont {Q.}~\bibnamefont {Pei}}, \bibinfo {author}
  {\bibfnamefont {W.}~\bibnamefont {He}}, \bibinfo {author} {\bibfnamefont
  {J.}~\bibnamefont {Bao}}, \bibinfo {author} {\bibfnamefont {H.}~\bibnamefont
  {Sun}}, \ and\ \bibinfo {author} {\bibfnamefont {X.}~\bibnamefont {Zhang}},\
  }\href {\doibase 10.1103/PhysRevLett.126.146802} {\bibfield  {journal}
  {\bibinfo  {journal} {Physical Review Letters}\ }\textbf {\bibinfo {volume}
  {126}},\ \bibinfo {pages} {146802} (\bibinfo {year} {2021})}\BibitemShut
  {NoStop}%
\bibitem [{\citenamefont {Schindler}\ \emph
  {et~al.}(2018{\natexlab{b}})\citenamefont {Schindler}, \citenamefont {Wang},
  \citenamefont {Vergniory}, \citenamefont {Cook}, \citenamefont {Murani},
  \citenamefont {Sengupta}, \citenamefont {Kasumov}, \citenamefont {Deblock},
  \citenamefont {Jeon}, \citenamefont {Drozdov}, \citenamefont {Bouchiat},
  \citenamefont {Gu{\'e}ron}, \citenamefont {Yazdani}, \citenamefont
  {Bernevig},\ and\ \citenamefont {Neupert}}]{Schindler2018Bismuth}%
  \BibitemOpen
  \bibfield  {author} {\bibinfo {author} {\bibfnamefont {F.}~\bibnamefont
  {Schindler}}, \bibinfo {author} {\bibfnamefont {Z.}~\bibnamefont {Wang}},
  \bibinfo {author} {\bibfnamefont {M.~G.}\ \bibnamefont {Vergniory}}, \bibinfo
  {author} {\bibfnamefont {A.~M.}\ \bibnamefont {Cook}}, \bibinfo {author}
  {\bibfnamefont {A.}~\bibnamefont {Murani}}, \bibinfo {author} {\bibfnamefont
  {S.}~\bibnamefont {Sengupta}}, \bibinfo {author} {\bibfnamefont {A.~Y.}\
  \bibnamefont {Kasumov}}, \bibinfo {author} {\bibfnamefont {R.}~\bibnamefont
  {Deblock}}, \bibinfo {author} {\bibfnamefont {S.}~\bibnamefont {Jeon}},
  \bibinfo {author} {\bibfnamefont {I.}~\bibnamefont {Drozdov}}, \bibinfo
  {author} {\bibfnamefont {H.}~\bibnamefont {Bouchiat}}, \bibinfo {author}
  {\bibfnamefont {S.}~\bibnamefont {Gu{\'e}ron}}, \bibinfo {author}
  {\bibfnamefont {A.}~\bibnamefont {Yazdani}}, \bibinfo {author} {\bibfnamefont
  {B.~A.}\ \bibnamefont {Bernevig}}, \ and\ \bibinfo {author} {\bibfnamefont
  {T.}~\bibnamefont {Neupert}},\ }\href {\doibase 10.1038/s41567-018-0224-7}
  {\bibfield  {journal} {\bibinfo  {journal} {Nature Physics}\ }\textbf
  {\bibinfo {volume} {14}},\ \bibinfo {pages} {918} (\bibinfo {year}
  {2018}{\natexlab{b}})}\BibitemShut {NoStop}%
\bibitem [{\citenamefont {Shumiya}\ \emph {et~al.}(2022)\citenamefont
  {Shumiya}, \citenamefont {Hossain}, \citenamefont {Yin}, \citenamefont
  {Wang}, \citenamefont {Litskevich}, \citenamefont {Yoon}, \citenamefont {Li},
  \citenamefont {Yang}, \citenamefont {Jiang}, \citenamefont {Cheng},
  \citenamefont {Lin}, \citenamefont {Zhang}, \citenamefont {Cheng},
  \citenamefont {Cochran}, \citenamefont {Multer}, \citenamefont {Yang},
  \citenamefont {Casas}, \citenamefont {Chang}, \citenamefont {Neupert},
  \citenamefont {Yuan}, \citenamefont {Jia}, \citenamefont {Lin}, \citenamefont
  {Yao}, \citenamefont {Balicas}, \citenamefont {Zhang}, \citenamefont {Yao},\
  and\ \citenamefont {Hasan}}]{Shumiya2022}%
  \BibitemOpen
  \bibfield  {author} {\bibinfo {author} {\bibfnamefont {N.}~\bibnamefont
  {Shumiya}}, \bibinfo {author} {\bibfnamefont {M.~S.}\ \bibnamefont
  {Hossain}}, \bibinfo {author} {\bibfnamefont {J.-X.}\ \bibnamefont {Yin}},
  \bibinfo {author} {\bibfnamefont {Z.}~\bibnamefont {Wang}}, \bibinfo {author}
  {\bibfnamefont {M.}~\bibnamefont {Litskevich}}, \bibinfo {author}
  {\bibfnamefont {C.}~\bibnamefont {Yoon}}, \bibinfo {author} {\bibfnamefont
  {Y.}~\bibnamefont {Li}}, \bibinfo {author} {\bibfnamefont {Y.}~\bibnamefont
  {Yang}}, \bibinfo {author} {\bibfnamefont {Y.-X.}\ \bibnamefont {Jiang}},
  \bibinfo {author} {\bibfnamefont {G.}~\bibnamefont {Cheng}}, \bibinfo
  {author} {\bibfnamefont {Y.-C.}\ \bibnamefont {Lin}}, \bibinfo {author}
  {\bibfnamefont {Q.}~\bibnamefont {Zhang}}, \bibinfo {author} {\bibfnamefont
  {Z.-J.}\ \bibnamefont {Cheng}}, \bibinfo {author} {\bibfnamefont {T.~A.}\
  \bibnamefont {Cochran}}, \bibinfo {author} {\bibfnamefont {D.}~\bibnamefont
  {Multer}}, \bibinfo {author} {\bibfnamefont {X.~P.}\ \bibnamefont {Yang}},
  \bibinfo {author} {\bibfnamefont {B.}~\bibnamefont {Casas}}, \bibinfo
  {author} {\bibfnamefont {T.-R.}\ \bibnamefont {Chang}}, \bibinfo {author}
  {\bibfnamefont {T.}~\bibnamefont {Neupert}}, \bibinfo {author} {\bibfnamefont
  {Z.}~\bibnamefont {Yuan}}, \bibinfo {author} {\bibfnamefont {S.}~\bibnamefont
  {Jia}}, \bibinfo {author} {\bibfnamefont {H.}~\bibnamefont {Lin}}, \bibinfo
  {author} {\bibfnamefont {N.}~\bibnamefont {Yao}}, \bibinfo {author}
  {\bibfnamefont {L.}~\bibnamefont {Balicas}}, \bibinfo {author} {\bibfnamefont
  {F.}~\bibnamefont {Zhang}}, \bibinfo {author} {\bibfnamefont
  {Y.}~\bibnamefont {Yao}}, \ and\ \bibinfo {author} {\bibfnamefont {M.~Z.}\
  \bibnamefont {Hasan}},\ }\href {\doibase 10.1038/s41563-022-01304-3}
  {\bibfield  {journal} {\bibinfo  {journal} {Nature Materials}\ }\textbf
  {\bibinfo {volume} {21}},\ \bibinfo {pages} {1111} (\bibinfo {year}
  {2022})}\BibitemShut {NoStop}%
\bibitem [{\citenamefont {Lee}\ \emph {et~al.}(2023)\citenamefont {Lee},
  \citenamefont {Kwon}, \citenamefont {Lee}, \citenamefont {Park},
  \citenamefont {Cha}, \citenamefont {Watanabe}, \citenamefont {Taniguchi},
  \citenamefont {Jo},\ and\ \citenamefont {Choi}}]{Lee2023}%
  \BibitemOpen
  \bibfield  {author} {\bibinfo {author} {\bibfnamefont {J.}~\bibnamefont
  {Lee}}, \bibinfo {author} {\bibfnamefont {J.}~\bibnamefont {Kwon}}, \bibinfo
  {author} {\bibfnamefont {E.}~\bibnamefont {Lee}}, \bibinfo {author}
  {\bibfnamefont {J.}~\bibnamefont {Park}}, \bibinfo {author} {\bibfnamefont
  {S.}~\bibnamefont {Cha}}, \bibinfo {author} {\bibfnamefont {K.}~\bibnamefont
  {Watanabe}}, \bibinfo {author} {\bibfnamefont {T.}~\bibnamefont {Taniguchi}},
  \bibinfo {author} {\bibfnamefont {M.-H.}\ \bibnamefont {Jo}}, \ and\ \bibinfo
  {author} {\bibfnamefont {H.}~\bibnamefont {Choi}},\ }\href {\doibase
  10.1038/s41467-023-37482-0} {\bibfield  {journal} {\bibinfo  {journal}
  {Nature Communications}\ }\textbf {\bibinfo {volume} {14}},\ \bibinfo {pages}
  {1801} (\bibinfo {year} {2023})}\BibitemShut {NoStop}%
\bibitem [{\citenamefont {L{\'{o}}io}\ \emph {et~al.}(2024)\citenamefont
  {L{\'{o}}io}, \citenamefont {Gon{\c{c}}alves}, \citenamefont {Ribeiro},\ and\
  \citenamefont {Castro}}]{hugo2023}%
  \BibitemOpen
  \bibfield  {author} {\bibinfo {author} {\bibfnamefont {H.}~\bibnamefont
  {L{\'{o}}io}}, \bibinfo {author} {\bibfnamefont {M.}~\bibnamefont
  {Gon{\c{c}}alves}}, \bibinfo {author} {\bibfnamefont {P.}~\bibnamefont
  {Ribeiro}}, \ and\ \bibinfo {author} {\bibfnamefont {E.~V.}\ \bibnamefont
  {Castro}},\ }\href {\doibase 10.1103/PhysRevB.109.014204} {\bibfield
  {journal} {\bibinfo  {journal} {Physical Review B}\ }\textbf {\bibinfo
  {volume} {109}},\ \bibinfo {pages} {014204} (\bibinfo {year}
  {2024})}\BibitemShut {NoStop}%
\bibitem [{\citenamefont {Wang}\ and\ \citenamefont
  {Wang}(2020)}]{PhysRevResearch.2.033521}%
  \BibitemOpen
  \bibfield  {author} {\bibinfo {author} {\bibfnamefont {C.}~\bibnamefont
  {Wang}}\ and\ \bibinfo {author} {\bibfnamefont {X.~R.}\ \bibnamefont
  {Wang}},\ }\href {\doibase 10.1103/PhysRevResearch.2.033521} {\bibfield
  {journal} {\bibinfo  {journal} {Phys. Rev. Res.}\ }\textbf {\bibinfo {volume}
  {2}},\ \bibinfo {pages} {033521} (\bibinfo {year} {2020})}\BibitemShut
  {NoStop}%
\bibitem [{\citenamefont {Wang}\ \emph
  {et~al.}(2021{\natexlab{b}})\citenamefont {Wang}, \citenamefont {Yang},
  \citenamefont {Dai},\ and\ \citenamefont {Xu}}]{PhysRevLett.126.206404}%
  \BibitemOpen
  \bibfield  {author} {\bibinfo {author} {\bibfnamefont {J.-H.}\ \bibnamefont
  {Wang}}, \bibinfo {author} {\bibfnamefont {Y.-B.}\ \bibnamefont {Yang}},
  \bibinfo {author} {\bibfnamefont {N.}~\bibnamefont {Dai}}, \ and\ \bibinfo
  {author} {\bibfnamefont {Y.}~\bibnamefont {Xu}},\ }\href {\doibase
  10.1103/PhysRevLett.126.206404} {\bibfield  {journal} {\bibinfo  {journal}
  {Phys. Rev. Lett.}\ }\textbf {\bibinfo {volume} {126}},\ \bibinfo {pages}
  {206404} (\bibinfo {year} {2021}{\natexlab{b}})}\BibitemShut {NoStop}%
\bibitem [{\citenamefont {Peng}\ \emph {et~al.}(2021)\citenamefont {Peng},
  \citenamefont {Hua}, \citenamefont {Chen}, \citenamefont {Liu}, \citenamefont
  {Xu},\ and\ \citenamefont {Zhou}}]{Peng2021}%
  \BibitemOpen
  \bibfield  {author} {\bibinfo {author} {\bibfnamefont {T.}~\bibnamefont
  {Peng}}, \bibinfo {author} {\bibfnamefont {C.-B.}\ \bibnamefont {Hua}},
  \bibinfo {author} {\bibfnamefont {R.}~\bibnamefont {Chen}}, \bibinfo {author}
  {\bibfnamefont {Z.-R.}\ \bibnamefont {Liu}}, \bibinfo {author} {\bibfnamefont
  {D.-H.}\ \bibnamefont {Xu}}, \ and\ \bibinfo {author} {\bibfnamefont
  {B.}~\bibnamefont {Zhou}},\ }\href {\doibase 10.1103/PhysRevB.104.245302}
  {\bibfield  {journal} {\bibinfo  {journal} {Physical Review B}\ }\textbf
  {\bibinfo {volume} {104}},\ \bibinfo {pages} {245302} (\bibinfo {year}
  {2021})}\BibitemShut {NoStop}%
\bibitem [{\citenamefont {Johansson}\ and\ \citenamefont
  {Riklund}(1991)}]{PhysRevB.43.13468}%
  \BibitemOpen
  \bibfield  {author} {\bibinfo {author} {\bibfnamefont {M.}~\bibnamefont
  {Johansson}}\ and\ \bibinfo {author} {\bibfnamefont {R.}~\bibnamefont
  {Riklund}},\ }\href {\doibase 10.1103/PhysRevB.43.13468} {\bibfield
  {journal} {\bibinfo  {journal} {Phys. Rev. B}\ }\textbf {\bibinfo {volume}
  {43}},\ \bibinfo {pages} {13468} (\bibinfo {year} {1991})}\BibitemShut
  {NoStop}%
\bibitem [{\citenamefont {Biddle}\ and\ \citenamefont {{Das
  Sarma}}(2010)}]{PhysRevLett.104.070601}%
  \BibitemOpen
  \bibfield  {author} {\bibinfo {author} {\bibfnamefont {J.}~\bibnamefont
  {Biddle}}\ and\ \bibinfo {author} {\bibfnamefont {S.}~\bibnamefont {{Das
  Sarma}}},\ }\href {\doibase 10.1103/PhysRevLett.104.070601} {\bibfield
  {journal} {\bibinfo  {journal} {Phys. Rev. Lett.}\ }\textbf {\bibinfo
  {volume} {104}},\ \bibinfo {pages} {70601} (\bibinfo {year}
  {2010})}\BibitemShut {NoStop}%
\bibitem [{\citenamefont {Bodyfelt}\ \emph {et~al.}(2014)\citenamefont
  {Bodyfelt}, \citenamefont {Leykam}, \citenamefont {Danieli}, \citenamefont
  {Yu},\ and\ \citenamefont {Flach}}]{PhysRevLett.113.236403}%
  \BibitemOpen
  \bibfield  {author} {\bibinfo {author} {\bibfnamefont {J.~D.}\ \bibnamefont
  {Bodyfelt}}, \bibinfo {author} {\bibfnamefont {D.}~\bibnamefont {Leykam}},
  \bibinfo {author} {\bibfnamefont {C.}~\bibnamefont {Danieli}}, \bibinfo
  {author} {\bibfnamefont {X.}~\bibnamefont {Yu}}, \ and\ \bibinfo {author}
  {\bibfnamefont {S.}~\bibnamefont {Flach}},\ }\href {\doibase
  10.1103/PhysRevLett.113.236403} {\bibfield  {journal} {\bibinfo  {journal}
  {Phys. Rev. Lett.}\ }\textbf {\bibinfo {volume} {113}},\ \bibinfo {pages}
  {236403} (\bibinfo {year} {2014})}\BibitemShut {NoStop}%
\bibitem [{\citenamefont {Liu}\ \emph {et~al.}(2015)\citenamefont {Liu},
  \citenamefont {Ghosh},\ and\ \citenamefont {Chong}}]{Liu2015}%
  \BibitemOpen
  \bibfield  {author} {\bibinfo {author} {\bibfnamefont {F.}~\bibnamefont
  {Liu}}, \bibinfo {author} {\bibfnamefont {S.}~\bibnamefont {Ghosh}}, \ and\
  \bibinfo {author} {\bibfnamefont {Y.~D.}\ \bibnamefont {Chong}},\ }\href
  {\doibase 10.1103/PhysRevB.91.014108} {\bibfield  {journal} {\bibinfo
  {journal} {Phys. Rev. B - Condens. Matter Mater. Phys.}\ }\textbf {\bibinfo
  {volume} {91}},\ \bibinfo {pages} {014108} (\bibinfo {year}
  {2015})}\BibitemShut {NoStop}%
\bibitem [{\citenamefont {Danieli}\ \emph {et~al.}(2015)\citenamefont
  {Danieli}, \citenamefont {Bodyfelt},\ and\ \citenamefont
  {Flach}}]{PhysRevB.91.235134}%
  \BibitemOpen
  \bibfield  {author} {\bibinfo {author} {\bibfnamefont {C.}~\bibnamefont
  {Danieli}}, \bibinfo {author} {\bibfnamefont {J.~D.}\ \bibnamefont
  {Bodyfelt}}, \ and\ \bibinfo {author} {\bibfnamefont {S.}~\bibnamefont
  {Flach}},\ }\href {\doibase 10.1103/PhysRevB.91.235134} {\bibfield  {journal}
  {\bibinfo  {journal} {Phys. Rev. B}\ }\textbf {\bibinfo {volume} {91}},\
  \bibinfo {pages} {235134} (\bibinfo {year} {2015})}\BibitemShut {NoStop}%
\bibitem [{\citenamefont {Ganeshan}\ \emph {et~al.}(2015)\citenamefont
  {Ganeshan}, \citenamefont {Pixley},\ and\ \citenamefont {{Das
  Sarma}}}]{PhysRevLett.114.146601}%
  \BibitemOpen
  \bibfield  {author} {\bibinfo {author} {\bibfnamefont {S.}~\bibnamefont
  {Ganeshan}}, \bibinfo {author} {\bibfnamefont {J.~H.}\ \bibnamefont
  {Pixley}}, \ and\ \bibinfo {author} {\bibfnamefont {S.}~\bibnamefont {{Das
  Sarma}}},\ }\href {\doibase 10.1103/PhysRevLett.114.146601} {\bibfield
  {journal} {\bibinfo  {journal} {Phys. Rev. Lett.}\ }\textbf {\bibinfo
  {volume} {114}},\ \bibinfo {pages} {146601} (\bibinfo {year}
  {2015})}\BibitemShut {NoStop}%
\bibitem [{\citenamefont {Gon{\c{c}}alves}\ \emph
  {et~al.}(2022{\natexlab{a}})\citenamefont {Gon{\c{c}}alves}, \citenamefont
  {Amorim}, \citenamefont {Castro},\ and\ \citenamefont
  {Ribeiro}}]{10.21468/SciPostPhys.13.3.046}%
  \BibitemOpen
  \bibfield  {author} {\bibinfo {author} {\bibfnamefont {M.}~\bibnamefont
  {Gon{\c{c}}alves}}, \bibinfo {author} {\bibfnamefont {B.}~\bibnamefont
  {Amorim}}, \bibinfo {author} {\bibfnamefont {E.~V.}\ \bibnamefont {Castro}},
  \ and\ \bibinfo {author} {\bibfnamefont {P.}~\bibnamefont {Ribeiro}},\ }\href
  {\doibase 10.21468/SciPostPhys.13.3.046} {\bibfield  {journal} {\bibinfo
  {journal} {SciPost Phys.}\ }\textbf {\bibinfo {volume} {13}},\ \bibinfo
  {pages} {046} (\bibinfo {year} {2022}{\natexlab{a}})}\BibitemShut {NoStop}%
\bibitem [{\citenamefont {Liu}\ \emph {et~al.}(2022)\citenamefont {Liu},
  \citenamefont {Xia}, \citenamefont {Longhi},\ and\ \citenamefont
  {Sanchez-Palencia}}]{anomScipost}%
  \BibitemOpen
  \bibfield  {author} {\bibinfo {author} {\bibfnamefont {T.}~\bibnamefont
  {Liu}}, \bibinfo {author} {\bibfnamefont {X.}~\bibnamefont {Xia}}, \bibinfo
  {author} {\bibfnamefont {S.}~\bibnamefont {Longhi}}, \ and\ \bibinfo {author}
  {\bibfnamefont {L.}~\bibnamefont {Sanchez-Palencia}},\ }\href {\doibase
  10.21468/SciPostPhys.12.1.027} {\bibfield  {journal} {\bibinfo  {journal}
  {SciPost Phys.}\ }\textbf {\bibinfo {volume} {12}},\ \bibinfo {pages} {27}
  (\bibinfo {year} {2022})}\BibitemShut {NoStop}%
\bibitem [{\citenamefont {Gon\ifmmode~\mbox{\c{c}}\else \c{c}\fi{}alves}\ \emph
  {et~al.}(2023{\natexlab{a}})\citenamefont {Gon\ifmmode~\mbox{\c{c}}\else
  \c{c}\fi{}alves}, \citenamefont {Amorim}, \citenamefont {Castro},\ and\
  \citenamefont {Ribeiro}}]{PhysRevB.108.L100201}%
  \BibitemOpen
  \bibfield  {author} {\bibinfo {author} {\bibfnamefont {M.}~\bibnamefont
  {Gon\ifmmode~\mbox{\c{c}}\else \c{c}\fi{}alves}}, \bibinfo {author}
  {\bibfnamefont {B.}~\bibnamefont {Amorim}}, \bibinfo {author} {\bibfnamefont
  {E.~V.}\ \bibnamefont {Castro}}, \ and\ \bibinfo {author} {\bibfnamefont
  {P.}~\bibnamefont {Ribeiro}},\ }\href {\doibase 10.1103/PhysRevB.108.L100201}
  {\bibfield  {journal} {\bibinfo  {journal} {Phys. Rev. B}\ }\textbf {\bibinfo
  {volume} {108}},\ \bibinfo {pages} {L100201} (\bibinfo {year}
  {2023}{\natexlab{a}})}\BibitemShut {NoStop}%
\bibitem [{\citenamefont {Gon\ifmmode~\mbox{\c{c}}\else \c{c}\fi{}alves}\ \emph
  {et~al.}(2023{\natexlab{b}})\citenamefont {Gon\ifmmode~\mbox{\c{c}}\else
  \c{c}\fi{}alves}, \citenamefont {Amorim}, \citenamefont {Castro},\ and\
  \citenamefont {Ribeiro}}]{PhysRevLett.131.186303}%
  \BibitemOpen
  \bibfield  {author} {\bibinfo {author} {\bibfnamefont {M.}~\bibnamefont
  {Gon\ifmmode~\mbox{\c{c}}\else \c{c}\fi{}alves}}, \bibinfo {author}
  {\bibfnamefont {B.}~\bibnamefont {Amorim}}, \bibinfo {author} {\bibfnamefont
  {E.~V.}\ \bibnamefont {Castro}}, \ and\ \bibinfo {author} {\bibfnamefont
  {P.}~\bibnamefont {Ribeiro}},\ }\href {\doibase
  10.1103/PhysRevLett.131.186303} {\bibfield  {journal} {\bibinfo  {journal}
  {Phys. Rev. Lett.}\ }\textbf {\bibinfo {volume} {131}},\ \bibinfo {pages}
  {186303} (\bibinfo {year} {2023}{\natexlab{b}})}\BibitemShut {NoStop}%
\bibitem [{\citenamefont {Huang}\ \emph {et~al.}(2016)\citenamefont {Huang},
  \citenamefont {Ye}, \citenamefont {Chen}, \citenamefont {Kartashov},
  \citenamefont {Konotop},\ and\ \citenamefont {Torner}}]{Huang2016a}%
  \BibitemOpen
  \bibfield  {author} {\bibinfo {author} {\bibfnamefont {C.}~\bibnamefont
  {Huang}}, \bibinfo {author} {\bibfnamefont {F.}~\bibnamefont {Ye}}, \bibinfo
  {author} {\bibfnamefont {X.}~\bibnamefont {Chen}}, \bibinfo {author}
  {\bibfnamefont {Y.~V.}\ \bibnamefont {Kartashov}}, \bibinfo {author}
  {\bibfnamefont {V.~V.}\ \bibnamefont {Konotop}}, \ and\ \bibinfo {author}
  {\bibfnamefont {L.}~\bibnamefont {Torner}},\ }\href {\doibase
  10.1038/srep32546} {\bibfield  {journal} {\bibinfo  {journal} {Scientific
  Reports}\ }\textbf {\bibinfo {volume} {6}},\ \bibinfo {pages} {32546}
  (\bibinfo {year} {2016})}\BibitemShut {NoStop}%
\bibitem [{\citenamefont {Pixley}\ \emph {et~al.}(2018)\citenamefont {Pixley},
  \citenamefont {Wilson}, \citenamefont {Huse},\ and\ \citenamefont
  {Gopalakrishnan}}]{PhysRevLett.120.207604}%
  \BibitemOpen
  \bibfield  {author} {\bibinfo {author} {\bibfnamefont {J.~H.}\ \bibnamefont
  {Pixley}}, \bibinfo {author} {\bibfnamefont {J.~H.}\ \bibnamefont {Wilson}},
  \bibinfo {author} {\bibfnamefont {D.~A.}\ \bibnamefont {Huse}}, \ and\
  \bibinfo {author} {\bibfnamefont {S.}~\bibnamefont {Gopalakrishnan}},\ }\href
  {\doibase 10.1103/PhysRevLett.120.207604} {\bibfield  {journal} {\bibinfo
  {journal} {Phys. Rev. Lett.}\ }\textbf {\bibinfo {volume} {120}},\ \bibinfo
  {pages} {207604} (\bibinfo {year} {2018})}\BibitemShut {NoStop}%
\bibitem [{\citenamefont {Park}\ \emph {et~al.}(2019)\citenamefont {Park},
  \citenamefont {Kim},\ and\ \citenamefont {Lee}}]{Park2018}%
  \BibitemOpen
  \bibfield  {author} {\bibinfo {author} {\bibfnamefont {M.~J.}\ \bibnamefont
  {Park}}, \bibinfo {author} {\bibfnamefont {H.~S.}\ \bibnamefont {Kim}}, \
  and\ \bibinfo {author} {\bibfnamefont {S.}~\bibnamefont {Lee}},\ }\href
  {\doibase 10.1103/PhysRevB.99.245401} {\bibfield  {journal} {\bibinfo
  {journal} {Phys. Rev. B}\ }\textbf {\bibinfo {volume} {99}},\ \bibinfo
  {pages} {245401} (\bibinfo {year} {2019})},\ \Eprint
  {http://arxiv.org/abs/1812.09170} {arXiv:1812.09170} \BibitemShut {NoStop}%
\bibitem [{\citenamefont {Bordia}\ \emph {et~al.}(2017)\citenamefont {Bordia},
  \citenamefont {L\"uschen}, \citenamefont {Scherg}, \citenamefont
  {Gopalakrishnan}, \citenamefont {Knap}, \citenamefont {Schneider},\ and\
  \citenamefont {Bloch}}]{PhysRevX.7.041047}%
  \BibitemOpen
  \bibfield  {author} {\bibinfo {author} {\bibfnamefont {P.}~\bibnamefont
  {Bordia}}, \bibinfo {author} {\bibfnamefont {H.}~\bibnamefont {L\"uschen}},
  \bibinfo {author} {\bibfnamefont {S.}~\bibnamefont {Scherg}}, \bibinfo
  {author} {\bibfnamefont {S.}~\bibnamefont {Gopalakrishnan}}, \bibinfo
  {author} {\bibfnamefont {M.}~\bibnamefont {Knap}}, \bibinfo {author}
  {\bibfnamefont {U.}~\bibnamefont {Schneider}}, \ and\ \bibinfo {author}
  {\bibfnamefont {I.}~\bibnamefont {Bloch}},\ }\href {\doibase
  10.1103/PhysRevX.7.041047} {\bibfield  {journal} {\bibinfo  {journal} {Phys.
  Rev. X}\ }\textbf {\bibinfo {volume} {7}},\ \bibinfo {pages} {041047}
  (\bibinfo {year} {2017})}\BibitemShut {NoStop}%
\bibitem [{\citenamefont {Huang}\ and\ \citenamefont
  {Liu}(2019)}]{PhysRevB.100.144202}%
  \BibitemOpen
  \bibfield  {author} {\bibinfo {author} {\bibfnamefont {B.}~\bibnamefont
  {Huang}}\ and\ \bibinfo {author} {\bibfnamefont {W.~V.}\ \bibnamefont
  {Liu}},\ }\href {\doibase 10.1103/PhysRevB.100.144202} {\bibfield  {journal}
  {\bibinfo  {journal} {Phys. Rev. B}\ }\textbf {\bibinfo {volume} {100}},\
  \bibinfo {pages} {144202} (\bibinfo {year} {2019})}\BibitemShut {NoStop}%
\bibitem [{\citenamefont {Fu}\ \emph {et~al.}(2020)\citenamefont {Fu},
  \citenamefont {K{\"{o}}nig}, \citenamefont {Wilson}, \citenamefont {Chou},\
  and\ \citenamefont {Pixley}}]{Fu2020}%
  \BibitemOpen
  \bibfield  {author} {\bibinfo {author} {\bibfnamefont {Y.}~\bibnamefont
  {Fu}}, \bibinfo {author} {\bibfnamefont {E.~J.}\ \bibnamefont {K{\"{o}}nig}},
  \bibinfo {author} {\bibfnamefont {J.~H.}\ \bibnamefont {Wilson}}, \bibinfo
  {author} {\bibfnamefont {Y.-Z.}\ \bibnamefont {Chou}}, \ and\ \bibinfo
  {author} {\bibfnamefont {J.~H.}\ \bibnamefont {Pixley}},\ }\href {\doibase
  10.1038/s41535-020-00271-9} {\bibfield  {journal} {\bibinfo  {journal} {npj
  Quantum Materials}\ }\textbf {\bibinfo {volume} {5}},\ \bibinfo {pages} {71}
  (\bibinfo {year} {2020})}\BibitemShut {NoStop}%
\bibitem [{\citenamefont {Chou}\ \emph {et~al.}(2020)\citenamefont {Chou},
  \citenamefont {Fu}, \citenamefont {Wilson}, \citenamefont {K{\"{o}}nig},\
  and\ \citenamefont {Pixley}}]{PhysRevB.101.235121}%
  \BibitemOpen
  \bibfield  {author} {\bibinfo {author} {\bibfnamefont {Y.-Z.}\ \bibnamefont
  {Chou}}, \bibinfo {author} {\bibfnamefont {Y.}~\bibnamefont {Fu}}, \bibinfo
  {author} {\bibfnamefont {J.~H.}\ \bibnamefont {Wilson}}, \bibinfo {author}
  {\bibfnamefont {E.~J.}\ \bibnamefont {K{\"{o}}nig}}, \ and\ \bibinfo {author}
  {\bibfnamefont {J.~H.}\ \bibnamefont {Pixley}},\ }\href {\doibase
  10.1103/PhysRevB.101.235121} {\bibfield  {journal} {\bibinfo  {journal}
  {Phys. Rev. B}\ }\textbf {\bibinfo {volume} {101}},\ \bibinfo {pages}
  {235121} (\bibinfo {year} {2020})}\BibitemShut {NoStop}%
\bibitem [{\citenamefont {Gon{\c{c}}alves}\ \emph {et~al.}(2021)\citenamefont
  {Gon{\c{c}}alves}, \citenamefont {Olyaei}, \citenamefont {Amorim},
  \citenamefont {Mondaini}, \citenamefont {Ribeiro},\ and\ \citenamefont
  {Castro}}]{Goncalves_2022_2DMat}%
  \BibitemOpen
  \bibfield  {author} {\bibinfo {author} {\bibfnamefont {M.}~\bibnamefont
  {Gon{\c{c}}alves}}, \bibinfo {author} {\bibfnamefont {H.~Z.}\ \bibnamefont
  {Olyaei}}, \bibinfo {author} {\bibfnamefont {B.}~\bibnamefont {Amorim}},
  \bibinfo {author} {\bibfnamefont {R.}~\bibnamefont {Mondaini}}, \bibinfo
  {author} {\bibfnamefont {P.}~\bibnamefont {Ribeiro}}, \ and\ \bibinfo
  {author} {\bibfnamefont {E.~V.}\ \bibnamefont {Castro}},\ }\href {\doibase
  10.1088/2053-1583/ac3259} {\bibfield  {journal} {\bibinfo  {journal} {2D
  Materials}\ }\textbf {\bibinfo {volume} {9}},\ \bibinfo {pages} {011001}
  (\bibinfo {year} {2021})}\BibitemShut {NoStop}%
\bibitem [{\citenamefont {Kraus}\ \emph
  {et~al.}(2012{\natexlab{a}})\citenamefont {Kraus}, \citenamefont {Lahini},
  \citenamefont {Ringel}, \citenamefont {Verbin},\ and\ \citenamefont
  {Zilberberg}}]{Kraus2012}%
  \BibitemOpen
  \bibfield  {author} {\bibinfo {author} {\bibfnamefont {Y.~E.}\ \bibnamefont
  {Kraus}}, \bibinfo {author} {\bibfnamefont {Y.}~\bibnamefont {Lahini}},
  \bibinfo {author} {\bibfnamefont {Z.}~\bibnamefont {Ringel}}, \bibinfo
  {author} {\bibfnamefont {M.}~\bibnamefont {Verbin}}, \ and\ \bibinfo {author}
  {\bibfnamefont {O.}~\bibnamefont {Zilberberg}},\ }\href {\doibase
  10.1103/PhysRevLett.109.106402} {\bibfield  {journal} {\bibinfo  {journal}
  {Phys. Rev. Lett.}\ }\textbf {\bibinfo {volume} {109}},\ \bibinfo {pages}
  {106402} (\bibinfo {year} {2012}{\natexlab{a}})}\BibitemShut {NoStop}%
\bibitem [{\citenamefont {Prodan}(2015)}]{PhysRevB.91.245104}%
  \BibitemOpen
  \bibfield  {author} {\bibinfo {author} {\bibfnamefont {E.}~\bibnamefont
  {Prodan}},\ }\href {\doibase 10.1103/PhysRevB.91.245104} {\bibfield
  {journal} {\bibinfo  {journal} {Phys. Rev. B}\ }\textbf {\bibinfo {volume}
  {91}},\ \bibinfo {pages} {245104} (\bibinfo {year} {2015})}\BibitemShut
  {NoStop}%
\bibitem [{\citenamefont {Zilberberg}(2021)}]{Zilberberg21}%
  \BibitemOpen
  \bibfield  {author} {\bibinfo {author} {\bibfnamefont {O.}~\bibnamefont
  {Zilberberg}},\ }\href {\doibase 10.1364/OME.416552} {\bibfield  {journal}
  {\bibinfo  {journal} {Opt. Mater. Express}\ }\textbf {\bibinfo {volume}
  {11}},\ \bibinfo {pages} {1143} (\bibinfo {year} {2021})}\BibitemShut
  {NoStop}%
\bibitem [{\citenamefont {Kraus}\ \emph {et~al.}(2013)\citenamefont {Kraus},
  \citenamefont {Ringel},\ and\ \citenamefont
  {Zilberberg}}]{PhysRevLett.111.226401}%
  \BibitemOpen
  \bibfield  {author} {\bibinfo {author} {\bibfnamefont {Y.~E.}\ \bibnamefont
  {Kraus}}, \bibinfo {author} {\bibfnamefont {Z.}~\bibnamefont {Ringel}}, \
  and\ \bibinfo {author} {\bibfnamefont {O.}~\bibnamefont {Zilberberg}},\
  }\href {\doibase 10.1103/PhysRevLett.111.226401} {\bibfield  {journal}
  {\bibinfo  {journal} {Phys. Rev. Lett.}\ }\textbf {\bibinfo {volume} {111}},\
  \bibinfo {pages} {226401} (\bibinfo {year} {2013})}\BibitemShut {NoStop}%
\bibitem [{\citenamefont {Fujimoto}\ \emph {et~al.}(2020)\citenamefont
  {Fujimoto}, \citenamefont {Koschke},\ and\ \citenamefont
  {Koshino}}]{PhysRevB.101.041112}%
  \BibitemOpen
  \bibfield  {author} {\bibinfo {author} {\bibfnamefont {M.}~\bibnamefont
  {Fujimoto}}, \bibinfo {author} {\bibfnamefont {H.}~\bibnamefont {Koschke}}, \
  and\ \bibinfo {author} {\bibfnamefont {M.}~\bibnamefont {Koshino}},\ }\href
  {\doibase 10.1103/PhysRevB.101.041112} {\bibfield  {journal} {\bibinfo
  {journal} {Phys. Rev. B}\ }\textbf {\bibinfo {volume} {101}},\ \bibinfo
  {pages} {041112} (\bibinfo {year} {2020})}\BibitemShut {NoStop}%
\bibitem [{\citenamefont {Koshino}\ and\ \citenamefont
  {Oka}(2022)}]{PhysRevResearch.4.013028}%
  \BibitemOpen
  \bibfield  {author} {\bibinfo {author} {\bibfnamefont {M.}~\bibnamefont
  {Koshino}}\ and\ \bibinfo {author} {\bibfnamefont {H.}~\bibnamefont {Oka}},\
  }\href {\doibase 10.1103/PhysRevResearch.4.013028} {\bibfield  {journal}
  {\bibinfo  {journal} {Phys. Rev. Research}\ }\textbf {\bibinfo {volume}
  {4}},\ \bibinfo {pages} {013028} (\bibinfo {year} {2022})}\BibitemShut
  {NoStop}%
\bibitem [{\citenamefont {Boers}\ \emph {et~al.}(2007)\citenamefont {Boers},
  \citenamefont {Goedeke}, \citenamefont {Hinrichs},\ and\ \citenamefont
  {Holthaus}}]{PhysRevA.75.063404}%
  \BibitemOpen
  \bibfield  {author} {\bibinfo {author} {\bibfnamefont {D.~J.}\ \bibnamefont
  {Boers}}, \bibinfo {author} {\bibfnamefont {B.}~\bibnamefont {Goedeke}},
  \bibinfo {author} {\bibfnamefont {D.}~\bibnamefont {Hinrichs}}, \ and\
  \bibinfo {author} {\bibfnamefont {M.}~\bibnamefont {Holthaus}},\ }\href
  {\doibase 10.1103/PhysRevA.75.063404} {\bibfield  {journal} {\bibinfo
  {journal} {Phys. Rev. A}\ }\textbf {\bibinfo {volume} {75}},\ \bibinfo
  {pages} {63404} (\bibinfo {year} {2007})}\BibitemShut {NoStop}%
\bibitem [{\citenamefont {Roati}\ \emph {et~al.}(2008)\citenamefont {Roati},
  \citenamefont {D'Errico}, \citenamefont {Fallani}, \citenamefont {Fattori},
  \citenamefont {Fort}, \citenamefont {Zaccanti}, \citenamefont {Modugno},
  \citenamefont {Modugno},\ and\ \citenamefont {Inguscio}}]{Roati2008}%
  \BibitemOpen
  \bibfield  {author} {\bibinfo {author} {\bibfnamefont {G.}~\bibnamefont
  {Roati}}, \bibinfo {author} {\bibfnamefont {C.}~\bibnamefont {D'Errico}},
  \bibinfo {author} {\bibfnamefont {L.}~\bibnamefont {Fallani}}, \bibinfo
  {author} {\bibfnamefont {M.}~\bibnamefont {Fattori}}, \bibinfo {author}
  {\bibfnamefont {C.}~\bibnamefont {Fort}}, \bibinfo {author} {\bibfnamefont
  {M.}~\bibnamefont {Zaccanti}}, \bibinfo {author} {\bibfnamefont
  {G.}~\bibnamefont {Modugno}}, \bibinfo {author} {\bibfnamefont
  {M.}~\bibnamefont {Modugno}}, \ and\ \bibinfo {author} {\bibfnamefont
  {M.}~\bibnamefont {Inguscio}},\ }\href {\doibase 10.1038/nature07071}
  {\bibfield  {journal} {\bibinfo  {journal} {Nature}\ }\textbf {\bibinfo
  {volume} {453}},\ \bibinfo {pages} {895} (\bibinfo {year} {2008})},\ \Eprint
  {http://arxiv.org/abs/0804.2609} {arXiv:0804.2609} \BibitemShut {NoStop}%
\bibitem [{\citenamefont {Modugno}(2009)}]{Modugno_2009}%
  \BibitemOpen
  \bibfield  {author} {\bibinfo {author} {\bibfnamefont {M.}~\bibnamefont
  {Modugno}},\ }\href {\doibase 10.1088/1367-2630/11/3/033023} {\bibfield
  {journal} {\bibinfo  {journal} {New Journal of Physics}\ }\textbf {\bibinfo
  {volume} {11}},\ \bibinfo {pages} {33023} (\bibinfo {year}
  {2009})}\BibitemShut {NoStop}%
\bibitem [{\citenamefont {Schreiber}\ \emph {et~al.}(2015)\citenamefont
  {Schreiber}, \citenamefont {Hodgman}, \citenamefont {Bordia}, \citenamefont
  {L{\"{u}}schen}, \citenamefont {Fischer}, \citenamefont {Vosk}, \citenamefont
  {Altman}, \citenamefont {Schneider},\ and\ \citenamefont
  {Bloch}}]{Schreiber842}%
  \BibitemOpen
  \bibfield  {author} {\bibinfo {author} {\bibfnamefont {M.}~\bibnamefont
  {Schreiber}}, \bibinfo {author} {\bibfnamefont {S.~S.}\ \bibnamefont
  {Hodgman}}, \bibinfo {author} {\bibfnamefont {P.}~\bibnamefont {Bordia}},
  \bibinfo {author} {\bibfnamefont {H.~P.}\ \bibnamefont {L{\"{u}}schen}},
  \bibinfo {author} {\bibfnamefont {M.~H.}\ \bibnamefont {Fischer}}, \bibinfo
  {author} {\bibfnamefont {R.}~\bibnamefont {Vosk}}, \bibinfo {author}
  {\bibfnamefont {E.}~\bibnamefont {Altman}}, \bibinfo {author} {\bibfnamefont
  {U.}~\bibnamefont {Schneider}}, \ and\ \bibinfo {author} {\bibfnamefont
  {I.}~\bibnamefont {Bloch}},\ }\href {\doibase 10.1126/science.aaa7432}
  {\bibfield  {journal} {\bibinfo  {journal} {Science}\ }\textbf {\bibinfo
  {volume} {349}},\ \bibinfo {pages} {842} (\bibinfo {year} {2015})},\ \Eprint
  {http://arxiv.org/abs/1501.05661} {arXiv:1501.05661} \BibitemShut {NoStop}%
\bibitem [{\citenamefont {L\"uschen}\ \emph {et~al.}(2018)\citenamefont
  {L\"uschen}, \citenamefont {Scherg}, \citenamefont {Kohlert}, \citenamefont
  {Schreiber}, \citenamefont {Bordia}, \citenamefont {Li}, \citenamefont
  {Das~Sarma},\ and\ \citenamefont {Bloch}}]{Luschen2018}%
  \BibitemOpen
  \bibfield  {author} {\bibinfo {author} {\bibfnamefont {H.~P.}\ \bibnamefont
  {L\"uschen}}, \bibinfo {author} {\bibfnamefont {S.}~\bibnamefont {Scherg}},
  \bibinfo {author} {\bibfnamefont {T.}~\bibnamefont {Kohlert}}, \bibinfo
  {author} {\bibfnamefont {M.}~\bibnamefont {Schreiber}}, \bibinfo {author}
  {\bibfnamefont {P.}~\bibnamefont {Bordia}}, \bibinfo {author} {\bibfnamefont
  {X.}~\bibnamefont {Li}}, \bibinfo {author} {\bibfnamefont {S.}~\bibnamefont
  {Das~Sarma}}, \ and\ \bibinfo {author} {\bibfnamefont {I.}~\bibnamefont
  {Bloch}},\ }\href {\doibase 10.1103/PhysRevLett.120.160404} {\bibfield
  {journal} {\bibinfo  {journal} {Phys. Rev. Lett.}\ }\textbf {\bibinfo
  {volume} {120}},\ \bibinfo {pages} {160404} (\bibinfo {year}
  {2018})}\BibitemShut {NoStop}%
\bibitem [{\citenamefont {Yao}\ \emph {et~al.}(2019)\citenamefont {Yao},
  \citenamefont {Khoudli}, \citenamefont {Bresque},\ and\ \citenamefont
  {Sanchez-Palencia}}]{PhysRevLett.123.070405}%
  \BibitemOpen
  \bibfield  {author} {\bibinfo {author} {\bibfnamefont {H.}~\bibnamefont
  {Yao}}, \bibinfo {author} {\bibfnamefont {H.}~\bibnamefont {Khoudli}},
  \bibinfo {author} {\bibfnamefont {L.}~\bibnamefont {Bresque}}, \ and\
  \bibinfo {author} {\bibfnamefont {L.}~\bibnamefont {Sanchez-Palencia}},\
  }\href {\doibase 10.1103/PhysRevLett.123.070405} {\bibfield  {journal}
  {\bibinfo  {journal} {Phys. Rev. Lett.}\ }\textbf {\bibinfo {volume} {123}},\
  \bibinfo {pages} {070405} (\bibinfo {year} {2019})}\BibitemShut {NoStop}%
\bibitem [{\citenamefont {Yao}\ \emph {et~al.}(2020)\citenamefont {Yao},
  \citenamefont {Giamarchi},\ and\ \citenamefont
  {Sanchez-Palencia}}]{PhysRevLett.125.060401}%
  \BibitemOpen
  \bibfield  {author} {\bibinfo {author} {\bibfnamefont {H.}~\bibnamefont
  {Yao}}, \bibinfo {author} {\bibfnamefont {T.}~\bibnamefont {Giamarchi}}, \
  and\ \bibinfo {author} {\bibfnamefont {L.}~\bibnamefont {Sanchez-Palencia}},\
  }\href {\doibase 10.1103/PhysRevLett.125.060401} {\bibfield  {journal}
  {\bibinfo  {journal} {Phys. Rev. Lett.}\ }\textbf {\bibinfo {volume} {125}},\
  \bibinfo {pages} {060401} (\bibinfo {year} {2020})}\BibitemShut {NoStop}%
\bibitem [{\citenamefont {Gautier}\ \emph {et~al.}(2021)\citenamefont
  {Gautier}, \citenamefont {Yao},\ and\ \citenamefont
  {Sanchez-Palencia}}]{PhysRevLett.126.110401}%
  \BibitemOpen
  \bibfield  {author} {\bibinfo {author} {\bibfnamefont {R.}~\bibnamefont
  {Gautier}}, \bibinfo {author} {\bibfnamefont {H.}~\bibnamefont {Yao}}, \ and\
  \bibinfo {author} {\bibfnamefont {L.}~\bibnamefont {Sanchez-Palencia}},\
  }\href {\doibase 10.1103/PhysRevLett.126.110401} {\bibfield  {journal}
  {\bibinfo  {journal} {Phys. Rev. Lett.}\ }\textbf {\bibinfo {volume} {126}},\
  \bibinfo {pages} {110401} (\bibinfo {year} {2021})}\BibitemShut {NoStop}%
\bibitem [{\citenamefont {An}\ \emph {et~al.}(2021)\citenamefont {An},
  \citenamefont {Padavi\ifmmode~\acute{c}\else \'{c}\fi{}}, \citenamefont
  {Meier}, \citenamefont {Hegde}, \citenamefont {Ganeshan}, \citenamefont
  {Pixley}, \citenamefont {Vishveshwara},\ and\ \citenamefont
  {Gadway}}]{PhysRevLett.126.040603}%
  \BibitemOpen
  \bibfield  {author} {\bibinfo {author} {\bibfnamefont {F.~A.}\ \bibnamefont
  {An}}, \bibinfo {author} {\bibfnamefont {K.}~\bibnamefont
  {Padavi\ifmmode~\acute{c}\else \'{c}\fi{}}}, \bibinfo {author} {\bibfnamefont
  {E.~J.}\ \bibnamefont {Meier}}, \bibinfo {author} {\bibfnamefont
  {S.}~\bibnamefont {Hegde}}, \bibinfo {author} {\bibfnamefont
  {S.}~\bibnamefont {Ganeshan}}, \bibinfo {author} {\bibfnamefont {J.~H.}\
  \bibnamefont {Pixley}}, \bibinfo {author} {\bibfnamefont {S.}~\bibnamefont
  {Vishveshwara}}, \ and\ \bibinfo {author} {\bibfnamefont {B.}~\bibnamefont
  {Gadway}},\ }\href {\doibase 10.1103/PhysRevLett.126.040603} {\bibfield
  {journal} {\bibinfo  {journal} {Phys. Rev. Lett.}\ }\textbf {\bibinfo
  {volume} {126}},\ \bibinfo {pages} {040603} (\bibinfo {year}
  {2021})}\BibitemShut {NoStop}%
\bibitem [{\citenamefont {Kohlert}\ \emph {et~al.}(2019)\citenamefont
  {Kohlert}, \citenamefont {Scherg}, \citenamefont {Li}, \citenamefont
  {L{\"{u}}schen}, \citenamefont {{Das Sarma}}, \citenamefont {Bloch},\ and\
  \citenamefont {Aidelsburger}}]{PhysRevLett.122.170403}%
  \BibitemOpen
  \bibfield  {author} {\bibinfo {author} {\bibfnamefont {T.}~\bibnamefont
  {Kohlert}}, \bibinfo {author} {\bibfnamefont {S.}~\bibnamefont {Scherg}},
  \bibinfo {author} {\bibfnamefont {X.}~\bibnamefont {Li}}, \bibinfo {author}
  {\bibfnamefont {H.~P.}\ \bibnamefont {L{\"{u}}schen}}, \bibinfo {author}
  {\bibfnamefont {S.}~\bibnamefont {{Das Sarma}}}, \bibinfo {author}
  {\bibfnamefont {I.}~\bibnamefont {Bloch}}, \ and\ \bibinfo {author}
  {\bibfnamefont {M.}~\bibnamefont {Aidelsburger}},\ }\href {\doibase
  10.1103/PhysRevLett.122.170403} {\bibfield  {journal} {\bibinfo  {journal}
  {Phys. Rev. Lett.}\ }\textbf {\bibinfo {volume} {122}},\ \bibinfo {pages}
  {170403} (\bibinfo {year} {2019})}\BibitemShut {NoStop}%
\bibitem [{\citenamefont {Lahini}\ \emph {et~al.}(2009)\citenamefont {Lahini},
  \citenamefont {Pugatch}, \citenamefont {Pozzi}, \citenamefont {Sorel},
  \citenamefont {Morandotti}, \citenamefont {Davidson},\ and\ \citenamefont
  {Silberberg}}]{Lahini2009}%
  \BibitemOpen
  \bibfield  {author} {\bibinfo {author} {\bibfnamefont {Y.}~\bibnamefont
  {Lahini}}, \bibinfo {author} {\bibfnamefont {R.}~\bibnamefont {Pugatch}},
  \bibinfo {author} {\bibfnamefont {F.}~\bibnamefont {Pozzi}}, \bibinfo
  {author} {\bibfnamefont {M.}~\bibnamefont {Sorel}}, \bibinfo {author}
  {\bibfnamefont {R.}~\bibnamefont {Morandotti}}, \bibinfo {author}
  {\bibfnamefont {N.}~\bibnamefont {Davidson}}, \ and\ \bibinfo {author}
  {\bibfnamefont {Y.}~\bibnamefont {Silberberg}},\ }\href {\doibase
  10.1103/PhysRevLett.103.013901} {\bibfield  {journal} {\bibinfo  {journal}
  {Phys. Rev. Lett.}\ }\textbf {\bibinfo {volume} {103}},\ \bibinfo {pages}
  {013901} (\bibinfo {year} {2009})}\BibitemShut {NoStop}%
\bibitem [{\citenamefont {Verbin}\ \emph {et~al.}(2013)\citenamefont {Verbin},
  \citenamefont {Zilberberg}, \citenamefont {Kraus}, \citenamefont {Lahini},\
  and\ \citenamefont {Silberberg}}]{Verbin2013}%
  \BibitemOpen
  \bibfield  {author} {\bibinfo {author} {\bibfnamefont {M.}~\bibnamefont
  {Verbin}}, \bibinfo {author} {\bibfnamefont {O.}~\bibnamefont {Zilberberg}},
  \bibinfo {author} {\bibfnamefont {Y.~E.}\ \bibnamefont {Kraus}}, \bibinfo
  {author} {\bibfnamefont {Y.}~\bibnamefont {Lahini}}, \ and\ \bibinfo {author}
  {\bibfnamefont {Y.}~\bibnamefont {Silberberg}},\ }\href {\doibase
  10.1103/PhysRevLett.110.076403} {\bibfield  {journal} {\bibinfo  {journal}
  {Phys. Rev. Lett.}\ }\textbf {\bibinfo {volume} {110}},\ \bibinfo {pages}
  {076403} (\bibinfo {year} {2013})}\BibitemShut {NoStop}%
\bibitem [{\citenamefont {Verbin}\ \emph {et~al.}(2015)\citenamefont {Verbin},
  \citenamefont {Zilberberg}, \citenamefont {Lahini}, \citenamefont {Kraus},\
  and\ \citenamefont {Silberberg}}]{PhysRevB.91.064201}%
  \BibitemOpen
  \bibfield  {author} {\bibinfo {author} {\bibfnamefont {M.}~\bibnamefont
  {Verbin}}, \bibinfo {author} {\bibfnamefont {O.}~\bibnamefont {Zilberberg}},
  \bibinfo {author} {\bibfnamefont {Y.}~\bibnamefont {Lahini}}, \bibinfo
  {author} {\bibfnamefont {Y.~E.}\ \bibnamefont {Kraus}}, \ and\ \bibinfo
  {author} {\bibfnamefont {Y.}~\bibnamefont {Silberberg}},\ }\href {\doibase
  10.1103/PhysRevB.91.064201} {\bibfield  {journal} {\bibinfo  {journal} {Phys.
  Rev. B}\ }\textbf {\bibinfo {volume} {91}},\ \bibinfo {pages} {64201}
  (\bibinfo {year} {2015})}\BibitemShut {NoStop}%
\bibitem [{\citenamefont {Wang}\ \emph {et~al.}(2020)\citenamefont {Wang},
  \citenamefont {Zheng}, \citenamefont {Chen}, \citenamefont {Huang},
  \citenamefont {Kartashov}, \citenamefont {Torner}, \citenamefont {Konotop},\
  and\ \citenamefont {Ye}}]{Wang2020}%
  \BibitemOpen
  \bibfield  {author} {\bibinfo {author} {\bibfnamefont {P.}~\bibnamefont
  {Wang}}, \bibinfo {author} {\bibfnamefont {Y.}~\bibnamefont {Zheng}},
  \bibinfo {author} {\bibfnamefont {X.}~\bibnamefont {Chen}}, \bibinfo {author}
  {\bibfnamefont {C.}~\bibnamefont {Huang}}, \bibinfo {author} {\bibfnamefont
  {Y.~V.}\ \bibnamefont {Kartashov}}, \bibinfo {author} {\bibfnamefont
  {L.}~\bibnamefont {Torner}}, \bibinfo {author} {\bibfnamefont {V.~V.}\
  \bibnamefont {Konotop}}, \ and\ \bibinfo {author} {\bibfnamefont
  {F.}~\bibnamefont {Ye}},\ }\href {\doibase 10.1038/s41586-019-1851-6}
  {\bibfield  {journal} {\bibinfo  {journal} {Nature}\ }\textbf {\bibinfo
  {volume} {577}},\ \bibinfo {pages} {42} (\bibinfo {year} {2020})}\BibitemShut
  {NoStop}%
\bibitem [{\citenamefont {Sinelnik}\ \emph {et~al.}(2020)\citenamefont
  {Sinelnik}, \citenamefont {Shishkin}, \citenamefont {Yu}, \citenamefont
  {Samusev}, \citenamefont {Belov}, \citenamefont {Limonov}, \citenamefont
  {Ginzburg},\ and\ \citenamefont
  {Rybin}}]{https://doi.org/10.1002/adom.202001170}%
  \BibitemOpen
  \bibfield  {author} {\bibinfo {author} {\bibfnamefont {A.~D.}\ \bibnamefont
  {Sinelnik}}, \bibinfo {author} {\bibfnamefont {I.~I.}\ \bibnamefont
  {Shishkin}}, \bibinfo {author} {\bibfnamefont {X.}~\bibnamefont {Yu}},
  \bibinfo {author} {\bibfnamefont {K.~B.}\ \bibnamefont {Samusev}}, \bibinfo
  {author} {\bibfnamefont {P.~A.}\ \bibnamefont {Belov}}, \bibinfo {author}
  {\bibfnamefont {M.~F.}\ \bibnamefont {Limonov}}, \bibinfo {author}
  {\bibfnamefont {P.}~\bibnamefont {Ginzburg}}, \ and\ \bibinfo {author}
  {\bibfnamefont {M.~V.}\ \bibnamefont {Rybin}},\ }\href {\doibase
  10.1002/adom.202001170} {\bibfield  {journal} {\bibinfo  {journal} {Advanced
  Optical Materials}\ }\textbf {\bibinfo {volume} {8}},\ \bibinfo {pages}
  {2001170} (\bibinfo {year} {2020})}\BibitemShut {NoStop}%
\bibitem [{\citenamefont {Wang}\ \emph
  {et~al.}(2022{\natexlab{a}})\citenamefont {Wang}, \citenamefont {Fu},
  \citenamefont {Peng}, \citenamefont {Kartashov}, \citenamefont {Torner},
  \citenamefont {Konotop},\ and\ \citenamefont {Ye}}]{Wang2022}%
  \BibitemOpen
  \bibfield  {author} {\bibinfo {author} {\bibfnamefont {P.}~\bibnamefont
  {Wang}}, \bibinfo {author} {\bibfnamefont {Q.}~\bibnamefont {Fu}}, \bibinfo
  {author} {\bibfnamefont {R.}~\bibnamefont {Peng}}, \bibinfo {author}
  {\bibfnamefont {Y.~V.}\ \bibnamefont {Kartashov}}, \bibinfo {author}
  {\bibfnamefont {L.}~\bibnamefont {Torner}}, \bibinfo {author} {\bibfnamefont
  {V.~V.}\ \bibnamefont {Konotop}}, \ and\ \bibinfo {author} {\bibfnamefont
  {F.}~\bibnamefont {Ye}},\ }\href {\doibase 10.1038/s41467-022-34394-3}
  {\bibfield  {journal} {\bibinfo  {journal} {Nature Communications}\ }\textbf
  {\bibinfo {volume} {13}},\ \bibinfo {pages} {6738} (\bibinfo {year}
  {2022}{\natexlab{a}})}\BibitemShut {NoStop}%
\bibitem [{\citenamefont {Apigo}\ \emph {et~al.}(2019)\citenamefont {Apigo},
  \citenamefont {Cheng}, \citenamefont {Dobiszewski}, \citenamefont {Prodan},\
  and\ \citenamefont {Prodan}}]{PhysRevLett.122.095501}%
  \BibitemOpen
  \bibfield  {author} {\bibinfo {author} {\bibfnamefont {D.~J.}\ \bibnamefont
  {Apigo}}, \bibinfo {author} {\bibfnamefont {W.}~\bibnamefont {Cheng}},
  \bibinfo {author} {\bibfnamefont {K.~F.}\ \bibnamefont {Dobiszewski}},
  \bibinfo {author} {\bibfnamefont {E.}~\bibnamefont {Prodan}}, \ and\ \bibinfo
  {author} {\bibfnamefont {C.}~\bibnamefont {Prodan}},\ }\href {\doibase
  10.1103/PhysRevLett.122.095501} {\bibfield  {journal} {\bibinfo  {journal}
  {Phys. Rev. Lett.}\ }\textbf {\bibinfo {volume} {122}},\ \bibinfo {pages}
  {095501} (\bibinfo {year} {2019})}\BibitemShut {NoStop}%
\bibitem [{\citenamefont {Ni}\ \emph {et~al.}(2019)\citenamefont {Ni},
  \citenamefont {Chen}, \citenamefont {Weiner}, \citenamefont {Apigo},
  \citenamefont {Prodan}, \citenamefont {Al{\`{u}}}, \citenamefont {Prodan},\
  and\ \citenamefont {Khanikaev}}]{Ni2019}%
  \BibitemOpen
  \bibfield  {author} {\bibinfo {author} {\bibfnamefont {X.}~\bibnamefont
  {Ni}}, \bibinfo {author} {\bibfnamefont {K.}~\bibnamefont {Chen}}, \bibinfo
  {author} {\bibfnamefont {M.}~\bibnamefont {Weiner}}, \bibinfo {author}
  {\bibfnamefont {D.~J.}\ \bibnamefont {Apigo}}, \bibinfo {author}
  {\bibfnamefont {C.}~\bibnamefont {Prodan}}, \bibinfo {author} {\bibfnamefont
  {A.}~\bibnamefont {Al{\`{u}}}}, \bibinfo {author} {\bibfnamefont
  {E.}~\bibnamefont {Prodan}}, \ and\ \bibinfo {author} {\bibfnamefont {A.~B.}\
  \bibnamefont {Khanikaev}},\ }\href {\doibase 10.1038/s42005-019-0151-7}
  {\bibfield  {journal} {\bibinfo  {journal} {Communications Physics}\ }\textbf
  {\bibinfo {volume} {2}},\ \bibinfo {pages} {55} (\bibinfo {year}
  {2019})}\BibitemShut {NoStop}%
\bibitem [{\citenamefont {Cheng}\ \emph {et~al.}(2020)\citenamefont {Cheng},
  \citenamefont {Prodan},\ and\ \citenamefont
  {Prodan}}]{PhysRevLett.125.224301}%
  \BibitemOpen
  \bibfield  {author} {\bibinfo {author} {\bibfnamefont {W.}~\bibnamefont
  {Cheng}}, \bibinfo {author} {\bibfnamefont {E.}~\bibnamefont {Prodan}}, \
  and\ \bibinfo {author} {\bibfnamefont {C.}~\bibnamefont {Prodan}},\ }\href
  {\doibase 10.1103/PhysRevLett.125.224301} {\bibfield  {journal} {\bibinfo
  {journal} {Phys. Rev. Lett.}\ }\textbf {\bibinfo {volume} {125}},\ \bibinfo
  {pages} {224301} (\bibinfo {year} {2020})}\BibitemShut {NoStop}%
\bibitem [{\citenamefont {Xia}\ \emph {et~al.}(2020)\citenamefont {Xia},
  \citenamefont {Erturk},\ and\ \citenamefont
  {Ruzzene}}]{PhysRevApplied.13.014023}%
  \BibitemOpen
  \bibfield  {author} {\bibinfo {author} {\bibfnamefont {Y.}~\bibnamefont
  {Xia}}, \bibinfo {author} {\bibfnamefont {A.}~\bibnamefont {Erturk}}, \ and\
  \bibinfo {author} {\bibfnamefont {M.}~\bibnamefont {Ruzzene}},\ }\href
  {\doibase 10.1103/PhysRevApplied.13.014023} {\bibfield  {journal} {\bibinfo
  {journal} {Phys. Rev. Applied}\ }\textbf {\bibinfo {volume} {13}},\ \bibinfo
  {pages} {014023} (\bibinfo {year} {2020})}\BibitemShut {NoStop}%
\bibitem [{\citenamefont {Chen}\ \emph {et~al.}(2021)\citenamefont {Chen},
  \citenamefont {Zhu}, \citenamefont {Tan}, \citenamefont {Wang},\ and\
  \citenamefont {Ma}}]{PhysRevX.11.011016}%
  \BibitemOpen
  \bibfield  {author} {\bibinfo {author} {\bibfnamefont {Z.-G.}\ \bibnamefont
  {Chen}}, \bibinfo {author} {\bibfnamefont {W.}~\bibnamefont {Zhu}}, \bibinfo
  {author} {\bibfnamefont {Y.}~\bibnamefont {Tan}}, \bibinfo {author}
  {\bibfnamefont {L.}~\bibnamefont {Wang}}, \ and\ \bibinfo {author}
  {\bibfnamefont {G.}~\bibnamefont {Ma}},\ }\href {\doibase
  10.1103/PhysRevX.11.011016} {\bibfield  {journal} {\bibinfo  {journal} {Phys.
  Rev. X}\ }\textbf {\bibinfo {volume} {11}},\ \bibinfo {pages} {011016}
  (\bibinfo {year} {2021})}\BibitemShut {NoStop}%
\bibitem [{\citenamefont {Gei}\ \emph {et~al.}(2020)\citenamefont {Gei},
  \citenamefont {Chen}, \citenamefont {Bosi},\ and\ \citenamefont
  {Morini}}]{doi:10.1063/5.0013528}%
  \BibitemOpen
  \bibfield  {author} {\bibinfo {author} {\bibfnamefont {M.}~\bibnamefont
  {Gei}}, \bibinfo {author} {\bibfnamefont {Z.}~\bibnamefont {Chen}}, \bibinfo
  {author} {\bibfnamefont {F.}~\bibnamefont {Bosi}}, \ and\ \bibinfo {author}
  {\bibfnamefont {L.}~\bibnamefont {Morini}},\ }\href {\doibase
  10.1063/5.0013528} {\bibfield  {journal} {\bibinfo  {journal} {Applied
  Physics Letters}\ }\textbf {\bibinfo {volume} {116}},\ \bibinfo {pages}
  {241903} (\bibinfo {year} {2020})}\BibitemShut {NoStop}%
\bibitem [{\citenamefont {Balents}\ \emph {et~al.}(2020)\citenamefont
  {Balents}, \citenamefont {Dean}, \citenamefont {Efetov},\ and\ \citenamefont
  {Young}}]{Balents2020}%
  \BibitemOpen
  \bibfield  {author} {\bibinfo {author} {\bibfnamefont {L.}~\bibnamefont
  {Balents}}, \bibinfo {author} {\bibfnamefont {C.~R.}\ \bibnamefont {Dean}},
  \bibinfo {author} {\bibfnamefont {D.~K.}\ \bibnamefont {Efetov}}, \ and\
  \bibinfo {author} {\bibfnamefont {A.~F.}\ \bibnamefont {Young}},\ }\href
  {\doibase 10.1038/s41567-020-0906-9} {\bibfield  {journal} {\bibinfo
  {journal} {Nat. Phys.}\ }\textbf {\bibinfo {volume} {16}},\ \bibinfo {pages}
  {725} (\bibinfo {year} {2020})}\BibitemShut {NoStop}%
\bibitem [{\citenamefont {Andrei}\ and\ \citenamefont
  {MacDonald}(2020)}]{Andrei2020}%
  \BibitemOpen
  \bibfield  {author} {\bibinfo {author} {\bibfnamefont {E.~Y.}\ \bibnamefont
  {Andrei}}\ and\ \bibinfo {author} {\bibfnamefont {A.~H.}\ \bibnamefont
  {MacDonald}},\ }\href {\doibase 10.1038/s41563-020-00840-0} {\bibfield
  {journal} {\bibinfo  {journal} {Nature Materials}\ }\textbf {\bibinfo
  {volume} {19}},\ \bibinfo {pages} {1265} (\bibinfo {year}
  {2020})}\BibitemShut {NoStop}%
\bibitem [{\citenamefont {Uri}\ \emph {et~al.}(2023)\citenamefont {Uri},
  \citenamefont {de~la Barrera}, \citenamefont {Randeria}, \citenamefont
  {Rodan-Legrain}, \citenamefont {Devakul}, \citenamefont {Crowley},
  \citenamefont {Paul}, \citenamefont {Watanabe}, \citenamefont {Taniguchi},
  \citenamefont {Lifshitz}, \citenamefont {Fu}, \citenamefont {Ashoori},\ and\
  \citenamefont {Jarillo-Herrero}}]{uri2023superconductivity}%
  \BibitemOpen
  \bibfield  {author} {\bibinfo {author} {\bibfnamefont {A.}~\bibnamefont
  {Uri}}, \bibinfo {author} {\bibfnamefont {S.~C.}\ \bibnamefont {de~la
  Barrera}}, \bibinfo {author} {\bibfnamefont {M.~T.}\ \bibnamefont
  {Randeria}}, \bibinfo {author} {\bibfnamefont {D.}~\bibnamefont
  {Rodan-Legrain}}, \bibinfo {author} {\bibfnamefont {T.}~\bibnamefont
  {Devakul}}, \bibinfo {author} {\bibfnamefont {P.~J.~D.}\ \bibnamefont
  {Crowley}}, \bibinfo {author} {\bibfnamefont {N.}~\bibnamefont {Paul}},
  \bibinfo {author} {\bibfnamefont {K.}~\bibnamefont {Watanabe}}, \bibinfo
  {author} {\bibfnamefont {T.}~\bibnamefont {Taniguchi}}, \bibinfo {author}
  {\bibfnamefont {R.}~\bibnamefont {Lifshitz}}, \bibinfo {author}
  {\bibfnamefont {L.}~\bibnamefont {Fu}}, \bibinfo {author} {\bibfnamefont
  {R.~C.}\ \bibnamefont {Ashoori}}, \ and\ \bibinfo {author} {\bibfnamefont
  {P.}~\bibnamefont {Jarillo-Herrero}},\ }\href {\doibase
  10.1038/s41586-023-06294-z} {\bibfield  {journal} {\bibinfo  {journal}
  {Nature}\ }\textbf {\bibinfo {volume} {620}},\ \bibinfo {pages} {762}
  (\bibinfo {year} {2023})}\BibitemShut {NoStop}%
\bibitem [{\citenamefont {Fu}\ \emph {et~al.}(2021)\citenamefont {Fu},
  \citenamefont {Wilson},\ and\ \citenamefont {Pixley}}]{Fu_2021}%
  \BibitemOpen
  \bibfield  {author} {\bibinfo {author} {\bibfnamefont {Y.}~\bibnamefont
  {Fu}}, \bibinfo {author} {\bibfnamefont {J.~H.}\ \bibnamefont {Wilson}}, \
  and\ \bibinfo {author} {\bibfnamefont {J.~H.}\ \bibnamefont {Pixley}},\
  }\href {\doibase 10.1103/physrevb.104.l041106} {\bibfield  {journal}
  {\bibinfo  {journal} {Physical Review B}\ }\textbf {\bibinfo {volume} {104}}
  (\bibinfo {year} {2021}),\ 10.1103/physrevb.104.l041106}\BibitemShut
  {NoStop}%
\bibitem [{\citenamefont {Cheng}\ \emph {et~al.}(2022)\citenamefont {Cheng},
  \citenamefont {Asgari},\ and\ \citenamefont {Xianlong}}]{cheng2022_december}%
  \BibitemOpen
  \bibfield  {author} {\bibinfo {author} {\bibfnamefont {S.}~\bibnamefont
  {Cheng}}, \bibinfo {author} {\bibfnamefont {R.}~\bibnamefont {Asgari}}, \
  and\ \bibinfo {author} {\bibfnamefont {G.}~\bibnamefont {Xianlong}},\ }\href
  {\doibase 10.48550/ARXIV.2212.04082} {\enquote {\bibinfo {title} {From
  topological phase to anderson localization in a two-dimensional quasiperiodic
  system},}\ } (\bibinfo {year} {2022})\BibitemShut {NoStop}%
\bibitem [{\citenamefont {Madeira}\ and\ \citenamefont
  {Sacramento}(2022)}]{PhysRevB.106.224505}%
  \BibitemOpen
  \bibfield  {author} {\bibinfo {author} {\bibfnamefont {M.~F.}\ \bibnamefont
  {Madeira}}\ and\ \bibinfo {author} {\bibfnamefont {P.~D.}\ \bibnamefont
  {Sacramento}},\ }\href {\doibase 10.1103/PhysRevB.106.224505} {\bibfield
  {journal} {\bibinfo  {journal} {Phys. Rev. B}\ }\textbf {\bibinfo {volume}
  {106}},\ \bibinfo {pages} {224505} (\bibinfo {year} {2022})}\BibitemShut
  {NoStop}%
\bibitem [{\citenamefont {Gon{\c{c}}alves}\ \emph
  {et~al.}(2022{\natexlab{b}})\citenamefont {Gon{\c{c}}alves}, \citenamefont
  {Gon{\c{c}}alves}, \citenamefont {Ribeiro},\ and\ \citenamefont
  {Amorim}}]{goncalves2022topological}%
  \BibitemOpen
  \bibfield  {author} {\bibinfo {author} {\bibfnamefont {T.~S.}\ \bibnamefont
  {Gon{\c{c}}alves}}, \bibinfo {author} {\bibfnamefont {M.}~\bibnamefont
  {Gon{\c{c}}alves}}, \bibinfo {author} {\bibfnamefont {P.}~\bibnamefont
  {Ribeiro}}, \ and\ \bibinfo {author} {\bibfnamefont {B.}~\bibnamefont
  {Amorim}},\ }\href@noop {} {\enquote {\bibinfo {title} {Topological phase
  transitions for any taste in 2d quasiperiodic systems},}\ } (\bibinfo {year}
  {2022}{\natexlab{b}}),\ \Eprint {http://arxiv.org/abs/2212.08024}
  {arXiv:2212.08024 [cond-mat.dis-nn]} \BibitemShut {NoStop}%
\bibitem [{\citenamefont {Liquito}\ \emph {et~al.}(2024)\citenamefont
  {Liquito}, \citenamefont {Gon{\c{c}}alves},\ and\ \citenamefont
  {Castro}}]{liquito2023fate}%
  \BibitemOpen
  \bibfield  {author} {\bibinfo {author} {\bibfnamefont {R.}~\bibnamefont
  {Liquito}}, \bibinfo {author} {\bibfnamefont {M.}~\bibnamefont
  {Gon{\c{c}}alves}}, \ and\ \bibinfo {author} {\bibfnamefont {E.~V.}\
  \bibnamefont {Castro}},\ }\href {\doibase 10.1103/PhysRevB.109.174202}
  {\bibfield  {journal} {\bibinfo  {journal} {Physical Review B}\ }\textbf
  {\bibinfo {volume} {109}},\ \bibinfo {pages} {174202} (\bibinfo {year}
  {2024})}\BibitemShut {NoStop}%
\bibitem [{\citenamefont {Chen}\ \emph {et~al.}(2020)\citenamefont {Chen},
  \citenamefont {Chen}, \citenamefont {Gao}, \citenamefont {Zhou},\ and\
  \citenamefont {Xu}}]{PhysRevLett.124.036803}%
  \BibitemOpen
  \bibfield  {author} {\bibinfo {author} {\bibfnamefont {R.}~\bibnamefont
  {Chen}}, \bibinfo {author} {\bibfnamefont {C.-Z.}\ \bibnamefont {Chen}},
  \bibinfo {author} {\bibfnamefont {J.-H.}\ \bibnamefont {Gao}}, \bibinfo
  {author} {\bibfnamefont {B.}~\bibnamefont {Zhou}}, \ and\ \bibinfo {author}
  {\bibfnamefont {D.-H.}\ \bibnamefont {Xu}},\ }\href {\doibase
  10.1103/PhysRevLett.124.036803} {\bibfield  {journal} {\bibinfo  {journal}
  {Phys. Rev. Lett.}\ }\textbf {\bibinfo {volume} {124}},\ \bibinfo {pages}
  {036803} (\bibinfo {year} {2020})}\BibitemShut {NoStop}%
\bibitem [{\citenamefont {Huang}\ \emph {et~al.}(2021)\citenamefont {Huang},
  \citenamefont {Fan}, \citenamefont {Li},\ and\ \citenamefont
  {Liu}}]{acs.nanolett.1c02661}%
  \BibitemOpen
  \bibfield  {author} {\bibinfo {author} {\bibfnamefont {H.}~\bibnamefont
  {Huang}}, \bibinfo {author} {\bibfnamefont {J.}~\bibnamefont {Fan}}, \bibinfo
  {author} {\bibfnamefont {D.}~\bibnamefont {Li}}, \ and\ \bibinfo {author}
  {\bibfnamefont {F.}~\bibnamefont {Liu}},\ }\href {\doibase
  10.1021/acs.nanolett.1c02661} {\bibfield  {journal} {\bibinfo  {journal}
  {Nano Letters}\ }\textbf {\bibinfo {volume} {21}},\ \bibinfo {pages} {7056}
  (\bibinfo {year} {2021})},\ \bibinfo {note} {pMID: 34350755},\ \Eprint
  {http://arxiv.org/abs/https://doi.org/10.1021/acs.nanolett.1c02661}
  {https://doi.org/10.1021/acs.nanolett.1c02661} \BibitemShut {NoStop}%
\bibitem [{\citenamefont {Hua}\ \emph {et~al.}(2020)\citenamefont {Hua},
  \citenamefont {Chen}, \citenamefont {Zhou},\ and\ \citenamefont
  {Xu}}]{PhysRevB.102.241102}%
  \BibitemOpen
  \bibfield  {author} {\bibinfo {author} {\bibfnamefont {C.-B.}\ \bibnamefont
  {Hua}}, \bibinfo {author} {\bibfnamefont {R.}~\bibnamefont {Chen}}, \bibinfo
  {author} {\bibfnamefont {B.}~\bibnamefont {Zhou}}, \ and\ \bibinfo {author}
  {\bibfnamefont {D.-H.}\ \bibnamefont {Xu}},\ }\href {\doibase
  10.1103/PhysRevB.102.241102} {\bibfield  {journal} {\bibinfo  {journal}
  {Phys. Rev. B}\ }\textbf {\bibinfo {volume} {102}},\ \bibinfo {pages}
  {241102} (\bibinfo {year} {2020})}\BibitemShut {NoStop}%
\bibitem [{\citenamefont {Varjas}\ \emph {et~al.}(2019)\citenamefont {Varjas},
  \citenamefont {Lau}, \citenamefont {P\"oyh\"onen}, \citenamefont {Akhmerov},
  \citenamefont {Pikulin},\ and\ \citenamefont
  {Fulga}}]{PhysRevLett.123.196401}%
  \BibitemOpen
  \bibfield  {author} {\bibinfo {author} {\bibfnamefont {D.}~\bibnamefont
  {Varjas}}, \bibinfo {author} {\bibfnamefont {A.}~\bibnamefont {Lau}},
  \bibinfo {author} {\bibfnamefont {K.}~\bibnamefont {P\"oyh\"onen}}, \bibinfo
  {author} {\bibfnamefont {A.~R.}\ \bibnamefont {Akhmerov}}, \bibinfo {author}
  {\bibfnamefont {D.~I.}\ \bibnamefont {Pikulin}}, \ and\ \bibinfo {author}
  {\bibfnamefont {I.~C.}\ \bibnamefont {Fulga}},\ }\href {\doibase
  10.1103/PhysRevLett.123.196401} {\bibfield  {journal} {\bibinfo  {journal}
  {Phys. Rev. Lett.}\ }\textbf {\bibinfo {volume} {123}},\ \bibinfo {pages}
  {196401} (\bibinfo {year} {2019})}\BibitemShut {NoStop}%
\bibitem [{\citenamefont {Wang}\ \emph
  {et~al.}(2022{\natexlab{b}})\citenamefont {Wang}, \citenamefont {Liu},\ and\
  \citenamefont {Huang}}]{PhysRevLett.129.056403}%
  \BibitemOpen
  \bibfield  {author} {\bibinfo {author} {\bibfnamefont {C.}~\bibnamefont
  {Wang}}, \bibinfo {author} {\bibfnamefont {F.}~\bibnamefont {Liu}}, \ and\
  \bibinfo {author} {\bibfnamefont {H.}~\bibnamefont {Huang}},\ }\href
  {\doibase 10.1103/PhysRevLett.129.056403} {\bibfield  {journal} {\bibinfo
  {journal} {Phys. Rev. Lett.}\ }\textbf {\bibinfo {volume} {129}},\ \bibinfo
  {pages} {056403} (\bibinfo {year} {2022}{\natexlab{b}})}\BibitemShut
  {NoStop}%
\bibitem [{\citenamefont {Spurrier}\ and\ \citenamefont
  {Cooper}(2020)}]{PhysRevResearch.2.033071}%
  \BibitemOpen
  \bibfield  {author} {\bibinfo {author} {\bibfnamefont {S.}~\bibnamefont
  {Spurrier}}\ and\ \bibinfo {author} {\bibfnamefont {N.~R.}\ \bibnamefont
  {Cooper}},\ }\href {\doibase 10.1103/PhysRevResearch.2.033071} {\bibfield
  {journal} {\bibinfo  {journal} {Phys. Rev. Res.}\ }\textbf {\bibinfo {volume}
  {2}},\ \bibinfo {pages} {033071} (\bibinfo {year} {2020})}\BibitemShut
  {NoStop}%
\bibitem [{Note1()}]{Note1}%
  \BibitemOpen
  \bibinfo {note} {In calculations with OBC we set the $\delta \approx 10^{-5}$
  to break the degeneracy of the corner states.}\BibitemShut {Stop}%
\bibitem [{Note2()}]{Note2}%
  \BibitemOpen
  \bibinfo {note} {For a lattice with geometric center at $(m_{x},m_{y})$,
  symmetry is retained for $\left (\phi _{x},\phi _{y}\right )=(\pi \beta
  (2m_{x}+1),\pi \beta (2m_{y}+1))\protect \allowbreak \mkern 12mu{\mathgroup
  \symoperators mod}\protect \,\protect \,2\pi $ (for even $L$), $\left (\phi
  _{x},\phi _{y}\right )=(2\pi \beta m_{x},2\pi \beta m_{y})\protect
  \allowbreak \mkern 12mu{\mathgroup \symoperators mod}\protect \,\protect
  \,2\pi $ (for odd $L$).}\BibitemShut {Stop}%
\bibitem [{Note3()}]{Note3}%
  \BibitemOpen
  \bibinfo {note} {These results include direct calculations of the spectral
  gap and topological properties}\BibitemShut {NoStop}%
\bibitem [{\citenamefont {Chaou}\ \emph {et~al.}(2024)\citenamefont {Chaou},
  \citenamefont {Moreno-Gonzalez}, \citenamefont {Altland},\ and\ \citenamefont
  {Brouwer}}]{chaou2024disordered}%
  \BibitemOpen
  \bibfield  {author} {\bibinfo {author} {\bibfnamefont {A.~Y.}\ \bibnamefont
  {Chaou}}, \bibinfo {author} {\bibfnamefont {M.}~\bibnamefont
  {Moreno-Gonzalez}}, \bibinfo {author} {\bibfnamefont {A.}~\bibnamefont
  {Altland}}, \ and\ \bibinfo {author} {\bibfnamefont {P.~W.}\ \bibnamefont
  {Brouwer}},\ }\href@noop {} {\bibfield  {journal} {\bibinfo  {journal} {arXiv
  preprint arXiv:2412.01883}\ } (\bibinfo {year} {2024})}\BibitemShut {NoStop}%
\bibitem [{\citenamefont {Yang}\ \emph {et~al.}(2021)\citenamefont {Yang},
  \citenamefont {Li}, \citenamefont {Duan},\ and\ \citenamefont
  {Xu}}]{Yang2021}%
  \BibitemOpen
  \bibfield  {author} {\bibinfo {author} {\bibfnamefont {Y.-B.}\ \bibnamefont
  {Yang}}, \bibinfo {author} {\bibfnamefont {K.}~\bibnamefont {Li}}, \bibinfo
  {author} {\bibfnamefont {L.-M.}\ \bibnamefont {Duan}}, \ and\ \bibinfo
  {author} {\bibfnamefont {Y.}~\bibnamefont {Xu}},\ }\href {\doibase
  10.1103/PhysRevB.103.085408} {\bibfield  {journal} {\bibinfo  {journal}
  {Physical Review B}\ }\textbf {\bibinfo {volume} {103}},\ \bibinfo {pages}
  {085408} (\bibinfo {year} {2021})}\BibitemShut {NoStop}%
\bibitem [{\citenamefont {Li}\ \emph {et~al.}(2020)\citenamefont {Li},
  \citenamefont {Fu}, \citenamefont {Hu}, \citenamefont {Li},\ and\
  \citenamefont {Shen}}]{PhysRevLett.125.166801}%
  \BibitemOpen
  \bibfield  {author} {\bibinfo {author} {\bibfnamefont {C.-A.}\ \bibnamefont
  {Li}}, \bibinfo {author} {\bibfnamefont {B.}~\bibnamefont {Fu}}, \bibinfo
  {author} {\bibfnamefont {Z.-A.}\ \bibnamefont {Hu}}, \bibinfo {author}
  {\bibfnamefont {J.}~\bibnamefont {Li}}, \ and\ \bibinfo {author}
  {\bibfnamefont {S.-Q.}\ \bibnamefont {Shen}},\ }\href {\doibase
  10.1103/PhysRevLett.125.166801} {\bibfield  {journal} {\bibinfo  {journal}
  {Phys. Rev. Lett.}\ }\textbf {\bibinfo {volume} {125}},\ \bibinfo {pages}
  {166801} (\bibinfo {year} {2020})}\BibitemShut {NoStop}%
\bibitem [{\citenamefont {Benalcazar}\ and\ \citenamefont
  {Cerjan}(2022)}]{Benalcazar2022}%
  \BibitemOpen
  \bibfield  {author} {\bibinfo {author} {\bibfnamefont {W.~A.}\ \bibnamefont
  {Benalcazar}}\ and\ \bibinfo {author} {\bibfnamefont {A.}~\bibnamefont
  {Cerjan}},\ }\href {\doibase 10.1103/PhysRevLett.128.127601} {\bibfield
  {journal} {\bibinfo  {journal} {Physical Review Letters}\ }\textbf {\bibinfo
  {volume} {128}},\ \bibinfo {pages} {127601} (\bibinfo {year}
  {2022})}\BibitemShut {NoStop}%
\bibitem [{\citenamefont {Kang}\ \emph {et~al.}(2019)\citenamefont {Kang},
  \citenamefont {Shiozaki},\ and\ \citenamefont {Cho}}]{Kang2019}%
  \BibitemOpen
  \bibfield  {author} {\bibinfo {author} {\bibfnamefont {B.}~\bibnamefont
  {Kang}}, \bibinfo {author} {\bibfnamefont {K.}~\bibnamefont {Shiozaki}}, \
  and\ \bibinfo {author} {\bibfnamefont {G.~Y.}\ \bibnamefont {Cho}},\ }\href
  {\doibase 10.1103/PhysRevB.100.245134} {\bibfield  {journal} {\bibinfo
  {journal} {Physical Review B}\ }\textbf {\bibinfo {volume} {100}},\ \bibinfo
  {pages} {245134} (\bibinfo {year} {2019})}\BibitemShut {NoStop}%
\bibitem [{\citenamefont {Wheeler}\ \emph {et~al.}(2019)\citenamefont
  {Wheeler}, \citenamefont {Wagner},\ and\ \citenamefont
  {Hughes}}]{Wheeler2019}%
  \BibitemOpen
  \bibfield  {author} {\bibinfo {author} {\bibfnamefont {W.~A.}\ \bibnamefont
  {Wheeler}}, \bibinfo {author} {\bibfnamefont {L.~K.}\ \bibnamefont {Wagner}},
  \ and\ \bibinfo {author} {\bibfnamefont {T.~L.}\ \bibnamefont {Hughes}},\
  }\href {\doibase 10.1103/PhysRevB.100.245135} {\bibfield  {journal} {\bibinfo
   {journal} {Physical Review B}\ }\textbf {\bibinfo {volume} {100}},\ \bibinfo
  {pages} {245135} (\bibinfo {year} {2019})}\BibitemShut {NoStop}%
\bibitem [{\citenamefont {Resta}(1998)}]{Resta1998}%
  \BibitemOpen
  \bibfield  {author} {\bibinfo {author} {\bibfnamefont {R.}~\bibnamefont
  {Resta}},\ }\href {\doibase 10.1103/PhysRevLett.80.1800} {\bibfield
  {journal} {\bibinfo  {journal} {Physical Review Letters}\ }\textbf {\bibinfo
  {volume} {80}},\ \bibinfo {pages} {1800} (\bibinfo {year}
  {1998})}\BibitemShut {NoStop}%
\bibitem [{\citenamefont {Peng}\ \emph {et~al.}(2017)\citenamefont {Peng},
  \citenamefont {Bao},\ and\ \citenamefont {von Oppen}}]{Peng2017}%
  \BibitemOpen
  \bibfield  {author} {\bibinfo {author} {\bibfnamefont {Y.}~\bibnamefont
  {Peng}}, \bibinfo {author} {\bibfnamefont {Y.}~\bibnamefont {Bao}}, \ and\
  \bibinfo {author} {\bibfnamefont {F.}~\bibnamefont {von Oppen}},\ }\href
  {\doibase 10.1103/PhysRevB.95.235143} {\bibfield  {journal} {\bibinfo
  {journal} {Physical Review B}\ }\textbf {\bibinfo {volume} {95}},\ \bibinfo
  {pages} {235143} (\bibinfo {year} {2017})}\BibitemShut {NoStop}%
\bibitem [{Note4()}]{Note4}%
  \BibitemOpen
  \bibinfo {note} {This holds for this particular model. It has been shown
  \protect \citep {Benalcazar2022} that in the case of chiral symmetric
  quadrupole insulators, topological phases defined by $q_{xy}=P=0$ can occur,
  where zero energy corner modes are present with much higher degeneracy.
  Although these are topologically non trivial, they display zero quadrupole
  moment.}\BibitemShut {Stop}%
\bibitem [{\citenamefont {Wei\ss{}e}\ \emph {et~al.}(2006)\citenamefont
  {Wei\ss{}e}, \citenamefont {Wellein}, \citenamefont {Alvermann},\ and\
  \citenamefont {Fehske}}]{Weisse2006}%
  \BibitemOpen
  \bibfield  {author} {\bibinfo {author} {\bibfnamefont {A.}~\bibnamefont
  {Wei\ss{}e}}, \bibinfo {author} {\bibfnamefont {G.}~\bibnamefont {Wellein}},
  \bibinfo {author} {\bibfnamefont {A.}~\bibnamefont {Alvermann}}, \ and\
  \bibinfo {author} {\bibfnamefont {H.}~\bibnamefont {Fehske}},\ }\href
  {\doibase 10.1103/RevModPhys.78.275} {\bibfield  {journal} {\bibinfo
  {journal} {Rev. Mod. Phys.}\ }\textbf {\bibinfo {volume} {78}},\ \bibinfo
  {pages} {275} (\bibinfo {year} {2006})}\BibitemShut {NoStop}%
\bibitem [{\citenamefont {MacKinnon}\ and\ \citenamefont
  {Kramer}(1983)}]{MacKinnon1983}%
  \BibitemOpen
  \bibfield  {author} {\bibinfo {author} {\bibfnamefont {A.}~\bibnamefont
  {MacKinnon}}\ and\ \bibinfo {author} {\bibfnamefont {B.}~\bibnamefont
  {Kramer}},\ }\href {\doibase 10.1007/BF01578242} {\bibfield  {journal}
  {\bibinfo  {journal} {Zeitschrift f{\{}{\"{u}}{\}}r Physik B Condensed
  Matter}\ }\textbf {\bibinfo {volume} {53}},\ \bibinfo {pages} {1} (\bibinfo
  {year} {1983})}\BibitemShut {NoStop}%
\bibitem [{\citenamefont {Mackinnon}\ and\ \citenamefont
  {Kramer}(1981)}]{MK81}%
  \BibitemOpen
  \bibfield  {author} {\bibinfo {author} {\bibfnamefont {A.}~\bibnamefont
  {Mackinnon}}\ and\ \bibinfo {author} {\bibfnamefont {B.}~\bibnamefont
  {Kramer}},\ }\href@noop {} {\bibfield  {journal} {\bibinfo  {journal} {Phys.
  Rev. Lett.}\ }\textbf {\bibinfo {volume} {47}},\ \bibinfo {pages} {1546}
  (\bibinfo {year} {1981})}\BibitemShut {NoStop}%
\bibitem [{\citenamefont {Wegner}(1980)}]{wegner1980}%
  \BibitemOpen
  \bibfield  {author} {\bibinfo {author} {\bibfnamefont {F.}~\bibnamefont
  {Wegner}},\ }\href {\doibase 10.1007/BF01325284} {\bibfield  {journal}
  {\bibinfo  {journal} {Zeitschrift f{\"u}r Physik B Condensed Matter}\
  }\textbf {\bibinfo {volume} {36}},\ \bibinfo {pages} {209} (\bibinfo {year}
  {1980})}\BibitemShut {NoStop}%
\bibitem [{Note5()}]{Note5}%
  \BibitemOpen
  \bibinfo {note} {Notice that for $\xi _{l}>30$, the localization length of
  the corner state is of the order of the linear size of the entire
  system.}\BibitemShut {Stop}%
\bibitem [{\citenamefont {Kraus}\ \emph
  {et~al.}(2012{\natexlab{b}})\citenamefont {Kraus}, \citenamefont {Lahini},
  \citenamefont {Ringel}, \citenamefont {Verbin},\ and\ \citenamefont
  {Zilberberg}}]{Kraus_2012}%
  \BibitemOpen
  \bibfield  {author} {\bibinfo {author} {\bibfnamefont {Y.~E.}\ \bibnamefont
  {Kraus}}, \bibinfo {author} {\bibfnamefont {Y.}~\bibnamefont {Lahini}},
  \bibinfo {author} {\bibfnamefont {Z.}~\bibnamefont {Ringel}}, \bibinfo
  {author} {\bibfnamefont {M.}~\bibnamefont {Verbin}}, \ and\ \bibinfo {author}
  {\bibfnamefont {O.}~\bibnamefont {Zilberberg}},\ }\href {\doibase
  10.1103/physrevlett.109.106402} {\bibfield  {journal} {\bibinfo  {journal}
  {Physical Review Letters}\ }\textbf {\bibinfo {volume} {109}} (\bibinfo
  {year} {2012}{\natexlab{b}}),\ 10.1103/physrevlett.109.106402}\BibitemShut
  {NoStop}%
\bibitem [{\citenamefont {Imhof}\ \emph {et~al.}(2018)\citenamefont {Imhof},
  \citenamefont {Berger}, \citenamefont {Bayer}, \citenamefont {Brehm},
  \citenamefont {Molenkamp}, \citenamefont {Kiessling}, \citenamefont
  {Schindler}, \citenamefont {Lee}, \citenamefont {Greiter}, \citenamefont
  {Neupert},\ and\ \citenamefont {Thomale}}]{Imhof2018}%
  \BibitemOpen
  \bibfield  {author} {\bibinfo {author} {\bibfnamefont {S.}~\bibnamefont
  {Imhof}}, \bibinfo {author} {\bibfnamefont {C.}~\bibnamefont {Berger}},
  \bibinfo {author} {\bibfnamefont {F.}~\bibnamefont {Bayer}}, \bibinfo
  {author} {\bibfnamefont {J.}~\bibnamefont {Brehm}}, \bibinfo {author}
  {\bibfnamefont {L.~W.}\ \bibnamefont {Molenkamp}}, \bibinfo {author}
  {\bibfnamefont {T.}~\bibnamefont {Kiessling}}, \bibinfo {author}
  {\bibfnamefont {F.}~\bibnamefont {Schindler}}, \bibinfo {author}
  {\bibfnamefont {C.~H.}\ \bibnamefont {Lee}}, \bibinfo {author} {\bibfnamefont
  {M.}~\bibnamefont {Greiter}}, \bibinfo {author} {\bibfnamefont
  {T.}~\bibnamefont {Neupert}}, \ and\ \bibinfo {author} {\bibfnamefont
  {R.}~\bibnamefont {Thomale}},\ }\href {\doibase 10.1038/s41567-018-0246-1}
  {\bibfield  {journal} {\bibinfo  {journal} {Nature Physics}\ }\textbf
  {\bibinfo {volume} {14}},\ \bibinfo {pages} {925} (\bibinfo {year}
  {2018})}\BibitemShut {NoStop}%
\bibitem [{\citenamefont {Mittal}\ \emph {et~al.}(2019)\citenamefont {Mittal},
  \citenamefont {Orre}, \citenamefont {Zhu}, \citenamefont {Gorlach},
  \citenamefont {Poddubny},\ and\ \citenamefont {Hafezi}}]{Mittal2019}%
  \BibitemOpen
  \bibfield  {author} {\bibinfo {author} {\bibfnamefont {S.}~\bibnamefont
  {Mittal}}, \bibinfo {author} {\bibfnamefont {V.~V.}\ \bibnamefont {Orre}},
  \bibinfo {author} {\bibfnamefont {G.}~\bibnamefont {Zhu}}, \bibinfo {author}
  {\bibfnamefont {M.~A.}\ \bibnamefont {Gorlach}}, \bibinfo {author}
  {\bibfnamefont {A.}~\bibnamefont {Poddubny}}, \ and\ \bibinfo {author}
  {\bibfnamefont {M.}~\bibnamefont {Hafezi}},\ }\href {\doibase
  10.1038/s41566-019-0452-0} {\bibfield  {journal} {\bibinfo  {journal} {Nature
  Photonics}\ }\textbf {\bibinfo {volume} {13}},\ \bibinfo {pages} {692}
  (\bibinfo {year} {2019})}\BibitemShut {NoStop}%
\end{thebibliography}%

\appendix

\section{Constant $W$ Cut\label{Sec:constat_W_cut}}

\begin{figure}[H]
\begin{centering}
\includegraphics[width=1\columnwidth]{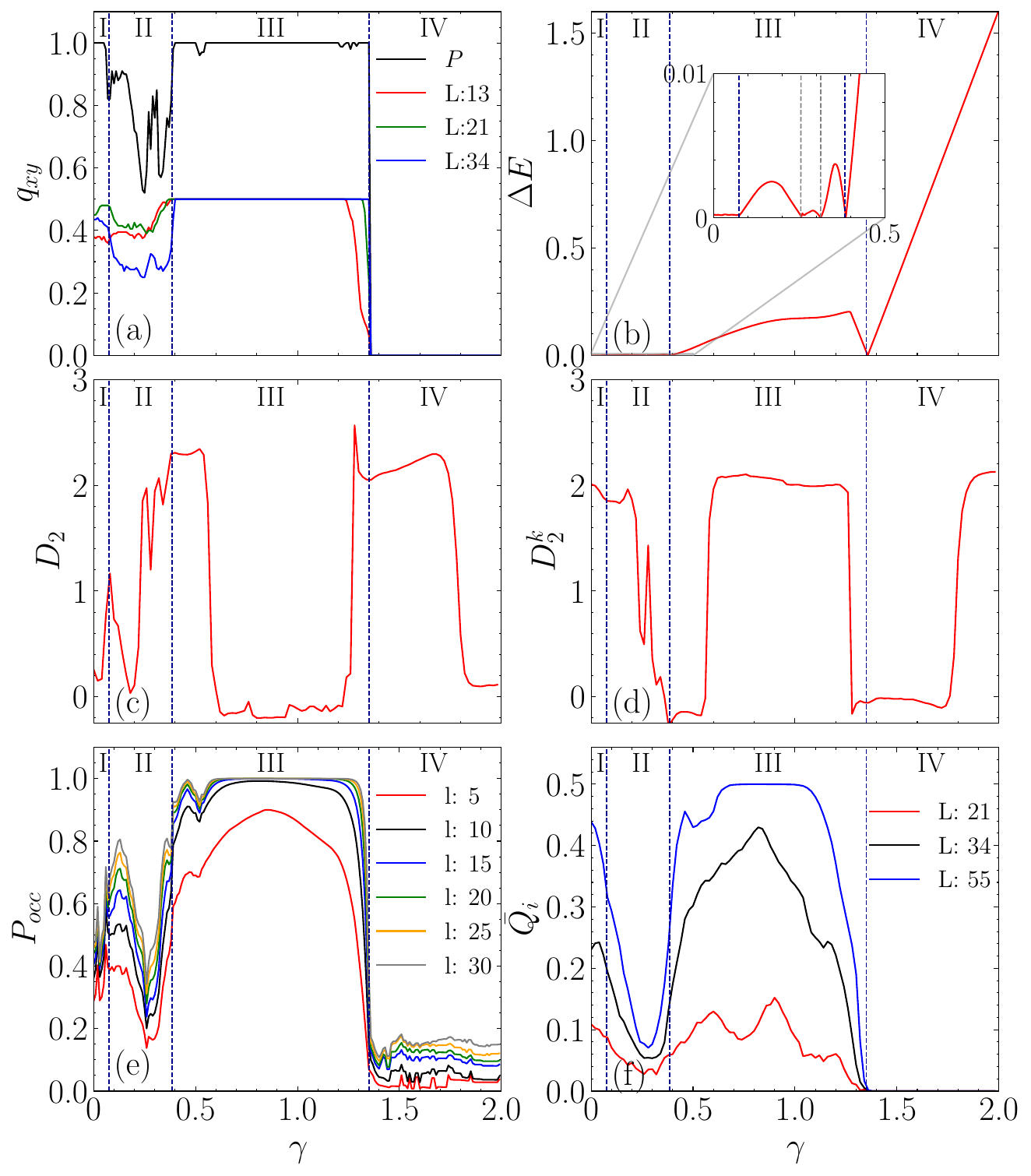}
\par\end{centering}
\caption{(a) $q_{xy}$ and $P$ as a function of $\gamma$. The $P$ invariant
was computed for a system size of $L=233$ with $100$ averages over
phase shifts. $q_{xy}$ was obtained via real space methods with $100$
averages over phase shifts. For even $L$, $\theta_{x}=\theta_{y}=0$;
for odd $L$ $\theta_{x}=\theta_{y}=0$. (b) Spectral gap as a function
of $\gamma$ computed for an even system size $L=144\protect\implies\theta_{i}=0$
and averaged over $50$ phase shifts realizations. (c-d) Fractal dimension
averaged out over $100$ phase twists, shifts and over the first $8$
eigenstates with energies closest to $E=0$. (e) Corner occupation
probability ($P_{occ}$) as a function of $\gamma$ for the first
$2$ zero energy states. The results were obtained for a system size
of $L=377$ and averaged over $100$ phase shifts. (f) Corner charge
($\bar{Q}$) as a function of $\gamma$ for several system sizes and
averaged over $100$ phase shifts and over the $4$ corners. (e-f)
In all plots we set $\delta=10^{-5}$ and $W=3$. \label{fig:W3}}
\end{figure}

In the main text we have focused our analysis in two constant $\gamma$
cuts. However, a constant $W=3$ cut was also performed as seen in
Fig.~\ref{fig:phase-diagram}. We plot several quantities as a function
of hopping strength ($\gamma$), such as the quadrupole moment and
boundary invariant (a), the bulk spectral gap (b), fractal dimensions
(c-d), corner occupation probability (e) and the corner charge (f).

The most interesting result is regarding phase $I$ which is a gapless
QPQI phase. Although this regime is gapless, the states at Fermi level
are localized which allows for the existence of localized topological
corner modes at zero energy. Despite the topological nature of this
regime, $P_{occ}$ does not quantize since numerical exact diagonalization
of the system Hamiltonian yield linear combinations of localized corner
modes and localized bulk states that live at zero energy. Phase $II$
displays several trivial gapped regimes with either a localized or
critical gap edge. From phase $II$ to $III$ the gap closes in a
ballistic gap-edge into a gapped QI phase with well quantized $q_{xy}$
and $P$. Regarding the corner occupation probability, it plateaus
for most of phase $III$, however at around $\gamma\approx0.5$ it
displays two dips that are associated with corner edge hybridization.
Furthermore if we consider the constant $\gamma=0.5$ cut and set
$W=3$ (end of phase $I$, Fig.~\ref{fig:qxy_P_vs_W_and spectral gap}(a)),
a small dip can also be observed in the $P$ invariant, in $P_{occ}$
and in $\bar{Q}$. Phase $IV$ is a band gap trivial insulator.

\section{Edge and Corner States via DOS\label{Sec:DOS-edge}}

\begin{figure*}[!t]
\begin{centering}
\includegraphics[width=2\columnwidth]{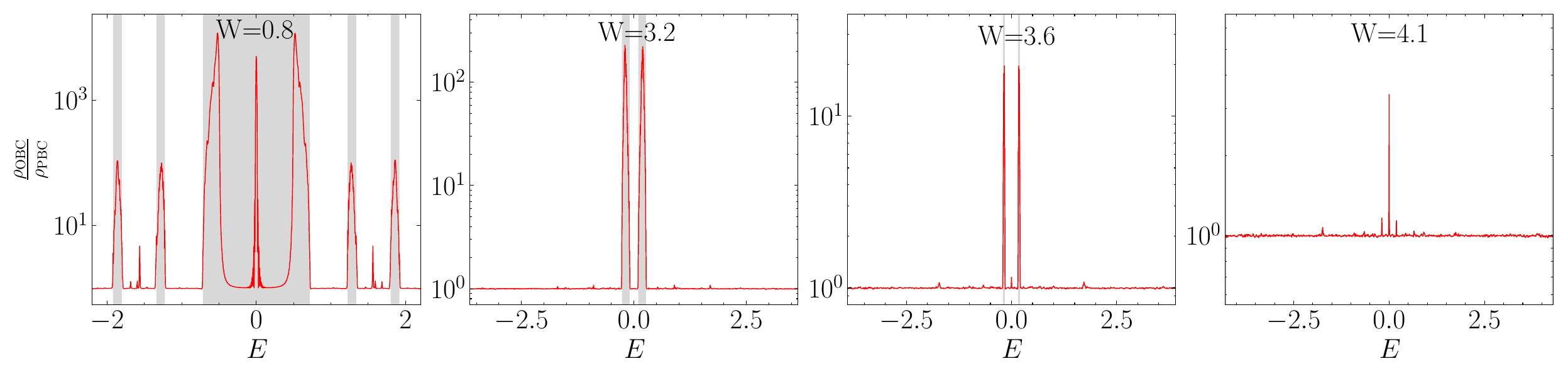}
\par\end{centering}
\caption{Ratio of open boundary DOS ($\rho_{OBC}$) and periodic boundaries
DOS ($\rho_{PBC}$) as a function of energy for $W$ in different
phases and $\gamma=0.5$. The greyed out regions correspond to bulk
gaps opened by the QP modulation. The following KPM parameters were
used: $L=987$, $M=5000$, $R=10$. \label{fig:ratio_DOS_0.5}}
\end{figure*}

\begin{figure*}
\begin{centering}
\includegraphics[width=2\columnwidth]{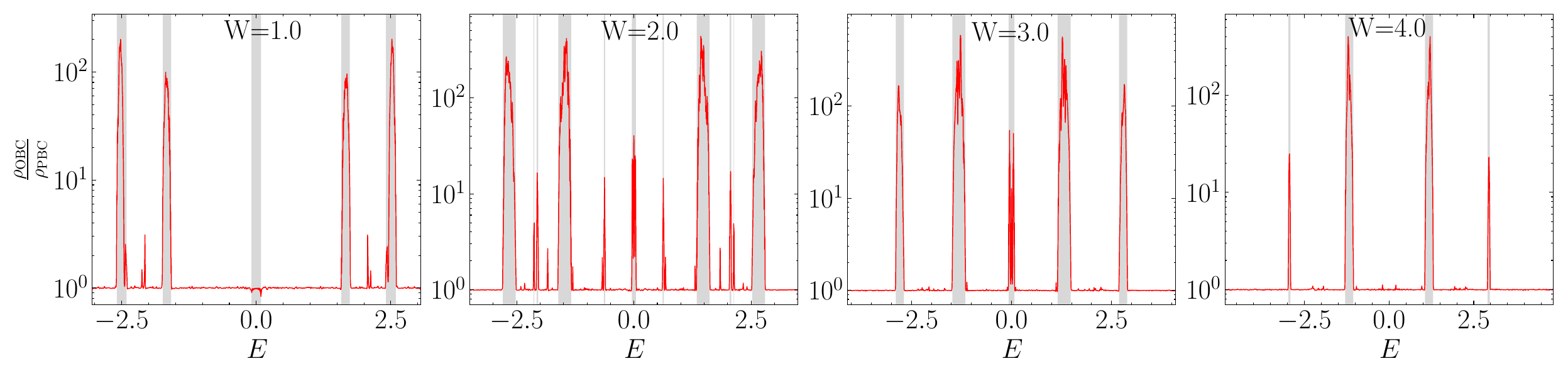}
\par\end{centering}
\caption{Ratio of open boundary DOS ($\rho_{OBC}$) and periodic boundaries
DOS ($\rho_{PBC}$) as a function of energy for $W$ in different
phases and $\gamma=1.1$. The greyed out regions correspond to bulk
gaps opened by the QP modulation. The following KPM parameters were
used: $L=987$, $M=5000$, $R=10$. \label{fig:ratio_DOS_1.1}}
\end{figure*}

Another quantity that is of use is the ratio of DOS defined as $\rho_{OBC}/\rho_{PBC}$.
In bulk regions $\rho_{OBC}/\rho_{PBC}\to1$, however in gapped regions
where corner and edge states may live it is expected that $\rho_{OBC}/\rho_{PBC}\gg1$.This
quantity is of particular interest since we can use it to look for
topological corner modes at zero energy. In this manner, in a topological
phase, a peak of $\rho_{OBC}/\rho_{PBC}$ at zero energy should be
observed. In Fig.~\ref{fig:ratio_DOS_0.5} and Fig.~\ref{fig:ratio_DOS_1.1}
we plot this quantity as a function of $E$ for several $W$'s in
each of the observed phases in the two different constant $\gamma$
cuts.

\section{Phase Twists and Spectral Gap\label{sec:Phase-Twist-and-spectral-gap}}

Introducing phase twists leads to shifts in the allowed $\mathbf{k}$'s,
such that if a twist $\theta_{i}$ is introduced then $\mathbf{k}_{i}=\left(-\frac{\pi}{L}+\frac{2\pi}{L}n\right)\mathbf{b}_{i}$.
We can add a twist that shifts all the allowed $k_{x}$'s in such
a way that for odd $L$, $k_{n}=0$ is a possible value of crystalline
momenta. This is achieved by considering $\theta_{x}=\pi$ for odd
$L$ while for even $L$ we use regular periodic boundary conditions
which is the same as considering twists with $\theta_{x}=0$. This
trick is also relevant for gap dependent results as topological phases
and TPT. A particular choice of twists is also relevant in some disordered
systems, although it is not clear that in the absence of translational
symmetry, using phase twists is advantageous. In fact, even for some
disordered system, phase twists solve spectral gap finite size effects.
quasiperiodic systems in the low $W$ regime maintain ballistic states
at Fermi level and so setting a non zero twist can improve gap calculations,
especially when the gap closes in a ballistic-ballistic transition.
In contrast regular uncorrelated disorder does not benefit from phase
twists, since all of the electronic states are Anderson localized
(for low dimensional systems).

\section{Phase Shift Dependence\label{sec:phase shift dependences}}

\begin{figure}
\centering{}\includegraphics[width=1\columnwidth]{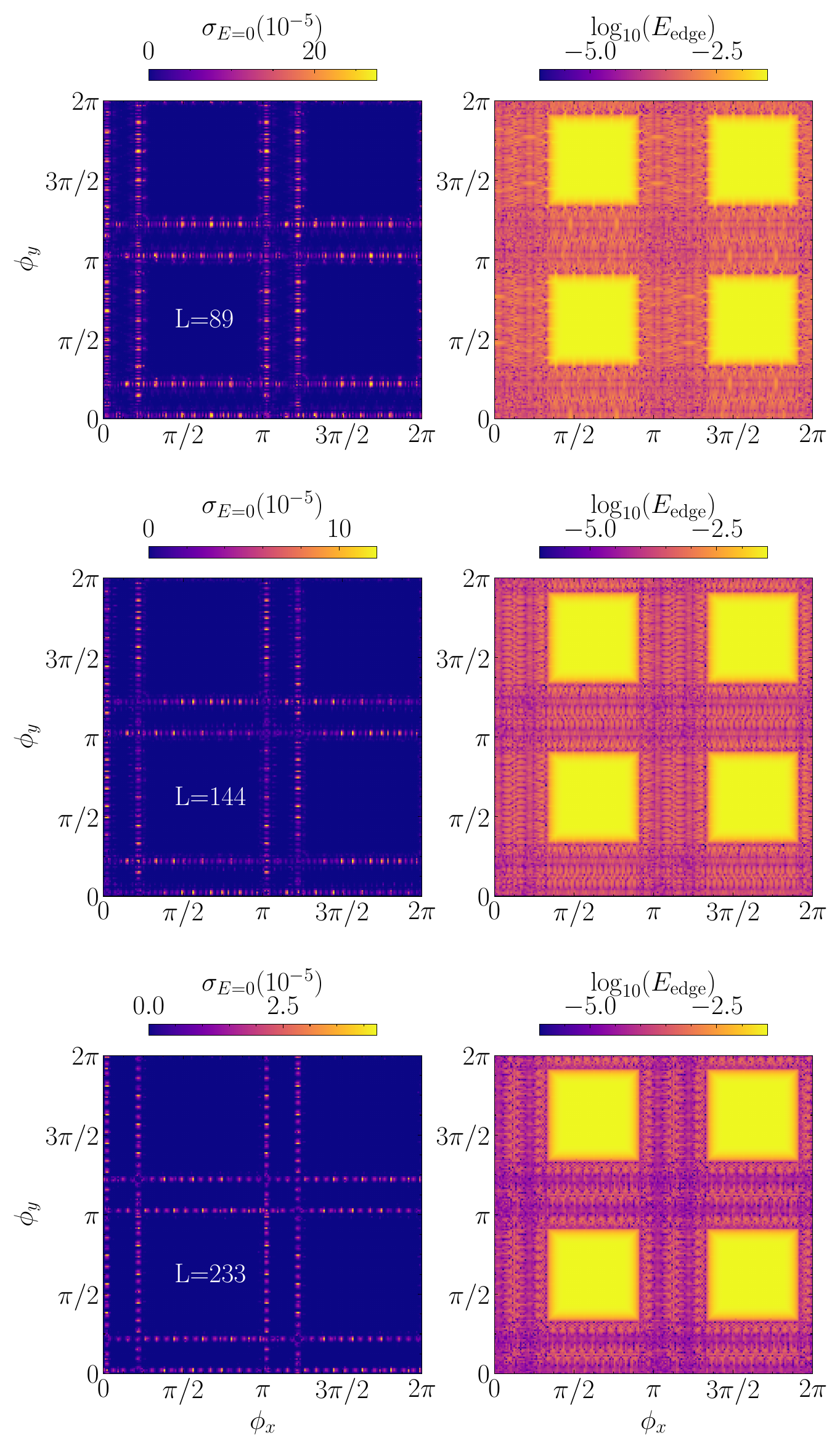}\caption{Standard deviation ($\sigma_{E=0}$) of the energies of the corner
modes as a function of the phase shifts ($\phi_{x},\phi_{y}$) and
logarithm of edge state energy ($\log_{10}(E_{edge})$) as a function
of the phase shifts ($\phi_{x},\phi_{y}$). Results obtained for $W=3.8$
and $\gamma=0.2$ in the QPQI phase for a system sizes $L=\{89,144,233\}$
as indicated.\label{fig:Standard-deviation_energy_expanded}}
\end{figure}

In Fig.~\ref{fig:Standard-deviation_energy_expanded} we expand on
the results of Fig.~\ref{fig:IPR_edge_outer_inner} by plotting the
standard deviation ($\sigma_{E=0}$) and logarithm of edge state energy
($\log_{10}(E_{edge})$) as a function of the phase shifts ($\phi_{x},\phi_{y}$)
for smaller system sizes.

\end{document}